\documentclass[prd,preprint,eqsecnum,nofootinbib,amsmath,amssymb,tightenlines]{revtex4}
\usepackage{graphicx}
\usepackage{bm}
\usepackage{amsfonts}

\def\alphas{\alpha_{\rm s}}
\def\alphaqed{\alpha_{\scriptscriptstyle\rm EM}}
\def\Nc{N_{\rm c}}
\def\Nf{N_{\rm f}}
\def\lstop{\ell_{\rm stop}}
\def\eps{\epsilon}
\def\qhat{\hat q}
\def\qhatA{\hat q_{\rm A}}
\def\eff{{\rm eff}}
\def\stop{{\rm stop}}
\def\qhatAeff{\qhatA^{\,\eff}}
\def\CA{C_{\rm A}}
\def\Re{\operatorname{Re}}

\def\fac{{\rm fac}}
\def\LO{{\rm LO}}
\def\NLO{{\rm NLO}}
\def\net{{\rm net}}
\def\min{{\rm min}}
\def\max{{\rm max}}
\def\net{{\rm net}}
\def\BIM{{\rm BIM}}

\def\virtI{{\rm virt\,I}}
\def\virtII{{\rm virt\,II}}
\def\classI{{\rm class\,I}}
\def\classII{{\rm class\,II}}
\def\MSbar{\overline{\mbox{MS}}}
\def\NLObar{\overline{\rm NLO}}
\def\renlog{{\rm ren\,log}}
\def\gammaE{\gamma_{\rm\scriptscriptstyle E}}

\def\b{{\bm b}}
\def\B{{\bm B}}
\def\Bbar{{\bar{\cal B}}}
\def\Bzero{{\cal B}_0}
\def\Avg{\operatorname{Avg}}
\def\dAvg{\operatorname{\delta Avg}}
\def\calW{{\cal W}}
\def\altx{\mathfrak x}
\def\baralphas{\overline{\alpha}_{\rm s}}

\begin {document}
%%%%%%%%%%%%%%%%%%%%%%%%%%%%%%%%%%%%%%%%%%%%%%%%%%%%%%%%%%%%%%%%%%%%%%%%%%%%%%%

\title
    {
      The LPM effect in sequential bremsstrahlung:
      gluon shower development
    }

\author{Peter Arnold}
\author{Omar Elgedawy}
\affiliation
    {%
    Department of Physics,
    University of Virginia,
    Charlottesville, Virginia 22904-4714, USA
    }%
\author{Shahin Iqbal}
\affiliation
    {%
    National Centre for Physics,
    Quaid-i-Azam University Campus,
    Islamabad, 45320 Pakistan
    }%

\date {\today}

\begin {abstract}%
{%
  We give details of our study
  %\cite{finale}
  of whether high-energy gluon showers inside a QCD medium can be treated
  as a sequence of individual splitting processes $g \to gg$, or whether
  there is significant quantum overlap between where one splitting ends and
  the next begins (neglecting effects that can be absorbed into an effective
  value of the jet quenching parameter $\hat q$ that characterizes the
  medium).  The study is carried out by imagining in-medium gluon shower
  development in the simplest theoretical situation,
  which includes imagining
  a very large, static, homogeneous medium and taking the large $\Nc$ limit.
  Along the way, we also show how in-medium shower evolution can be written
  in terms of a ``net'' splitting rate $[d\Gamma/dx]_\net$,
  and we provide a moderately
  simple analytic fit to our numerical results for
  the overlap effects included in that rate, which we
  hope may be of use to others wishing to study possible consequences of
  overlapping splittings.
}%
\end {abstract}

\maketitle

\thispagestyle {empty}

{\def\boldmath{}\tableofcontents}
\newpage

%%%%%%%%%%%%%%%%%%%%%%%%%%%%%%%%%%%%%%%%%%%%%%%%%%%%%%%%%%%%%%%%%%%%%%%%%%%%%%%

\section{Introduction}

When passing through matter, high energy particles lose energy by
showering, via the splitting processes of hard bremsstrahlung and pair
production.  At very high energy, the quantum mechanical duration of
each splitting process, known as the formation time, exceeds the mean
free time for collisions with the medium, leading to a significant
reduction in the splitting rate known as the Landau-Pomeranchuk-Migdal
(LPM) effect \cite{LP1,LP2,Migdal}.%
\footnote{
  The papers of Landau and Pomeranchuk \cite{LP1,LP2} are also available in
  English translation \cite{LPenglish}.
}
A long-standing problem in field theory has
been to understand how to implement this effect in cases where
the formation times of two consecutive splittings overlap.
Several authors \cite{Blaizot,Iancu,Wu} previously analyzed this issue
for QCD at leading-log order, which arises from the limit where one
bremsstrahlung gluon is soft compared to the other very-high energy
partons.
They found large effects at high energy, but those effects could be
absorbed into an effective value $\qhat_{\rm eff}$ of the medium parameter
$\qhat$ that encodes the rate of transverse momentum kicks to a high-energy
particle by the medium.
In a short companion paper \cite{finale}, which should be read first,
we motivated and outlined a method for
investigating the size of overlapping formation time effects that
{\it cannot}\/ be absorbed into $\qhat$, and we presented selected results.
The purpose of the current paper is to provide details of the methods and
derivations used in ref.\ \cite{finale}, and to provide a more complete
exposition of results.

As described in ref.\ \cite{finale}, our focus will be on computing
the statistically averaged distribution $\eps(z)$ of energy deposited
in the medium by a gluon shower initiated by a very high-energy gluon with
energy $E_0$ that starts at the origin traveling in the $z$ direction.
We will be particularly focused on overlapping
formation time corrections to the {\it shape} of that distribution,
\begin{equation}
   S(Z) \equiv
   \frac{\langle z \rangle}{E_0} \, \eps\bigl( \langle z \rangle Z \bigr)
   ,
\label {eq:shape}
\end{equation}
where
\begin {equation}
  \langle z \rangle \equiv \frac{1}{E_0} \int_0^\infty dz\> z \, \eps(z)
\end {equation}
is the characteristic length of the shower (of parametric order
$\alphas^{-1} \sqrt{E_0/\qhat}$\,), and $Z \equiv z/\langle z \rangle$.

Our results will all be derived in terms of what we call the net
rate $[d\Gamma/dx]_{\rm net}$ for splitting \cite{qcd}, defined as the
rate for splittings (including the case of two overlapping splittings)
to produce one
daughter of energy $xE$ plus any other daughters from a parent of
energy $E$.  Formulas for overlapping formation time effects appearing in
the net rate, developed in
refs.\ \cite{2brem,seq,dimreg,4point,QEDnf,qcd,qcdI},
are extremely long and
complicated.  They are also time-consuming to evaluate numerically.
In this paper, we will
present a relatively simple function that
fits well our numerical results (at first order in overlap effects)
for $[d\Gamma/dx]_{\rm net}$.
We need this quick-to-evaluate fit function to make our analysis of
the shape function $S(Z)$ numerically practical, but perhaps
others may find the fit function useful as well.

% --------------------------------------------------------------------------

\subsection {Assumptions}

For the sake of theoretical simplicity, we make the assumptions
outlined in ref.\ \cite{finale}, which mostly follow those of the
underlying rate calculations developed in
refs.\ \cite{2brem,seq,dimreg,4point,QEDnf,qcd,qcdI}.
We assume a homogeneous, static medium large enough to stop the shower;%
\footnote{
  The underlying rate calculations of 
  refs.\ \cite{2brem,seq,dimreg,4point,QEDnf,qcd,qcdI}
  only assumed that that medium was approximately static and homogeneous over
  the formation time and corresponding formation length.
  The analysis in this paper is made simpler by assuming that it's
  static and homogeneous over the entire development of the shower.
}
a nearly on-shell initial gluon;
transverse momentum transfer from the medium described by the
multiple-scattering ($\qhat$) approximation;
the large-$\Nc$ limit, and so purely gluonic showers.

There was yet another simplifying
assumption, made implicitly in ref.\ \cite{finale},
which we should be explicit about here.
To first order in high-energy radiative corrections,
write the effective value of $\qhat$ as
$\qhat_\eff = \qhat_{(0)} + \delta q$.
Here, $\qhat_{(0)}$ is what we might call the bare value of $\hat q$ ---
the value from scatterings of a high-energy parton
with the medium that are not accompanied by high-energy splitting.
In our analysis, we will treat $\qhat_{(0)}$ as a constant, independent
of energy.
There are caveats and counter-caveats concerning
logarithmic dependence of that approximation,
which we will simply ignore in this paper.%
\footnote{
  For example, for fixed-coupling calculations for a weakly-coupled
  medium, the large-$q_\perp$
  Rutherford tail $d\Gamma_{\rm el}/d(q_\perp^2) \propto \alphas^2 n/q_\perp^4$
  of the elastic scattering cross-section
  causes logarithmic dependence of $\langle q_\perp^2 \rangle$ on
  the upper scale of $q_\perp$ relevant to the process under
  consideration.  On the other hand, including running of $\alphas$
  as $d\Gamma_{\rm el}/d(q_\perp^2) \propto \alphas^2(q_\perp) n/q_\perp^4$
  is enough to eventually tame that dependence if the relevant
  upper scale $Q_\perp$ for $q_\perp$ is large enough that
  $\alphas(Q_\perp)$ is small compared to the strength of
  $\alphas$ at the scale of the medium.
  (See, for example, section VI.B of ref.\ \cite{DeepLPM}, which
  combined earlier observations of refs.\ \cite{BDMPS3} and \cite{Peshier}.)
}

In principle, the analysis of this paper can be applied to any
sufficiently thick QCD medium where the $\hat q$ approximation is
appropriate.  However, our own interest is ultimately motivated by quark-gluon
plasmas (QGPs), and so we will sometimes use that language.
In that context, we are making no assumption about whether the
coupling $\alphas(T)$ of the QGP is large or small --- all of the
details of the QGP are hidden away in the value of $\hat q_{(0)}$.
We will, however, work perturbatively in the size of the $\alphas(\mu)$
associated with a high-energy splitting vertex, for which the transverse
momentum scale is parametrically $\mu \sim (\hat q\omega)^{1/4}$,
where $\omega$ is the energy of the softest daughter.

Throughout this paper, we will only focus on the high-energy particles
($E \gg T$) in showers.  We ignore thermal gluon masses for
the high-energy gluons in our (purely gluonic) showers.

% --------------------------------------------------------------------------

\subsection {Outline}

The next section briefly summarizes the calculation of overlapping splitting
rates, previously worked out in
refs.\ \cite{2brem,seq,dimreg,4point,QEDnf,qcd,qcdI}, and explains
how the results of that work are packaged into results for different
types of rates (\ref{eq:rates}).

Section \ref{sec:dGnet} describes, and presents results for, the net
rate $[d\Gamma/dx]_\net$ that will be used throughout the rest of the
paper.  We first review how rates can be combined into the net rate.
The net rate is split into leading-order (BDMPS-Z) and next-to-leading-order
(overlap) pieces.
We review logarithmic infrared divergences of the net rate,
due to soft radiative corrections to hard splittings $g{\to}gg$, and
then factorize out those soft radiative corrections as described
in ref.\ \cite{finale}.
Numerical results, and an analytic fit, are presented
for overlap corrections to $[d\Gamma/dx]_\net$.
The section concludes with discussion of how to convert
$[d\Gamma/dx]_\net$ between different choices of factorization scale.

In principle, the factorized soft radiative corrections should be
resummed and absorbed into an effective value $\qhat_\eff$ of $\qhat$,
and that change will affect the effective ``leading-order'' development of
the shower.  Section \ref{sec:LOvEff} argues that this complication can
be ignored in our calculation.  This point is somewhat non-trivial and
requires partial discussion of resumming soft radiative corrections
to $\qhat$ at
next-to-leading-log order (NLLO);
the current state of the art is leading-log order.

Section \ref{sec:epsEquation} provides the starting point for
our analysis of shower energy deposition by showing that the
deposited energy distribution $\eps(z)$ satisfies an integro-differential
equation (\ref{eq:epseq})
in terms of the net splitting rate $[d\Gamma/dx]_\net$.
Since our goal is to study aspects of showers that are as insensitive as
possible to physics that can be absorbed into the effective value of
$\hat q$, our ultimate interest will be to follow ref.\ \cite{finale}
and study the shape $S(Z)$ of $\eps(z)$ given by (\ref{eq:shape}).

Numerically, the features of $\eps(z)$ that are easiest to calculate
are its moments $\langle z^n \rangle$.
Section \ref{sec:moments} presents a recursion relation
(\ref{eq:zn}) for those moments in terms of integrals of
$[d\Gamma/dx]_\net$.
These are then converted to various moments $\langle Z^n \rangle$ of
the shape function $S(Z)$.  Our interest lies in the relative size of
overlap corrections to those moments, which will be presented in
table \ref{tab:shape}.  We will find that most overlap corrections
are very small, but the fourth cumulant of $S(Z)$ turns out to be
very sensitive to overlap effects.

In order to convince ourselves that overlap effects on the shape function
are very small, regardless of the sensitivity of the fourth cumulant,
section \ref{sec:shape} turns away from moments and
takes on the more numerically complicated task
of directly calculating the size of overlap corrections to
the full $S(Z)$ as a function of $Z$, summarized in fig.\ \ref{fig:S}.
As prequel to this next-to-leading-order calculation, we also
provide what, as far as we know, are
the first full {\it leading}-order (BDMPS-Z) numerical
calculations of $\eps(z)$ and $S(Z)$,
and we compare those to what they would be in the
instructive Blaizot/Iancu/Mehtar-Tani analytic model for (leading-order)
showers \cite{BIM1,BIM2}.

Section \ref{sec:time} demonstrates that the ability to
analyze showers in terms of $[d\Gamma/dx]_\net$ is not restricted to
just energy deposition but also applies more generally to the
time development of the gluon distribution of the shower.
This generalizes leading-order versions of shower
evolution equations used by others \cite{BIM1,BIM2}.
But we have not made any attempt to simulate our evolution equation.

The results we find are that overlap effects on $S(Z)$ are very small ---
much smaller than related effects previously computed for
large-$\Nf$ QED \cite{qedNfstop}.
Section \ref{sec:why} attempts to give some crude, incomplete,
after-the-fact analysis of why the results of the two calculations are
so qualitatively different, which generates questions for future work.

In section \ref{sec:error}, we discuss what cross-checks are available
for our calculation of overlap effects.  Then we offer short concluding remarks
in section \ref{sec:conclusion}.

% ============================================================================

\section{Review of the building blocks: Splitting rates}

\subsection {Diagrams}

\begin {figure}[t]
\begin {center}
  \includegraphics[scale=0.6]{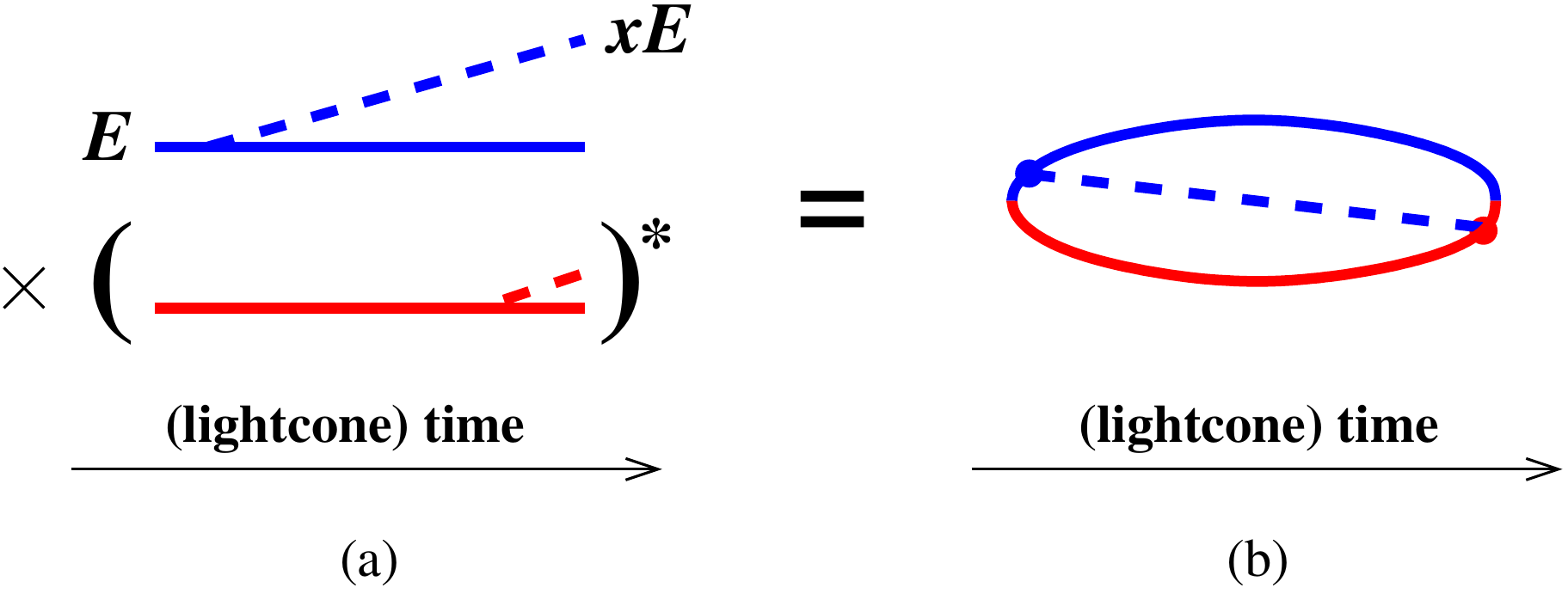}
  \caption{
     \label{fig:LO}
     (a) A time-ordered contribution to the LO rate for single splitting
     $g \to gg$, with
     amplitude in blue and conjugate amplitude in red.
     (b) A single diagram representing this contribution to the rate.
     In both cases, all lines implicitly interact
     with the medium.  We need not follow
     particles after the emission has occurred in both the amplitude
     and conjugate amplitude because we will consider only the
     $p_\perp$-integrated rate.
     (See, for example, section 4.1 of
     ref.\ \cite{2brem} for a more explicit argument,
     although applied there to a more complicated diagram.)
     Nor need we follow them before
     the first emission because we approximate the initial particle
     as on-shell.
     Only one of the two time orderings that contribute to the
     LO rate is shown above.
  }
\end {center}
\end {figure}

The calculation of the LPM effect was generalized from QED to QCD by
Baier, Dokshitzer, Mueller, Peigne, and Schiff \cite{BDMPS1,BDMPS2,BDMPS3}
and Zakharov \cite{Zakharov1,Zakharov2} (BDMPS-Z).  When specialized to an
infinite medium in the $\hat q$ approximation, their formalism gives the
in-medium $g{\to}gg$ splitting rate%
\footnote{
  It's difficult to figure out whom to reference for the first appearance
  of (\ref{eq:LOrate0}).  BDMS \cite{BDMS} give the $q{\to}qg$ formula
  in their eq.\ (42b) [with the relevant limit here being the infinite
  volume limit $\tau_0 \to \infty$ for their time $\tau_0$].  They then
  discuss elements of the $g{\to}gg$ case after that but don't quite
  give an explicit formula for the entire rate.  (They are not explicit
  about the formula for $\omega_0$.)
  Zakharov makes a
  few general statements about the $g{\to}gg$ case
  after eq.\ (75) of ref.\ \cite{Zakharov3}.
  As an example from ten years later, the explicit formula is
  given by eqs.\ (2.26) and (4.6) of ref.\ \cite{simple} in the
  case where $s$ represents a gluon.
}
\begin {equation}
  \left[ \frac{d\Gamma}{dx} \right]^\LO
  = \frac{\alphas P_{g\to gg}(x)}{2\pi}
    \sqrt{ \frac{(1{-}x{+}x^2) \qhatA}{x(1{-}x)E} }
\label {eq:LOrate0}
\end {equation}
for energies $E \to xE+(1{-}x)E$.
The subscript on $\qhatA$ indicates the $\hat q$ appropriate
for the adjoint color representation, i.e.\ for gluons, and
$\CA{=}\Nc$ is the adjoint-representation quadratic Casimir.
$P_{g\to gg}(x)$ is the
Dokshitzer-Gribov-Lipatov-Altarelli-Parisi (DGLAP)
splitting function.%
\footnote{
  Our
  $P_{g{\to}gg}(x) = 2 \CA (1 - x + x^2)^2/x(1-x)$
  does ${\it not}$\/ contain the pieces of the usual DGLAP splitting function
  used to include the effect of virtual diagrams.
  In particular, the $1/(1{-}x)$ in our formula for $P_{g\to gg}$
  is just the ordinary function
  $1/(1{-}x)$ and not the
  distribution $1/(1{-}x)_+$, and our $P_{g\to gg}$ does not contain a
  $\delta$-function term $\delta(1{-}x)$.
  When we need to deal with virtual diagrams in this paper, we will do so
  explicitly.
}
We refer to (\ref{eq:LOrate0}) as the ``leading-order'' (LO) result for
$g{\to}gg$.  For us, leading order means leading order in the number of
high-energy splitting vertices and includes the effects of
an arbitrary number of interactions with the medium.
In the following discussion,
we will adopt Zakharov's picture \cite{Zakharov1,Zakharov2}
of LPM rate calculations, which is to think of the rate for $g{\to}gg$
as time-ordered diagrams, such as fig.\ \ref{fig:LO}, combining
the amplitude for $g{\to}gg$ (blue) with the conjugate amplitude (red).
Zakharov then thought of fig.\ \ref{fig:LO}b as three particles propagating
forward in time which, in the high-energy limit, could be described
(between the splitting vertices) as a 3-particle, two-dimensional
quantum mechanics problem in the transverse plane.  The
medium-averaged effect of interactions with the medium can
be described by a non-Hermitian, effective ``potential energy''
between the three particles in the quantum mechanics problem.
In this language, the $\hat q$ approximation corresponds to a
harmonic oscillator problem (with imaginary-valued spring constants).
For a discussion and review in the particular context of our problem
with our notation, see, for example, refs.\ \cite{2brem} and \cite{logs2}.

\begin {figure}[t]
\begin {center}
  \includegraphics[scale=0.6]{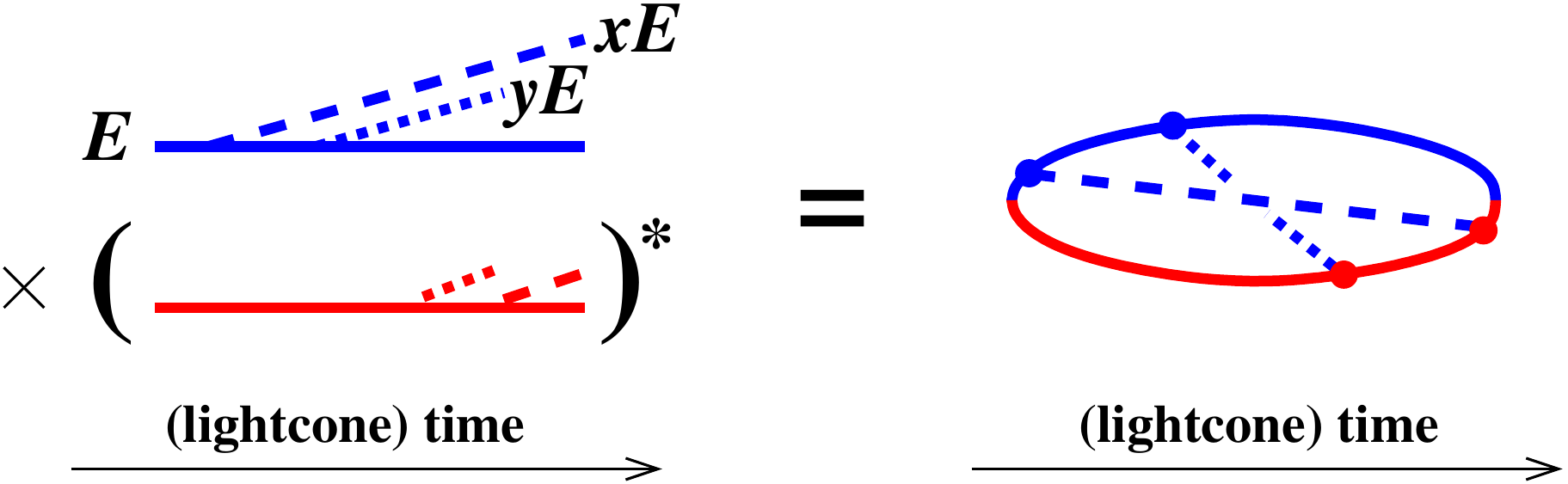}
  \caption{
     \label{fig:overlap}
     A particular example of two overlapping splittings.
  }
\end {center}
\end {figure}

\begin {figure}[t]
\begin {center}
  \includegraphics[scale=0.5]{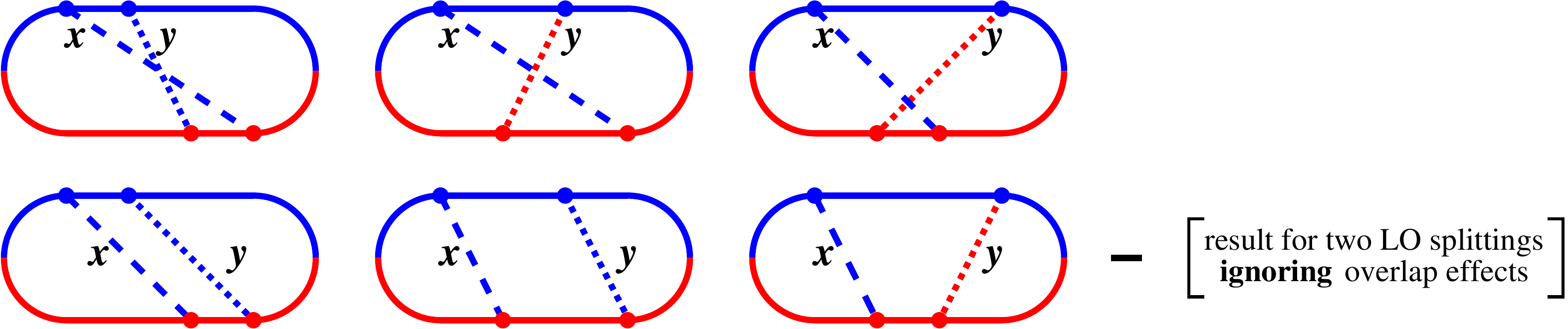}
  \caption{
     \label{fig:RealExamples}
     Examples of diagrams contributing to the effects of
     overlapping formation times for two splittings $g {\to} gg {\to} ggg$.
     The first and second rows (when combined with their conjugates and
     appropriate permutations of the daughters) were analyzed in
     refs.\ \cite{2brem} and \cite{seq}, respectively.
  }
\end {center}
\end {figure}

\begin {figure}[t]
\begin {center}
  \includegraphics[scale=0.5]{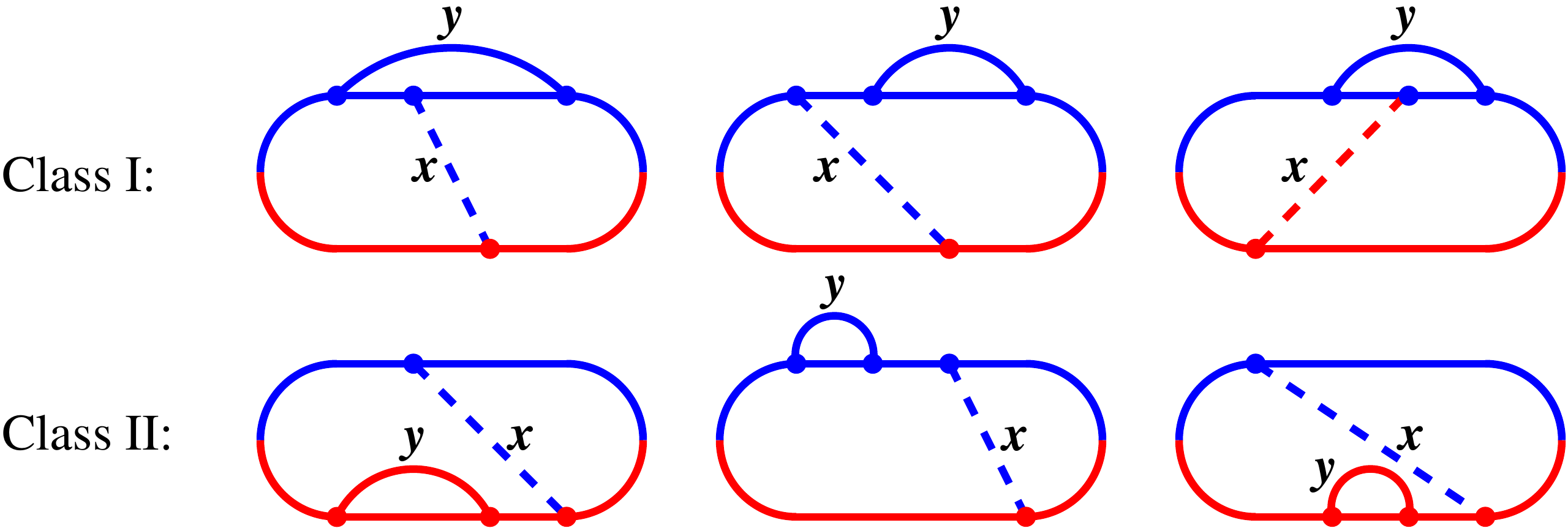}
  \caption{
     \label{fig:VirtExamples}
     Some examples from ref.\ \cite{qcd}
     of NLO virtual corrections to single splitting $g {\to} gg$.
  }
\end {center}
\end {figure}

\begin {figure}[t]
\begin {center}
  \includegraphics[scale=0.5]{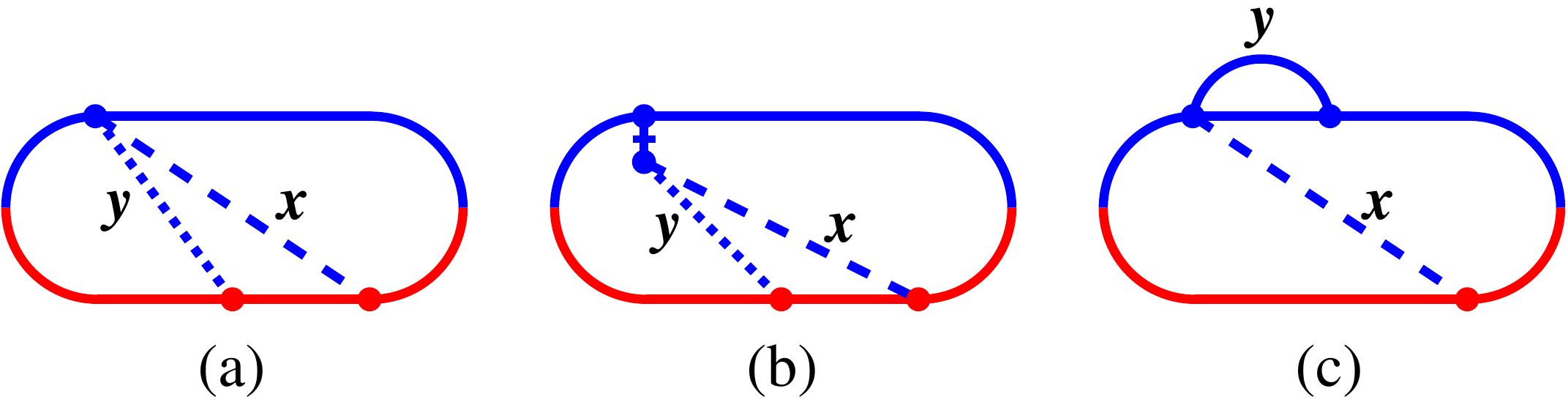}
  \caption{
     \label{fig:Fexamples}
     Some examples from ref.\ \cite{qcdI}
     that involve (a,c) a 4-gluon vertex or (b) exchange
     of a longitudinally polarized gluon (denoted by the vertical line
     crossed by a bar) in Light Cone Perturbation Theory (LCPT).
  }
\end {center}
\end {figure}

We refer to the effects of two overlapping $g{\to}gg$ splittings, such as
fig.\ \ref{fig:overlap}, as
one type of next-to-leading-order (NLO) effect.  Since there are four
high-energy splitting vertices in this rate diagram, it is suppressed
by one power of high-energy $\alphas(\mu)$ compared to the leading-order
splitting of fig.\ \ref{fig:LO}.
Fig.\ \ref{fig:RealExamples}
shows examples of diagrams
contributing to the rate,
drawn in the style of fig.\ \ref{fig:LO}b.
The subtraction in fig.\ \ref{fig:RealExamples} means that our
rates represent the {\it difference} between (i) a full calculation of
(potentially overlapping) $g \to gg \to ggg$ and (ii) approximating
a double splitting as two independent, consecutive single splittings
$g {\to} gg$
that each occur with the LO single splitting rate (\ref{eq:LOrate0}).%
\footnote{
  The key importance
  of this subtraction is explained in section 1.1 of
  ref.\ \cite{seq}.
}
At the same order in $\alphas(\mu)$, there are also NLO virtual corrections
to single splitting $g{\to}gg$, for which we show a few examples in
fig.\ \ref{fig:VirtExamples}.
Fig.\ \ref{fig:Fexamples} shows examples of
some more-direct $g{\to}ggg$ processes
that also contribute at the same order in $\alphas(\mu)$.
A complete list of all diagrams contained in our calculation may be
found in refs.\ \cite{qcd,qcdI}.

Throughout this paper, $\alphas$ will refer to high-energy $\alphas(\mu)$
unless stated otherwise.

% ---------------------------------------------------------------------------

\subsection {Notation for Rates}

Following ref.\ \cite{qcd}, we will refer to the leading-order $g{\to}gg$ rate,
its NLO correction, and the $g{\to}ggg$ rate as
\begin {equation}
   \left[ \frac{d\Gamma}{dx} \right]^{\rm LO} , 
   \qquad
   \left[ \Delta \frac{d\Gamma}{dx} \right]_{g \to gg}^{\rm NLO} ,
   \qquad
   \left[ \Delta \frac{d\Gamma}{dx\,dy} \right]_{g \to ggg} .
\label {eq:rates}
\end {equation}
The last one, $[ \Delta\,d\Gamma/dx\,dy ]_{g\to ggg}$,
represents both (i) overlap corrections to two consecutive
splittings, such as in fig.\ \ref{fig:RealExamples},
and (ii) processes involving direct $g{\to}ggg$, such as
figs.\ \ref{fig:Fexamples}a,b.
In both cases, energy is being split as $E \to xE + yE + (1{-}x{-}y)E$.
The symbol ``$\Delta$'' in front
of that rate is a reminder that it represents a {\it correction} to an
LO-based calculation of double splitting as two, consecutive,
independent $g{\to}gg$ splitting events.
$[ \Delta\,d\Gamma/dx ]_{g \to gg}^{\rm NLO}$ similarly represents
the corresponding virtual corrections to single splitting,
such as in figs.\ \ref{fig:VirtExamples} and \ref{fig:Fexamples}c.
In this case, energy is being split as $E \to xE + (1{-}x)E$.

Formulas for the rates (\ref{eq:rates})
are presented in refs.\ \cite{qcd,qcdI},%
\footnote{
  More specifically, see Appendix A of ref.\ \cite{qcd}, but supplement
  the formulas there as explained in Appendix A of ref.\ \cite{qcdI}
  in order to include diagrams like fig.\ \ref{fig:Fexamples}.
  Various pieces of these formulas are taken from earlier papers
  \cite{2brem,seq,dimreg,4point,QEDnf}.
}
which carried out the calculation in
Light Cone Perturbation Theory (LCPT).
We will be slightly sloppy with our terminology in this paper.
Technically, we should define $x$ and $y$ by the splitting of
lightcone longitudinal momentum:
$P^+ \to xP^+ + yP^+ + (1{-}x{-}y)P^+$ for $g \to ggg$ and
$P^+ \to xP^+ + (1{-}x)P^+$ for $g {\to} gg$.
But the splittings relevant to shower
development are high energy and nearly collinear, and so we may
also refer to $x$ and $y$ simply as ``energy fractions'' in our
applications.%
\footnote{
  More specifically, the difference between $p^+/P^+$ and $p^0/E$ is
  suppressed by
  $p_\perp^2/E^2 \sim \qhat t_{\rm form}/E^2 \sim \qhat^{1/2}/E^{3/2}$,
  and in all of our analysis we ignore effects that are suppressed by
  powers of $E$.
}

In the case of the virtual diagrams, the rate calculation involves integration
over the lightcone longitudinal momentum fraction
$y$ of one of the loop lines, as labeled in
figs.\ \ref{fig:VirtExamples} and \ref{fig:Fexamples}c.
One consequence of LCPT is that the $p^+$ of every (transverse-polarized)
gluon must be non-negative, which imposes constraints on the
allowed range of $y$ in the virtual diagrams.
Refs.\ \cite{qcd,qcdI} divide virtual diagrams into two classes.
Class I (such as the top line of fig.\ \ref{fig:VirtExamples}) means
that (i) $y$ should be integrated over $0 < y < 1{-}x$ and (ii) the
substitution $x \to 1{-}x$ generates a distinct set of diagrams that
must also be included.
Class II (such as the bottom line of fig.\ \ref{fig:VirtExamples}) means
that (i) $y$ should be integrated over $0 < y < 1$ and (ii) the
substitution $x \to 1{-}x$ does not generate any new diagrams.
With this nomenclature,
\begin {align}
   \left[ \Delta \frac{d\Gamma}{dx} \right]^\NLO_{g\to gg}
   &=
     \biggl(
       \left[ \Delta \frac{d\Gamma}{dx} \right]^\NLO_\classI
     \biggr)
     + (x \to 1{-}x)
   + \left[ \Delta \frac{d\Gamma}{dx} \right]^\NLO_\classII
\nonumber\\
   &=
     \biggl(
       \int_0^{1-x} dy \>
       \left[ \Delta \frac{d\Gamma}{dx\,dy} \right]^\NLO_\classI
     \biggr)
     + (x \to 1{-}x)
   + \int_0^1 dy \> \left[ \Delta \frac{d\Gamma}{dx\,dy} \right]^\NLO_\classII
   \,,
\label {eq:virt}
\end {align}
where the subscripts refer to Class I and
Class II virtual diagrams.%
\footnote{
  Following ref.\ \cite{qcd},
  our convention is that, when there is a loop in the amplitude (or a loop in
  the conjugate amplitude), the
  loop symmetry factor (if any) is already accounted for in the formulas for
  $[\Delta d\Gamma/dx\,dy]^\NLO_\classI$ and
  $[\Delta d\Gamma/dx\,dy]^\NLO_\classII$.
}
The virtual diagrams were computed with $\MSbar$ ultraviolet (UV)
renormalization, and so $\alphas(\mu)$ will refer to
the $\MSbar$ coupling in our work.

In this paper, we will need to do $y$ integrals numerically.
Ref.\ \cite{qcd} found it convenient to separate out
from the integrals in (\ref{eq:virt}) a piece containing the
renormalization scale $\mu$ dependence and to integrate that
piece analytically.
That's a choice, and a detail, that we leave to
appendix \ref{app:NLObar}, where the reader may find
the exact connection with the rate formulas as they are presented in
refs.\ \cite{qcd,qcdI}.%
\footnote{
  \label{foot:class}
  We've intentionally used subscript names
  ``$\classI$'' and ``$\classII$'' in (\ref{eq:virt}) that are
  different from those used in ref.\ \cite{qcd} to avoid confusing
  the formulas given there, where some pieces have been separated out,
  with the integrands in (\ref{eq:virt}),
  where they have not.  See appendix \ref{app:NLObar}.
}

In what follows, we will consider the shower as being made up of
$1{\to}2$ splittings and effective $1{\to}3$ splittings.
In that context, we find it convenient to use the notation
\begin {subequations}
\label {eq:dG1to23}
\begin {align}
  \left[ \frac{d\Gamma}{dx} \right]_{1\to2}
    \equiv
    \left[ \frac{d\Gamma}{dx} \right]^{\rm LO}
      + \left[ \Delta \frac{d\Gamma}{dx} \right]_{g \to gg}^{\rm NLO} ,
\\
  \left[ \frac{d\Gamma}{dx\,dy} \right]_{1\to3}
    \equiv
    \left[ \Delta \frac{d\Gamma}{dx\,dy} \right]_{g \to ggg} .
\label {eq:dG1to3}
\end {align}
\end {subequations}
Remember that, for simplicity, we are only considering purely gluonic
showers, and so the daughters of every splitting are identical particles.
Our convention is to not include final-state identical particle
factors in differential rates.  So, formally,
the total rate for any sort of $1{\to}2$ or $1{\to}3$ splittings
would be
\begin {equation}
  \Gamma =
  \frac{1}{2!} \int_0^1 dx \> \left[ \frac{d\Gamma}{dx} \right]_{1\to2}
  +
  \frac{1}{3!} \int_0^1 dx \int_0^{1-x} dy \>
    \left[ \frac{d\Gamma}{dx\,dy} \right]_{1\to3} ,
\label {eq:Gamma}
\end {equation}
or, equivalently,
\begin {equation}
  \Gamma =
  \int_{x<1-x} \hspace{-1.1em} dx \> \left[ \frac{d\Gamma}{dx} \right]_{1\to2}
  +
  \int_{y<x<1-x-y} \hspace{-2.3em} dx \> dy \>
    \left[ \frac{d\Gamma}{dx\,dy} \right]_{1\to3} ,
\end {equation}
We say ``formally'' because the total rate is infrared divergent.

We should note that the ``$1{\to}3$'' rate (\ref{eq:dG1to3})
can have either sign \cite{seq} because,
as mentioned earlier, part of it represents an overlap {\it correction} to
a shower of LO $1{\to}2$ splittings, and corrections may have either sign.

% ============================================================================

\section{\boldmath$[d\Gamma/dx]_\net$ and its factorization}
\label{sec:dGnet}

\subsection {Definition and Properties}

As mentioned earlier, we define
the ``net'' rate $[d\Gamma/dx]_\net$
as the probability per unit time that splittings of a
parent with energy $E$ create a daughter with energy $xE$ (along with any
other daughters).  For a shower
made up of $1{\to}2$ and $1{\to}3$
splittings,
\begin {equation}
  \left[ \frac{d\Gamma}{dx} \right]_\net =
  \left[ \frac{d\Gamma}{dx} \right]_{1\to2}
  +
  \frac{1}{2!} \int_0^{1-x} dy \>
    \left[ \frac{d\Gamma}{dx\,dy} \right]_{1\to3}
\label {eq:dGnet}
\end {equation}
if all the particles are identical (i.e.\ gluons in our case).
The reason for the $1/2!$ factor on the $1{\to}3$ terms is that one of
the three daughters has been distinguished as having energy
$xE$, but we don't want to double count the integration over the energies of
the other two (identical) daughters.

Note that the total rate (\ref{eq:Gamma}) is {\it not}\/ equal to
$\int dx \> [d\Gamma/dx]_\net$.  But one may show that
\begin {equation}
   \Gamma = \int_0^1 dx \> x \left[ \frac{d\Gamma}{dx} \right]_\net .
\label {eq:GammaAlt}
\end {equation}
To see this, use (\ref{eq:dGnet}) to write the right-hand side as
\begin {equation}
  \int_0^1 dx \> x \left[ \frac{d\Gamma}{dx} \right]_\net
  =
  \int_0^1 dx \> x \left[ \frac{d\Gamma}{dx} \right]_{1\to2}
  +
  \frac{1}{2!} \int_0^1 dx \> x \int_0^{1-x} dy \>
    \left[ \frac{d\Gamma}{dx\,dy} \right]_{1\to3} .
\label {eq:Gamma2}
\end {equation}
For the $1{\to}2$ integral in (\ref{eq:Gamma2}),
average (i) the integral with (ii) itself after the change of integration
variable $x \to 1{-}x$.  Since the daughters $(x,1{-}x)$ of the
splitting are identical particles,
$[d\Gamma/dx]_{1\to 2}$ does not change under $x \to 1{-}x$, and so
\begin {equation}
  \int_0^1 dx \> x \left[ \frac{d\Gamma}{dx} \right]_{1\to2}
  =
  \int_0^1 dx \> \frac{x + (1{-}x)}{2} \left[ \frac{d\Gamma}{dx} \right]_{1\to2}
  =
  \frac{1}{2} \int_0^1 dx \> \left[ \frac{d\Gamma}{dx} \right]_{1\to2} .
\label {eq:Gamma1to2}
\end {equation}
Do the same for the $1{\to}3$ integral in (\ref{eq:Gamma2}) except
average over (i) the original integral, (ii) $x \leftrightarrow y$, and
(iii) $x \leftrightarrow 1{-}x{-}y$.  These are just certain permutations of
the three identical daughters $(x,y,1{-}x{-}y)$,
and so $[d\Gamma/dx\,dy]_{1\to3}$ does not
change.  Comparing the resulting rewriting of (\ref{eq:Gamma2}) to
(\ref{eq:Gamma}) gives (\ref{eq:GammaAlt}).

% ---------------------------------------------------------------------------

\subsection {IR divergences and factorization}
\label{sec:fac}

As written, the definition (\ref{eq:dGnet}) of $[d\Gamma/dx]_\net$,
when applied to the
$1{\to}2$ and $1{\to}3$ processes (\ref{eq:dG1to23}), is plagued with
infrared divergences.
First, there are {\it power-law} infrared divergences associated with
the different boundaries ($0$, $1{-}x$, and $1$)
of the $y$ integrations in (\ref{eq:dGnet}) and
(\ref{eq:virt}), but these divergences cancel each other when all
added together.
It is possible to re-arrange the $y$ integrals so that
(i) the IR divergences (for fixed $x$) all become associated
with $y \to 0$ and (ii) the terms which generate power-law IR
divergences all cancel in the integrand.
Specifically, ref.\ \cite{qcd} showed that (\ref{eq:dGnet}) could
be rewritten as
\begin {equation}
  \left[ \frac{d\Gamma}{dx} \right]_\net =
  \left[ \frac{d\Gamma}{dx} \right]^\LO
  +
  \left[ \frac{d\Gamma}{dx} \right]^\NLO_\net
\end {equation}
with%
\footnote{
  See section 1.2 of ref.\ \cite{qcd}.
  Here we use a capital letter for the function $V$ to distinguish it
  from the lower-case function $v$ of ref.\ \cite{qcd}.  This is
  a technical point arising from our use of the full NLO virtual
  rates $[\Delta d\Gamma/dx\,dy]^\NLO_\classI$ and
  $[\Delta d\Gamma/dx\,dy]^\NLO_\classII$ in our discussion
  here, instead of
  their $\NLObar$ counterparts in ref.\ \cite{qcd}
  (where a piece including the renormalization
  scale dependence has been separated out).
  See footnote \ref{foot:class} and appendix \ref{app:NLObar}.
  We've also capitalized the function name $R$ for consistency of notation,
  but it is identical to the function $r$ in ref.\ \cite{qcd}.
}  
\begin {equation}
   \left[ \frac{d\Gamma}{dx} \right]_{\rm net}^{\NLO} =
   \int_0^{1/2} dy \>
   \Bigl\{
      V(x,y) \, \theta(y<\tfrac{1-x}{2})
      + V(1{-}x,y) \, \theta(y<\tfrac{x}{2})
      + R(x,y) \, \theta(y<\tfrac{1-x}{2})
   \Bigr\}
 ,
\label {eq:dGnetNLO}
\end {equation}
where
\begin {subequations}
\label {eq:VRdef}
\begin {align}
  V(x,y) &\equiv 
  \left(
     \left[ \Delta \frac{d\Gamma}{dx\,dy} \right]^\NLO_\classI
     + \left[ \Delta \frac{d\Gamma}{dx\,dy} \right]^\NLO_\classII
  \right)
  + ( y \leftrightarrow 1{-}x{-}y ) ,
\label {eq:Vdef}
\\
  R(x,y) & \equiv
  \left[ \Delta \frac{d\Gamma}{dx\,dy} \right]_{g\to ggg} .
\label {eq:Rdef}
\end {align}
\end {subequations}
The $\theta(\cdots)$ in (\ref{eq:dGnetNLO}) represent unit step
functions [$\theta(\mbox{true})=1$ and $\theta(\mbox{false})=0$],
and they just implement upper limits on the $y$ integration.
The advantage of using the $\theta$ functions is so that all
the integrals can be combined: the
integrals for the separate terms each have power-law
IR divergences, but their sum does not.

The explicit upper limit $1/2$ on the $y$ integral sign
$\int dy$ in (\ref{eq:dGnetNLO}) could just as well be replaced
by $\infty$ because
the {\it actual}\/ limits on various terms in the integrand are implemented by
the $\theta$ functions.  $1/2$ is simply the largest any of those
limits on $y$ could ever be.

  Though IR power-law divergences cancel, there remains an uncanceled
IR double-log divergence associated with $y\to 0$ in (\ref{eq:dGnetNLO}).
This is a double logarithm \cite{Blaizot,Iancu,Wu}
associated with soft radiative corrections
to an underlying, {\it hard}\/ single-splitting
process $[d\Gamma/dx]^\LO$.  It is essentially
the same double logarithm that was
originally discovered by considering radiative corrections to
$\hat q$ \cite{LMW}.  Physically, this double logarithm is cut off in
the infrared where the $\hat q$ approximation breaks down.
If one works exclusively in the $\hat q$ approximation, however,
the double log manifests as an infrared divergence that must be
regularized and/or subtracted.
Eq.\ (\ref{eq:dGnetNLO}) also generates a sub-leading, single logarithm
IR divergence that was extracted analytically in ref.\ \cite{logs}
and alternatively derived from the known
radiative corrections to $\hat q$ in
ref.\ \cite{logs2}.  The small-$y$ behavior of the integral in
(\ref{eq:dGnetNLO}) was found to be
\begin{equation}
   -\frac{\CA\alphas}{4\pi}
   \left[ \frac{d\Gamma}{dx} \right]^{\rm LO}
   \int_{y\ll \min(x,1-x)} \frac{dy}{y} \, \bigl[ \ln y + \bar s(x) \bigr]
\label {eq:IRlogs0}
\end{equation}
for fixed $x$, where
\begin {equation}
  \bar s(x) =
      - \ln\bigl(16\,x(1{-}x)(1{-}x{+}x^2)\bigr)
      + 2 \, \frac{ \bigl[
                 x^2 \bigl( \ln x - \frac{\pi}{8} \bigr)
                 + (1{-}x)^2 \bigl( \ln(1{-}x) - \frac{\pi}{8} \bigr)
               \bigr] }
             { (1-x+x^2) }
      .
\label {eq:sbar}
\end {equation}
For us, ``soft'' radiation means soft compared to both high-energy
daughters of the underlying LO splitting $E \to xE+(1{-}x)E$,
and so the small-$y$ approximation
used in (\ref{eq:IRlogs0}) is only valid for
$y \ll \min(x,1{-}x)$, which is parametrically equivalent to
$y \ll x(1{-}x)$.

$\bar s(x)$ diverges proportional to $\ln\bigl(x(1{-}x)\bigr)$ for
$x \to 0$ or $x \to 1$.  It's natural to rewrite the
$\ln y + \bar s(x)$ in
a way that combines the $\ln y$ and $\ln\bigl(x(1{-}x)\bigr)$ behavior:
\begin{equation}
   \ln y + \bar s(x)
   = \ln\Bigl( \frac{y}{x(1{-}x)} \Bigr) + \hat s(x)
\end{equation}
with
\begin {equation}
  \hat s(x) =
      - \ln\bigl(16(1{-}x{+}x^2)\bigr)
      + 2 \, \frac{ \bigl[
                 x^2 \bigl( \ln x - \frac{\pi}{8} \bigr)
                 + (1{-}x)^2 \bigl( \ln(1{-}x) - \frac{\pi}{8} \bigr)
               \bigr] }
             { (1-x+x^2) }
      .
\label {eq:shat}
\end {equation}
$\hat s(x)$ remains finite for $x{\to}0$ and $x{\to}1$.
It will also sometimes be useful to think of the integral (\ref{eq:IRlogs0})
in terms of energy and so rewrite it as
\begin{equation}
   -\frac{\CA\alphas}{4\pi}
   \left[ \frac{d\Gamma}{dx} \right]^{\rm LO}
   \int_{\omega_y \ll \min(x,1-x) E} \frac{d\omega_y}{\omega_y} \,
   \Bigl[
     \ln\Bigl( \frac{\omega_y}{x(1{-}x)E} \Bigr)
     + \hat s(x)
   \Bigr] ,
\label {eq:IRlogs}
\end{equation}
where $\omega_y \equiv yE$ is the energy of the soft $y$ daughter.

By itself, the integral in (\ref{eq:IRlogs}) is IR divergent and so
ultimately depends on the IR physics or IR regulator that cuts off
those divergences.  We will not be sensitive to the IR details
because we intend to
study infrared-safe
characteristics of the shower,
namely the {\it shape} (\ref{eq:shape}) of the energy deposition
distribution $\eps(z)$.  To this end, we will introduce an energy
factorization scale $\Lambda_\fac$ and separate the NLO contribution
to the net rate into
\begin {equation}
  \left[ \frac{d\Gamma}{dx} \right]^\NLO_\net =
  \left[ \frac{d\Gamma}{dx} \right]^{\NLO,\fac}_\net
  -
  \frac{\CA\alphas}{4\pi}
  \left[ \frac{d\Gamma}{dx} \right]^{\rm LO}
   \int_0^{\Lambda_\fac} \frac{d\omega_y}{\omega_y} \,
   \Bigl[
     \ln\Bigl( \frac{\omega_y}{x(1{-}x)E} \Bigr)
     + \hat s(x)
   \Bigr] ,
\label {eq:factorization}
\end {equation}
where the superscript ``fac'' above stands for ``factorized.''
The IR-subtracted net rate
\begin {multline}
   \left[ \frac{d\Gamma}{dx} \right]_{\rm net}^{\NLO,\fac} \equiv
   \int_0^\infty dy \>
   \biggl\{
      V(x,y) \, \theta(y<\tfrac{1-x}{2})
      + V(1{-}x,y) \, \theta(y<\tfrac{x}{2})
      + R(x,y) \, \theta(y<\tfrac{1-x}{2})
\\
      + \frac{\CA\alphas}{4\pi}
          \left[ \frac{d\Gamma}{dx} \right]^{\rm LO}
          \frac{\ln y + \bar s(x)}{y} \,
        \theta(yE < \Lambda_\fac)
   \biggr\}
\label {eq:dGfac}
\end {multline}
is then finite, and it can be computed numerically.

Our program is to absorb the last (IR-sensitive) term of
(\ref{eq:factorization}) into an effective value
$\hat q_\eff$ of $\hat q$ and so into an
effective value $[d\Gamma/dx]^{\LO}_\eff$ of the
leading-order $g{\to}gg$ splitting rate.
In principle, this simply shuffles the problem of IR-sensitive physics
to $[d\Gamma/dx]^\LO_\eff$.  Moreover, in principle,
the large double and single IR logarithms in $[d\Gamma/dx]^{\LO}_\eff$ would
then have to be tamed by a next-to-leading-log order (NLLO) resummation
of IR logarithms to all orders in $\alphas(\mu)$.
In practice, we will find that we can ignore the replacement of
$[d\Gamma/dx]^\LO$ by $[d\Gamma/dx]^\LO_\eff$ in evaluating
whether those overlap effects that cannot be absorbed into $\hat q$
are large or small.  In part, this is because constant shifts
$\delta \hat q$ to the value of $\hat q$ will, by design, have no effect
on the shape function (\ref{eq:shape}) --- that's precisely why we choose
to study the shape function.
In other part, it's because we will later show that changes that
could affect the leading-order shape function do not affect the
relative sizes NLO/LO of overlap effects at the order of our calculation.
For now, the upshot is that we will focus on the IR-subtracted version
(\ref{eq:dGfac}) of the net splitting rate.

Note that we've written the integral as $\int_0^\infty dy$ in
(\ref{eq:dGfac}).
However, the largest $y$ for which the integrand is non-zero
is $\max\bigl( x/2, (1{-}x)/2, \Lambda_\fac/E \bigr)$.

% ---------------------------------------------------------------------------

\subsection {Choice of factorization and renormalization scales}

\subsubsection{Our usual choice}

As previously noted, IR logarithms result from soft radiation with
energies $\omega_y$ up to the parametric scale
$\min\bigl(x,1{-}x\bigr)E$.  The choice of factorization scale that
subtracts as much of the IR logarithms as possible is then
$\Lambda_\fac \sim \min\bigl(x,1{-}x\bigr)E$, and our usual choice
will be
\begin {equation}
   \Lambda_\fac = \kappa x(1{-}x) E ,
\label {eq:Lambdafac1}
\end {equation}
where $\kappa$ is an $O(1)$ constant that we will canonically choose to be 1,
but which we will vary later.

Our UV renormalization scale $\mu$ should be chosen so that
the explicit $\alphas(\mu)$ in the leading-order splitting rate
$[d\Gamma/dx]^\LO$ (the $\alphas$ associated with the high-energy
splitting vertex) is evaluated at an appropriate physics scale to
account for anti-screening from virtual particle pairs present in
the vacuum.  During a formation time, the transverse separation $b$ of
the daughters of a $g{\to}gg$ splitting is of order
$(\qhat \omega)^{-1/4}$, where $\omega = \min(x,1{-}x)E$.
(Note that this is parametrically small compared to medium scales in
the high-energy limit.)  So we want $\alphas(1/b)$, which is
$\alphas(\mu)$ with $\mu \sim (\qhat\omega)^{1/4}$.
In terms of our choice (\ref{eq:Lambdafac1}), this is
$\mu \sim (\qhat\Lambda_\fac)^{1/4}$.
Rather than varying the exact choices of $\mu$ and $\Lambda_\fac$
separately, we will simply combine the two by choosing
\begin {equation}
   \Lambda_\fac = \kappa x(1{-}x) E ,
   \qquad
   \mu = (\qhatA\Lambda_\fac)^{1/4} .
\label {eq:Lambdafac}
\end {equation}

% -------------------------------------------------------------------------

\subsubsection{An alternate choice}
\label{sec:rE}

We will also consider another choice for comparison.
In our theorist's limit of arbitrarily high energy showers (and
an infinite-size medium), an underlying LO single splitting process
$g{\to}gg$, with $E \to xE+(1{-}x)E$,
should not affect where energy is deposited in the $z$ direction
in the limit
that the radiated energy fraction
$x$ (or $1{-}x$) is extremely small, since that soft $x$ gluon
deposits negligible energy.  So it won't matter if we make a poor
estimate of the size of the IR logarithms for the even-softer
radiative corrections to such an already-very-soft process.
{\it Parametrically}, we only
need do a reasonable job with choosing the factorization scale
for the case where $\min(x,1{-}x) \sim 1$.
So, though (\ref{eq:Lambdafac}) is a more physically sensible
choice,
one should in principle,
for the purpose of calculating $\eps(z)$ and then
its shape $S(Z)$,
be able to get away with choosing
\begin {equation}
   \Lambda_\fac = r E,
   \qquad
   \mu = (\qhatA \Lambda_\fac)^{1/4}
\label {eq:Lambdafacr}
\end {equation}
instead, where $r$ is an $O(1)$ constant.

We will later compare results using (\ref{eq:Lambdafac}) and
(\ref{eq:Lambdafacr}) to check the robustness of our conclusions about
the impact of overlap corrections that cannot be absorbed into
$\qhat$.
Note that, for a perfectly democratic splitting with $x=\tfrac12$,
our two different choices (\ref{eq:Lambdafac}) and (\ref{eq:Lambdafacr})
match up when $r = \kappa/4$.

% -------------------------------------------------------------------------

\subsection{Numerical results and fits for \boldmath$\Lambda_\fac = x(1-x)E$}
\label{sec:dGfacNums}

\begin {figure}[t]
\begin {center}
  \includegraphics[scale=0.5]{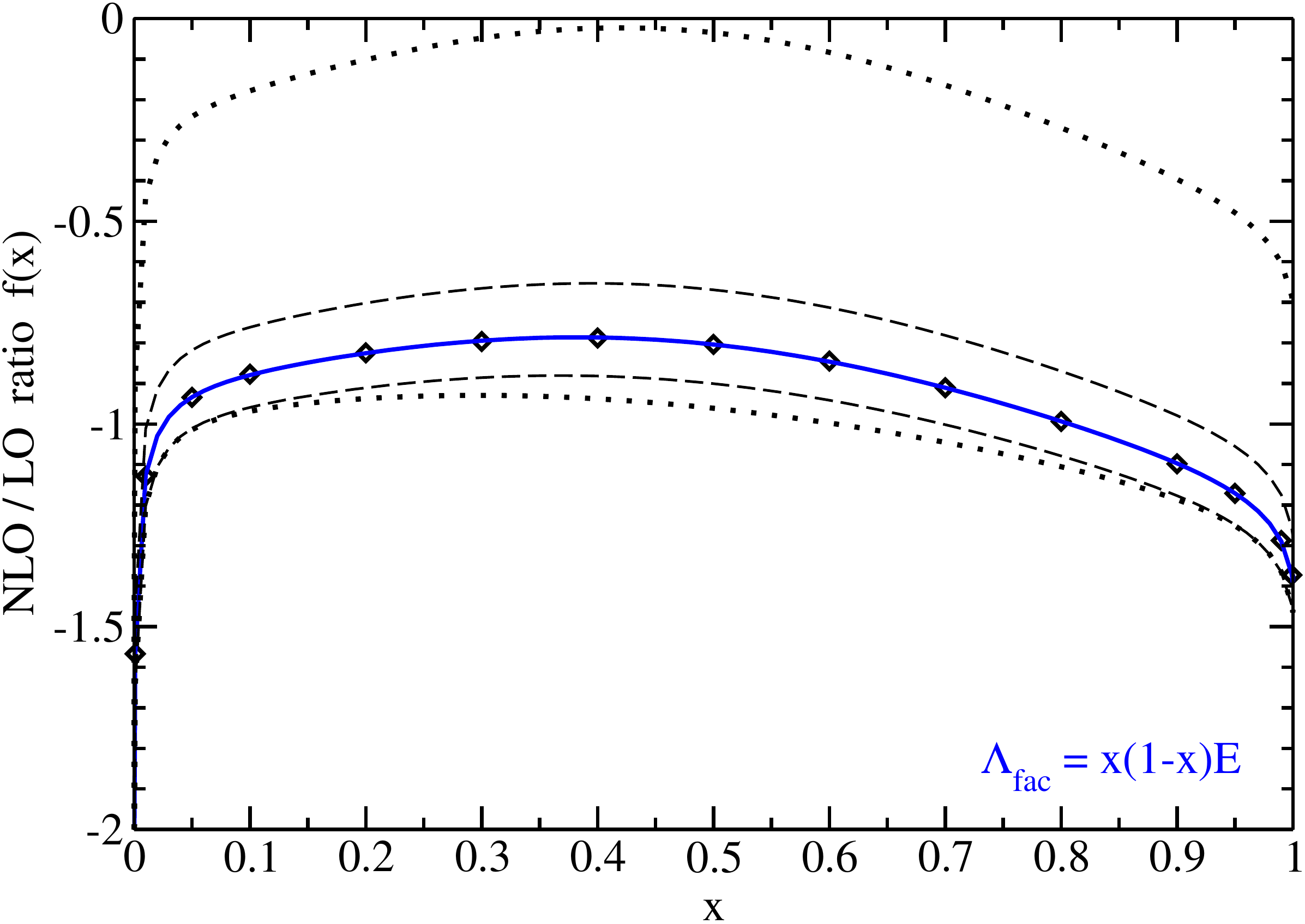}
  \caption{
     \label{fig:dGnet}
     Plot of the ratio (\ref{eq:dGnetRatio}) vs.\ $x$
     for $\Lambda_\fac = \kappa x(1-x)E$ and $\mu=(\qhatA \Lambda_\fac)^{1/4}$.
     The diamonds are numerically-computed data points for $\kappa=1$,
     and the solid curve is a fit (\ref{eq:fit}) to those points.
     For the sake of later discussion, the dashed lines show the results for
     $\kappa=\tfrac12$ (upper) and $\kappa=2$ (lower), and the dotted lines
     for $\kappa=\tfrac{1}{16}$ (upper) and $\kappa=16$ (lower).
  }
\end {center}
\end {figure}

Using (\ref{eq:dGfac}), with the rate formulas of
refs.\ \cite{qcd,qcdI} as described in appendix \ref{app:NLObar} of this
paper, and choosing $\Lambda_\fac = x(1{-}x)E$, 
we have numerically computed%
\footnote{
   See appendix \ref{app:dGfacNumerics} for some information on our
   numerical methods.
}
the values of $[d\Gamma/dx]_\net^{\NLO,\fac}$
represented by the data points in fig.\ \ref{fig:dGnet}
and in the last column of table \ref{tab:dGnet}.%
\footnote{
   The data points in Table \ref{tab:dGnet} and fig.\ \ref{fig:dGnetLog}
   that have extremely tiny $x$ or $1{-}x$
   are not intended to be relevant
   to any actual phenomenological situation,
   since our high-energy approximations
   fail when $xE$ or $(1{-}x)E$ are $\lesssim T$.
   They are included just for the
   purpose of understanding the asymptotic behavior of our formulas.
}
More specifically, the figure and table show the values of
\begin {equation}
   f(x) \equiv
   \frac{ \left[\frac{d\Gamma}{dx}\right]_\net^{\NLO,\fac} }
        { \CA\alphas \left[\frac{d\Gamma}{dx}\right]^\LO } \,,
\label{eq:dGnetRatio}
\end {equation}
where $[d\Gamma/dx]^\LO$ is given by (\ref{eq:LOrate0}).
It's convenient to plot this ratio not only to
see the relative size (in units of $\CA\alphas$) of the NLO
correction compared to the leading-order rate,
but also because both
the numerator and denominator blow up proportional to
$[x(1{-}x)]^{-3/2}$ (up to logarithms) as $x\to 0$ or $x\to 1$,
and so $f(x)$ is a smoother function than
$[d\Gamma/dx]_\net^{\NLO,\fac}$.

\begin {table}[tp]

\setlength{\tabcolsep}{7pt}
\begin{tabular}{llll}
\toprule
  \multicolumn{1}{c}{$x$} & \multicolumn{3}{c}{$f(x)$} \\
\cline{2-4}
      & \multicolumn{1}{c}{non-F}
      & \multicolumn{1}{c}{F diags}
      & \multicolumn{1}{c}{total} \\
\hline
%% resolving the +-1 in last digit of F diags using Shahin's results
0.0001  &-2.087 \\
0.001	&-1.525 	&-0.0425	&-1.568 \\
0.01	&-1.081 	&-0.0470	&-1.128 \\
0.05	&-0.8787	&-0.0551	&-0.9339 \\
0.1	&-0.8178	&-0.0586	&-0.8764 \\
0.2	&-0.7673	&-0.0571	&-0.8245 \\
0.3	&-0.7455	&-0.0509	&-0.7965 \\
0.4	&-0.7422	&-0.0459	&-0.7881 \\
0.5	&-0.7573	&-0.0463	&-0.8037 \\
0.6	&-0.7924	&-0.0530	&-0.8453 \\
0.7	&-0.8477	&-0.0625	&-0.9102 \\
0.8	&-0.9237	&-0.0697	&-0.9935 \\
0.9	&-1.0276	&-0.0697	&-1.0974 \\
0.95	&-1.1057	&-0.0653	&-1.1710 \\
0.99	&-1.228 	&-0.0577	&-1.286 \\
0.999	&-1.319 	&-0.0542	&-1.374 \\
0.9999  &-1.361 \\
\botrule
\end{tabular}
\caption{
   \label{tab:dGnet}
   Our numerical results for $f(x)$ for
   $\Lambda_\fac = x(1{-}x)E$ and $\mu = (\qhatA\Lambda_\fac)^{1/4}$.
   The last column shows values for the ratio
   (\ref{eq:dGnetRatio}), as plotted by the diamonds in fig.\ \ref{fig:dGnet}.
   The second column breaks out the contribution from only diagrams \cite{qcd}
   {\it without}\/ F=4+I vertices.  The third column is the contribution
   from diagrams \cite{qcdI} {\it with}\/ F=4+I vertices,
   which are shown by diamonds in fig.\ \ref{fig:dGnetF}.
   We estimate our numerical error in these results to be roughly
   $\pm1$ in the last digit for all entries
   {\it except}\/ the entries for $x=0.0001$ and $0.9999$
   [where we estimate $\pm$(a few) in the last digit].
   We expended computational effort
   to get the second-column entries for $x=0.0001$ and $0.9999$ in order
   to capture and fit the log behavior of (\ref{eq:fnonF}),
   but we did not see
   a need to expend similar effort for corresponding entries in the
   third column, which have been left blank.
}
\end{table}

The first thing to note about these results is that the relative size
of the (factorized) NLO contribution to $[d\Gamma/dx]_\net$ is a
roughly $\CA\alphas \times 100\%$ correction to $[d\Gamma/dx]^\LO$.
One would need $\CA\alphas = \Nc\alphas$ to be small for this to be
a small correction.
But remember that our motivation is to study overlap effects that
cannot be absorbed into $\hat q$.  If $f(x)$ were independent of $x$,
then, no matter how large $f$ was, the NLO corrections would simply
rescale the size of $[d\Gamma/dx]^\LO$, which could be absorbed by
rescaling the size of $\hat q$, which would have no effect on,
for example, the shape $S(Z)$ of the energy deposition distribution.
So what will be important about fig.\ \ref{fig:dGnet} is how it
varies with $x$, not its overall value.  We must wait until
we compute the NLO effect on the shape before we can draw conclusions.

The leading-order rate $[d\Gamma/dx]^\LO$ for $g{\to}gg$ is symmetric
under swapping the two daughters via $x \leftrightarrow 1{-}x$.
The second thing to note about fig.\ \ref{fig:dGnet} is that
$f(x)$ and so $[d\Gamma/dx]^\NLO_\fac$
are {\it not}\/ symmetric in $x \leftrightarrow 1{-}x$.
In general, $[d\Gamma/dx]_\net$ is not symmetric because
$1{\to}3$ processes are not.
Those processes (such as overlapping
$g {\to} gg {\to} ggg$) have three daughters; they are symmetric
under permutations of $(x,y,1{-}x{-}y)$ but not under
$x \leftrightarrow 1{-}x$.

\begin {figure}[t]
\begin {center}
  \includegraphics[scale=0.5]{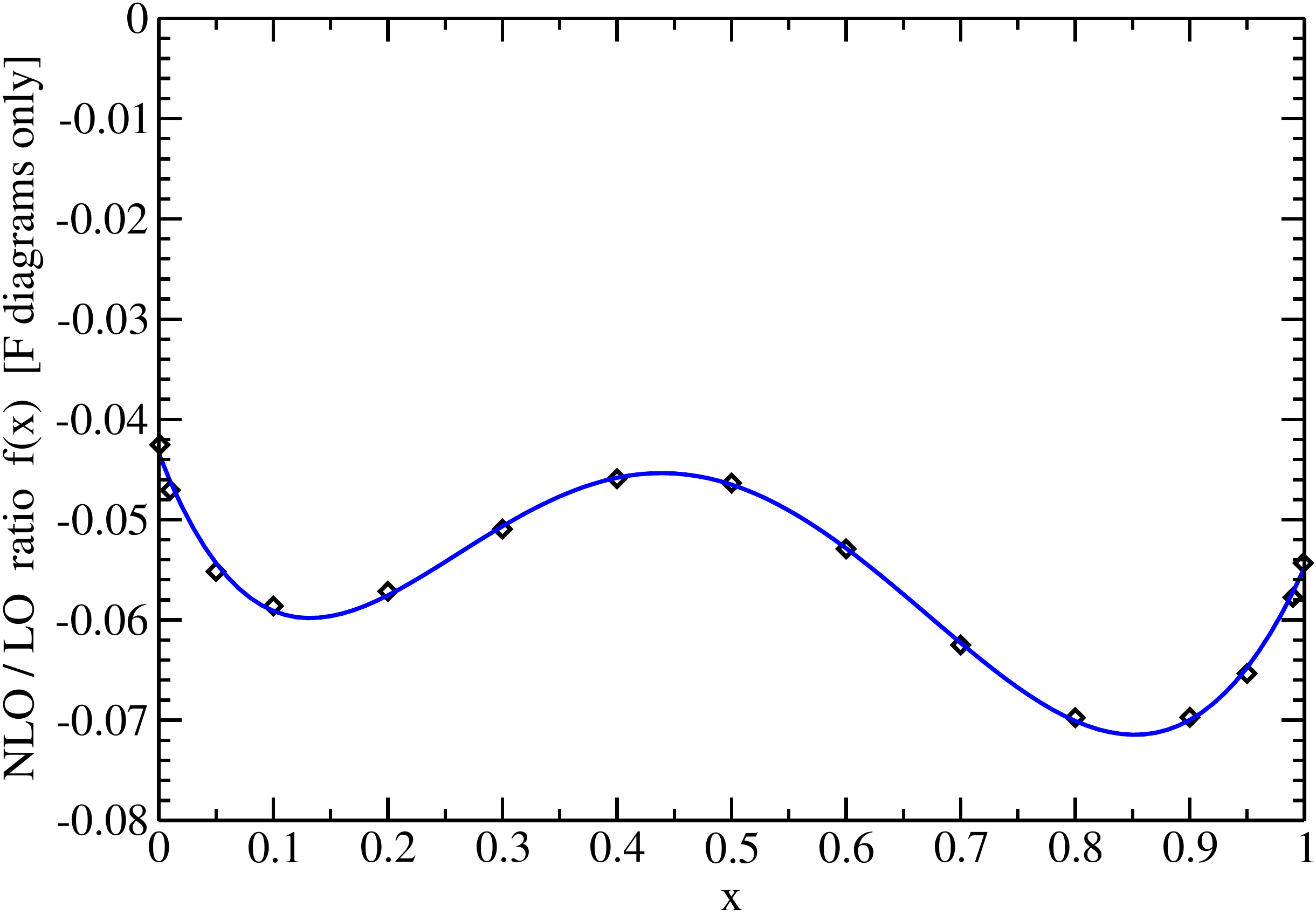}
  \caption{
     \label{fig:dGnetF}
     Like fig.\ \ref{fig:dGnet} but here showing {\it only} the
     contribution from diagrams that contain at least one
     F=4+I interaction \cite{qcdI}, like the examples in
     fig.\ \ref{fig:Fexamples}.  These diagrams do not have IR divergences
     and so do not require factorization, and so they do not affect
     the infrared subtraction in (\ref{eq:dGfac}) and are not
     sensitive to the choice of $\Lambda_\fac$.  These diagrams are also
     UV convergent and are not sensitive to the choice of renormalization
     scale $\mu$.  The solid curve corresponds to the fit
     (\ref{eq:fF}).
  }
\end {center}
\end {figure}

We will be curious later to understand
the relative importance or unimportance
of processes involving fundamental or effective 4-gluon interactions such as
fig.\ \ref{fig:Fexamples} on the shape properties that we will calculate.
Following ref.\ \cite{qcdI}, we refer to such interactions as
``F=4{+}I'' interactions, where ``F'' is meant to be evocative of the
word ``four''; ``4'' stands for fundamental 4-gluon vertices; and
``I'' stands for interactions via longitudinally polarized gluon
exchange, which are ``instantaneous'' in LCPT.
Fig.\ \ref{fig:dGnetF} shows our result for
the piece of fig.\ \ref{fig:dGnet}
that comes from processes involving F interactions \cite{qcdI}.

Since data points like those of table \ref{tab:dGnet} are slow to
compute numerically, and since we will later need to use
$[d\Gamma/dx]_\net$ both in integrals and in integro-differential
equations, we need a reasonable alternative that is quick to evaluate.
We've therefore fit the data of table \ref{tab:dGnet} to a fairly accurate
functional form.  We will continue to distinguish the contribution of
the $F$ diagrams,
and so we write
\begin {subequations}
\label {eq:fit}
\begin {equation}
   f(x) = f_{\text{non-F}}(x) + f_{\rm F}(x) .
\label {eq:fdecompose}
\end {equation}
We have found a good fit to the non-F contributions (the second column of
table \ref{tab:dGnet}) by the function
\begin {multline}
 f_{\text{non-F}}(x) =
   0.26873 \ln x + 0.00745 \ln(1{-}x)
   -3.92750 + 8.96222\, x - 1.69021\, x^2
\\
   - 2.93372\, x^{1/2} - 1.71625\, x^{3/2}
   + 1.26448\, (1{-}x)^{1/2} + 3.08068\, (1{-}x)^{3/2} .
\label {eq:fnonF}
\end {multline}
This fits all the non-F data of the table with at most 0.003 absolute
error and better than 0.3\% relative error.
The presence of $\ln x$ behavior as $x{\to}0$ is clear
from the log-linear plot of the non-F data in fig.\ \ref{fig:dGnetLog}a.
In contrast, fig.\ \ref{fig:dGnetLog}b does not convincingly demonstrate
$\ln(1{-}x)$ behavior as $x{\to}1$, and so for now the non-zero
coefficient of the $\ln(1{-}x)$ term in our fit (\ref{eq:fnonF}) should
not be taken too seriously.  (We have not made the numerical
effort to push our calculations to even smaller values of $1{-}x$.)
For the rest of (\ref{eq:fnonF}), we found the use of half powers of
$x$ and $1{-}x$ necessary to fit the data well with a relatively few
number of terms.  This is a possibility that might have been
anticipated: the somewhat-related
experience of ref.\ \cite{logs} was that small-$y$ expansions of
overlapping real splittings (and their virtual counterparts) were expansions
in powers of $y^{1/2}$ rather than integer powers of $y$
(where $y$ was the softest gluon).

\begin {figure}[t]
\begin {center}
  \includegraphics[scale=0.3]{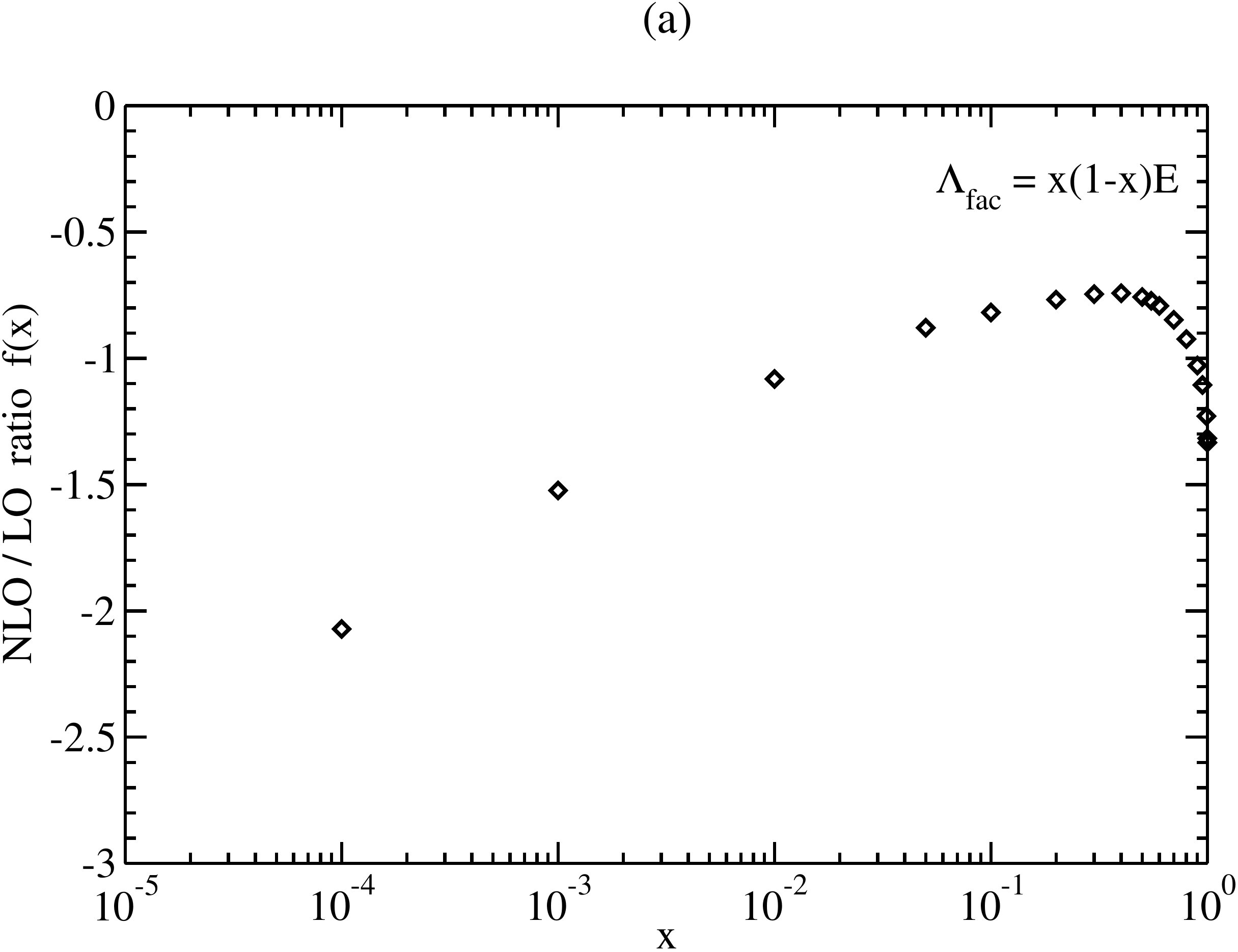}
  \hspace{3em}
  \includegraphics[scale=0.3]{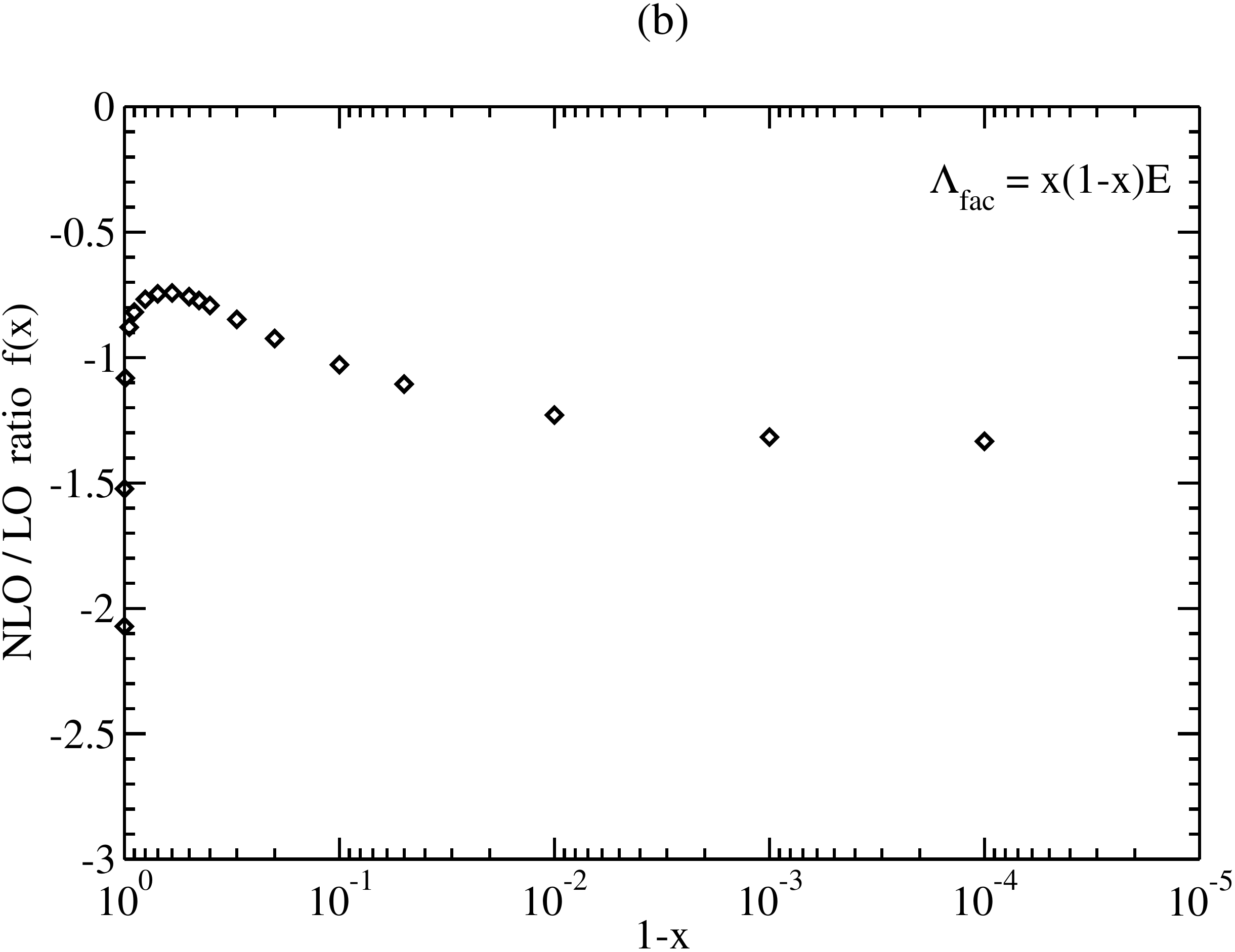}
  \caption{
     \label{fig:dGnetLog}
     (a) A log-linear plot of the non-F contributions to the ratio $f(x)$
     of (\ref{eq:dGnetRatio}).
     (b) The same data plotted vs. $1{-}x$ instead of $x$.
     Note that we've arranged both plots so that $x\to0$ is on the left
     and $x\to1$ is on the right.
  }
\end {center}
\end {figure}

For the F diagram contributions of fig.\ \ref{fig:dGnetF} (the third
column of table \ref{tab:dGnet}), we found that
a simple polynomial fit worked well enough:
\begin {equation}
 f_{\rm F}(x) =
 -0.04338 - 0.29586\, x + 1.69249\, x^2 - 3.29499\, x^3
   + 2.38669\, x^4 - 0.49977\, x^5 ,
\label {eq:fF}
\end {equation}
\end {subequations}
which is the solid curve plotted in fig.\ \ref{fig:dGnetF}.
This fits the data points with at most 0.001 absolute error, which is small
when combined with the non-F diagrams.
The solid curve plotted in fig.\ \ref{fig:dGnet} is the total
ratio (\ref{eq:fdecompose}).

% -------------------------------------------------------------------------

\subsection{Converting between different choices of \boldmath$\Lambda_\fac$}
\label{sec:convert}

\subsubsection{Overview}

To understand how our results for $[d\Gamma/dx]_\net^{\NLO,\fac}$ will
change if one changes the factorization scale $\Lambda_\fac$ and
renormalization scale $\mu$, we just need to know how our results
depend on those two scales.  We can read the $\Lambda_\fac$ dependence
from the last term of (\ref{eq:dGfac}):%
\footnote{
  The fact that the explicit integral shown in the first line of
  (\ref{eq:LambdaDependence}) is infrared divergent does not matter,
  since (i) that divergence does not depend on $\Lambda_\fac$ and (ii)
  the divergence cancels, by construction, against the other
  $\Lambda_\fac$-independent terms in (\ref{eq:dGfac}).
}
\begin {align}
   \left[ \frac{d\Gamma}{dx} \right]_{\rm net}^{\NLO,\fac}
   &=
   (\mbox{$\Lambda_\fac$ independent}) +
      \frac{\CA\alphas}{4\pi}
      \left[ \frac{d\Gamma}{dx} \right]^{\rm LO}
      \int_0^\infty dy \>
        \frac{\ln y + \bar s(x)}{y} \,
        \theta(yE < \Lambda_\fac)
\nonumber\\
   &=
   (\mbox{$\Lambda_\fac$ independent})
   +
      \frac{\CA\alphas}{4\pi}
      \left[ \frac{d\Gamma}{dx} \right]^{\rm LO}
      \left\{
        \frac12 \ln^2\Bigl( \frac{\Lambda_\fac}{E} \Bigr)
        + \bar s(x) \, \ln\Bigl( \frac{\Lambda_\fac}{E} \Bigr)
      \right\} .
\label {eq:LambdaDependence}
\end {align}
The renormalization scale $\mu$ dependence is even easier to isolate.
The explicit $\ln\mu$ dependence of the NLO result must cancel the
implicit dependence in the coupling $\alphas(\mu)$ in the leading-order
rate (\ref{eq:LOrate0}), and so
\begin {equation}
   \left[ \frac{d\Gamma}{dx} \right]_{\rm net}^{\NLO,\fac}
   =
   (\mbox{$\mu$ independent}) -
      \beta_0 \alphas
      \left[ \frac{d\Gamma}{dx} \right]^{\rm LO} \ln\mu ,
\label {eq:MuDependence}
\end {equation}
where $\beta_0$ is
the leading-order coefficient of the renormalization group
$\beta$ function for $\alphas$.
Since we are investigating purely gluonic showers in the large-$\Nc$ limit,
only the gluonic contribution matters:
\begin {equation}
  \beta_0 = - \frac{11\CA}{6\pi} \,.
\label {eq:beta0}
\end {equation}
Putting together (\ref{eq:LambdaDependence}) and (\ref{eq:MuDependence}),
the change
$\delta[d\Gamma/dx]$
in the net rate due to changing $\Lambda_\fac$ and/or $\mu$ is
\begin {equation}
  \delta \left[ \frac{d\Gamma}{dx} \right]_{\rm net}^{\NLO,\fac}
  =
  \frac{\CA\alphas}{4\pi}
  \left[ \frac{d\Gamma}{dx} \right]^{\rm LO}
  \times \delta
  \left\{
      \frac12 \ln^2\Bigl( \frac{\Lambda_\fac}{E} \Bigr)
      + \bar s(x) \, \ln\Bigl( \frac{\Lambda_\fac}{E} \Bigr)
      - \frac{4\pi \beta_0}{\CA} \ln\mu
  \right\} .
\end {equation}
A change from $\Lambda_\fac = x(1{-}x)E$ to $\Lambda_\fac = \kappa x(1-x)E$,
with $\mu = (\qhatA \Lambda_\fac)^{1/4}$ in both cases,
then gives
\begin {multline}
  \left[ \frac{d\Gamma}{dx} \right]_{\rm net}^{\NLO,\fac}
  \Biggr|_{\begin{subarray}{l} \Lambda_\fac{=}\kappa x(1{-}x)E \\
                            \mu{=}(\qhatA \Lambda_\fac)^{1/4} \end{subarray}}
  =
\\
  \left[ \frac{d\Gamma}{dx} \right]_{\rm net}^{\NLO,\fac}
  \Biggr|_{\kappa=1}
  + \frac{\CA\alphas}{4\pi}
  \left[ \frac{d\Gamma}{dx} \right]^{\rm LO}
  \biggl\{
      \tfrac12 \ln^2\kappa
      + \left( \hat s(x) - \frac{\pi\beta_0}{\CA} \right) \ln\kappa
  \biggr\}.
\label {eq:convert}
\end {multline}
The dashed curves in fig.\ \ref{fig:dGnet} show the variation in
the ratio $f(x)$ of (\ref{eq:dGnetRatio}) from increasing the choice
of $\kappa$ up or down by a factor of 2.
In estimating factorization scale dependence, one may reasonably
wonder whether it's more physically
relevant to vary the energy scale $\Lambda_\fac$ by a factor of 2 or so,
or to vary the associated transverse momentum scale
$(\qhat \Lambda_\fac)^{1/4}$ by a factor of 2 or so.
The latter corresponds to varying $\Lambda_\fac$ up or down by a factor of
16, shown by the dotted curves in fig.\ \ref{fig:dGnet}.
The conservative conclusion is that $f(x)$ and so
$[d\Gamma/dx]^{\NLO\,\fac}_\net$ are potentially very sensitive to the
choice of factorization scale.
Fortunately, our final results concerning overlap corrections to the
shape function $S(Z)$ will be {\it dramatically}\/ less sensitive.

Note that the $x$-independent terms in the factor $\{\cdots\}$
in the rescaling (\ref{eq:convert}) could be absorbed
into a constant shift in $\hat q$ and so will not affect the shape
function $S(Z)$.  Only the $x$-dependent
pieces will change the shape function.
Note also that in this case the change in renormalization scale $\mu$
has no explicit
effect on the size of the NLO correction to $S(Z)$.

% -------------------------------------------------------------------------

\subsubsection{An alternate choice}
\label{sec:convertr}

As mentioned earlier, we will eventually also examine how our results
turn out if one chooses
(more simply but more unphysically)
an $x$-independent factorization
scale $\Lambda_\fac = r E$ as in (\ref{eq:Lambdafacr}).
In that case, the relation to our
numerical results for $\Lambda_\fac = x(1{-}x)E$ is just
(\ref{eq:convert}) with $\kappa$ replaced by $r/x(1{-}x)$:
\begin {multline}
  \left[ \frac{d\Gamma}{dx} \right]_{\rm net}^{\NLO,\fac}
  \Biggr|_{\begin{subarray}{l} \Lambda_\fac{=}rE \\
                            \mu{=}(\qhatA \Lambda_\fac)^{1/4} \end{subarray}}
  =
  \left[ \frac{d\Gamma}{dx} \right]_{\rm net}^{\NLO,\fac}
  \Biggr|_{\begin{subarray}{l} \Lambda_\fac{=}x(1-x)E \\
                            \mu{=}(\qhatA \Lambda_\fac)^{1/4} \end{subarray}}
  + \frac{\CA\alphas}{4\pi}
  \left[ \frac{d\Gamma}{dx} \right]^{\rm LO}
  \biggl\{
      \tfrac12 \ln^2\Bigl( \frac{r}{x(1-x)} \Bigr)
\\
      + \left( \hat s(x) - \frac{\pi\beta_0}{\CA} \right)
        \ln\Bigl( \frac{r}{x(1-x)} \Bigr)
  \biggr\}.
\label{eq:convertr}
\end {multline}
We note that, because of the double log in (\ref{eq:convertr}),
the NLO/LO ratio $f(x)$ will diverge
like $\ln^2\bigl(x(1{-}x)\bigr)$ for $\Lambda_\fac=r E$
as $x \to 0$ or $x{\to}1$, instead of the milder $\ln x$ divergence
as $x \to 0$ (and perhaps no divergence for $x\to 1$)
that we found numerically for $\Lambda_\fac = x(1{-}x)$.
The worse divergence of $\Lambda_\fac = r E$
is an indication that $\Lambda_\fac = x(1{-}x)$
better captures the physics of $x{\to}0$ and $x{\to}1$, as we supposed.

% -------------------------------------------------------------------------

\subsubsection{Yet another choice}
\label{sec:convertE0}

Though we will not use it for numerics, it will be convenient
in some of our later discussion to also consider the choice
\begin {equation}
   \Lambda_\fac = r E_0,
   \qquad
   \mu = (\qhatA \Lambda_\fac)^{1/4} ,
\label {eq:LambdaE0}
\end {equation}
where $E_0$ is the energy of the original particle that initiates the
shower, and $r$ is again a fixed, $O(1)$ constant.
At first sight, a seeming failure of this choice is that it is the
wrong scale late in the development of the shower (or any part
of the shower), when particle energies have dropped to $E \ll E_0$.
In that case, however, those particles are already effectively stopped,
since their remaining stopping distance
$\ell_{\rm stop}(E) \sim \alphas^{-1} \sqrt{E/\hat q}$
is then parametrically small compared to the overall stopping distance
$\ell_{\rm stop}(E_0) \sim \alphas^{-1} \sqrt{E_0/\hat q}$.
Having chosen $\Lambda_\fac$ poorly for those $E \ll E_0$ splittings will
not have a significant effect on the energy deposition distribution
$\eps(z)$.  As to the lack of $x$ dependence in (\ref{eq:LambdaE0}),
the argument that was made in the case of
(\ref{eq:Lambdafacr}) applies here as well.

For later reference, the conversion is
\begin {multline}
  \left[ \frac{d\Gamma}{dx} \right]_{\rm net}^{\NLO,\fac}
  \Biggr|_{\begin{subarray}{l} \Lambda_\fac{=}rE_0 \\
                            \mu{=}(\qhatA \Lambda_\fac)^{1/4} \end{subarray}}
  =
  \left[ \frac{d\Gamma}{dx} \right]_{\rm net}^{\NLO,\fac}
  \Biggr|_{\begin{subarray}{l} \Lambda_\fac{=}x(1-x)E \\
                            \mu{=}(\qhatA \Lambda_\fac)^{1/4} \end{subarray}}
  + \frac{\CA\alphas}{4\pi}
  \left[ \frac{d\Gamma}{dx} \right]^{\rm LO}
  \biggl\{
      \tfrac12 \ln^2\Bigl( \frac{r E_0}{x(1-x)E} \Bigr)
\\
      + \left( \hat s(x) - \frac{\pi\beta_0}{\CA} \right)
        \ln\Bigl( \frac{r E_0}{x(1-x) E} \Bigr)
  \biggr\}.
\label{eq:convertE0}
\end {multline}

% -------------------------------------------------------------------------

\subsection{Scaling of \boldmath$[d\Gamma/dx]^\fac_\net$ with energy $E$}

The only dimensionful scales in the original NLO differential rates
$\Delta\,d\Gamma/dx\,dy$ are $\qhat$ and the parent energy $E$.
Like the leading-order rate (\ref{eq:LOrate0}), those differential rates
are proportional to $\sqrt{\qhat/E}$ and so scale like $E^{-1/2}$ for
fixed $x$ and $y$.  However, the integration over $y$ in
(\ref{eq:dGnetNLO}) to get
$[d\Gamma/dx]^{\NLO}_\net$ produced IR log divergences.
To factorize out those divergences, we introduced a new energy scale
$\Lambda_\fac$ to define $[d\Gamma/dx]^{\NLO,\fac}_\net$ in (\ref{eq:dGfac}).
If we take our canonical choice $\Lambda_\fac = \kappa x(1{-}x) E$ or
the alternate choice $\Lambda_\fac = r E$, then we are not introducing
a new dimensionful parameter, and $[d\Gamma/dx]^{\NLO,\fac}_\net$ will scale as
$E^{-1/2}$.  But this is not the case if we instead choose
$\Lambda_\fac = r E_0$ as in (\ref{eq:LambdaE0}).
Specifically, (\ref{eq:convertE0}) shows that this choice would
introduce a term into $[d\Gamma/dx]^{\NLO,\fac}_\net$ that scales as
$E^{-1/2} \ln^2(E_0/E)$.
Later, in sections \ref{sec:epsEquation} and beyond, we make use
of simplifications that occur when $[d\Gamma/dx]^\fac_\net$ scales
exactly as $E^{-1/2}$.  At that time, we will only consider choices where
$\Lambda_\fac \propto E$, like $\Lambda_\fac = x(1{-}x)E$ or
$\Lambda_\fac = rE$, and not $\Lambda_\fac \propto E_0$.

% =========================================================================

\section{LO vs.\ effective LO rates}
\label {sec:LOvEff}

In defining the factorized net rate (\ref{eq:dGfac}), we subtracted
the IR log divergences from the net rate and imagined absorbing those
divergences into an effective leading-order $g{\to}gg$ splitting rate
$[d\Gamma/dx]^{\LO}_\eff$.  {\it Formally}, within our approximations so far,%
\footnote{
  In (\ref{eq:LOeff0}), we are using the version of the integral from
  (\ref{eq:factorization}).
}
\begin {equation}
   \left[ \frac{d\Gamma}{dx} \right]^\LO_\eff =
   \left[ \frac{d\Gamma}{dx} \right]^\LO
   \left\{
     1
     - \frac{\CA\alphas}{4\pi}
     \int_0^{\Lambda_\fac} \frac{d\omega_y}{\omega_y} \,
     \Bigl[
       \ln\Bigl( \frac{\omega_y}{x(1{-}x)E} \Bigr)
       + \hat s(x)
     \Bigr]
   \right\} .
\label {eq:LOeff0}
\end {equation}
However, to really compute $[d\Gamma/dx]^\LO_\eff$, one would have to
correctly account for the infrared physics that cuts off the IR divergence
of the integral above.  Parametrically, the result at leading-log order is
\begin {equation}
   \left[ \frac{d\Gamma}{dx} \right]^\LO_\eff \approx
   \left[ \frac{d\Gamma}{dx} \right]^\LO
   \left\{
     1
     - \frac{\CA\alphas}{8\pi} \, \ln^2\Bigl( \frac{\Lambda_\fac}{T} \Bigr)
   \right\} .
\end {equation}
In the high-energy limit, the double logarithm becomes large since
we choose $\Lambda_\fac \propto E$.  That means that
$\alphas \ln^2(\Lambda_\fac/T)$ is not small at high energy, and
one must resum logarithms to all orders in $\alphas$
to get a usable result for $[d\Gamma/dx]^\LO_\eff$.

Let's ignore that complication for just a moment to give a very crude
preview of the type of argument we will eventually make.
Imagine, just for a moment, that the logarithms were not large
and that $\alphas  \ln^2(\Lambda_\fac/T)$ had size $O(\alphas)$.
In this paper, we want to explore the relative size of NLO corrections
that cannot be absorbed into $\qhat$, as measured by the shape function $S(Z)$.
That means that we will look at the ratio of the
factorized NLO correction
to the effective LO result for $S(Z)$.
But (if logarithms were not large), this ratio
would be
\begin {equation}
   \frac{\NLO_\fac}{\LO_\eff}
   =
   \frac{\NLO_\fac}{\LO \times [1 + O(\alphas)]}
   =
   \frac{\NLO_\fac}{\LO} \times [1 + O(\alphas)] .
\label {eq:ratio0}
\end {equation}
The desired ratio $\NLO_\fac/\LO_\eff$ is itself $O(\alphas)$, but
(\ref{eq:ratio0}) means that the
difference between using $\LO$ and $\LO_\eff$ in the denominator is
a {\it yet-higher} order correction to the ratio and so can be
ignored.  At the order of our calculation, we
can simply calculate $\NLO_\fac/\LO$ instead of $\NLO_\fac/\LO_\eff$.
Unfortunately, the logic of (\ref{eq:ratio0}) fails because the
accompanying logarithms are large.%
\footnote{
  In fact, such logarithms have to be large if we wish to treat
  our high-energy $\alphas(\mu)$ as smaller than the
  $\alphas(T)$ of the medium.
}

So think schematically about resumming the large logarithms
in $[d\Gamma/dx]^\LO_\eff$ to all orders in $\alphas$.
At first order in $\alphas$, (\ref{eq:LOeff0}) absorbs not only
a leading, double log but also a sub-leading, single log.
To be consistent, we must then consider NLLO resummation of
large logarithms.  We do not know how to do the full NLLO resummation.
Fortunately, we do not need it because the shape function $S(Z)$ and its
moments are completely insensitive to any {\it constant}\/ shift in $\hat q$,
which corresponds to any constant (i.e.\ $x$ and $E$ independent) contributions
to the braces $\{\cdots\}$ in (\ref{eq:LOeff0}).
Understanding the $x$ and $E$ dependence of the NLLO resummation
is much easier than understanding the full NLLO resummation.
To preview the result of this section: We will argue that, for large
logarithms, the resummed version of (\ref{eq:ratio0}) is
\begin {equation}
   \frac{\NLO_\fac}{\LO_\eff}
   =
   \frac{\NLO_\fac}{\LO \times [1 + O(\sqrt{\alphas}\,)]}
   =
   \frac{\NLO_\fac}{\LO} \times [1 + O(\sqrt{\alphas}\,)]
\label {eq:ratio}
\end {equation}
provided the LO quantity is (like the shape function)
insensitive to constant shifts of $\hat q$.

The following discussion may be a little clearer if we first
remove any
$x$ and $E$ dependence from our choice of factorization scale, taking
$\Lambda_\fac = r E_0$ as in (\ref{eq:LambdaE0}) for the purpose
of this argument.
The conversion (\ref{eq:convertE0}) between this scale and
our usual choice $\Lambda_\fac = x(1{-}x)E$ is finite and
is free of large logarithms unless $x(1{-}x) \ll 1$ or $E \ll E_0$.
As discussed in sections \ref{sec:rE} and \ref{sec:convertE0},
those limiting cases will not
significantly affect the calculation of the shower energy deposition
distribution $\eps(z)$ and its shape,
and so the conversion (\ref{eq:convertE0}) does not need to be resummed.

% --------------------------------------------------------------------------

\subsection{Origin of the IR double and single logs in (\ref{eq:LOeff0})}

\begin {figure}[t]
\begin {center}
  \includegraphics[scale=0.6]{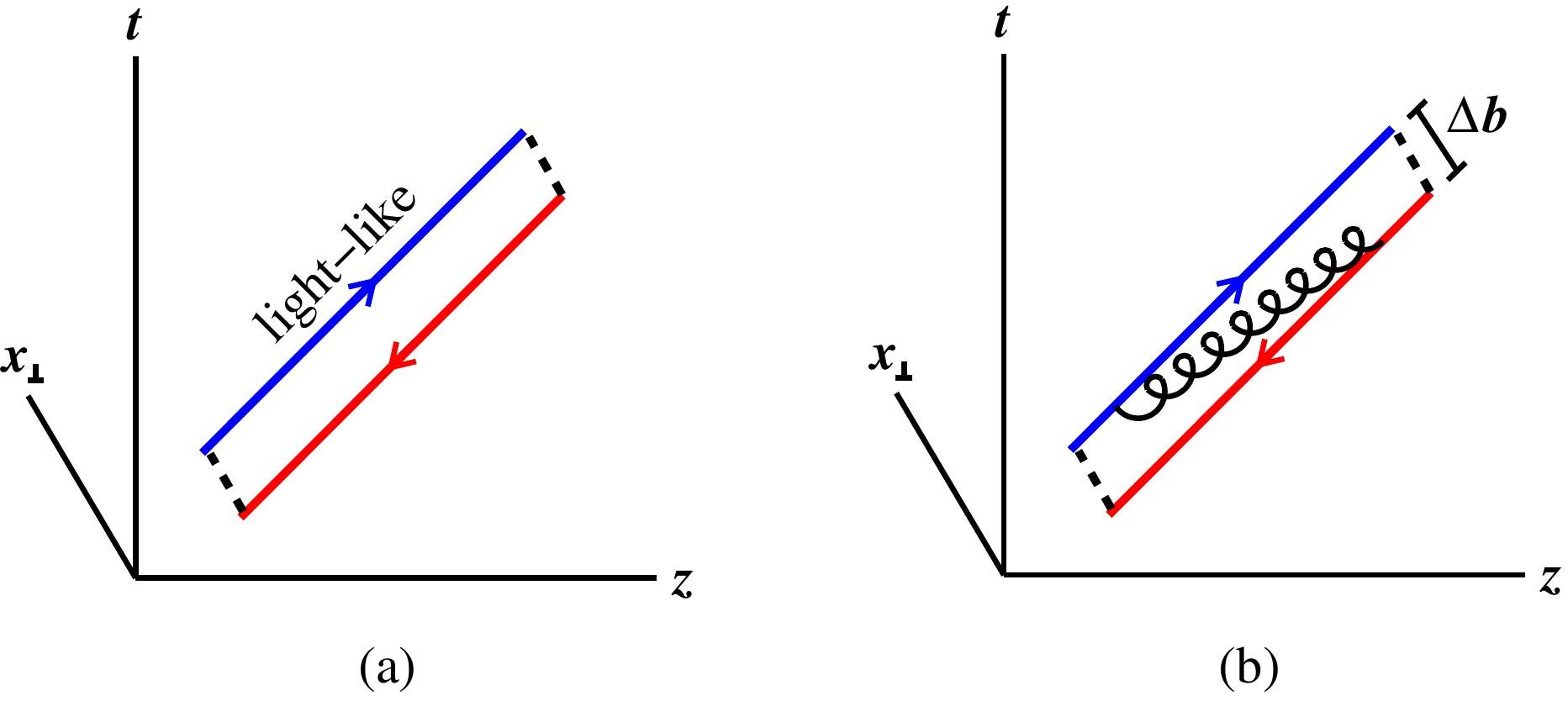}
  \caption{
     \label{fig:Wilson}
     (a) A Wilson loop with long, light-like sides and transverse spatial
     width $\Delta b$, whose expectation gives
     $\exp\bigl(-\frac14 \qhat(\Delta b) \,\mathbb{T}\, (\Delta b)^2 \bigr)$
     for
     small $\Delta b$ and large extent $\mathbb{T}$ in time $t$.
     (b) An example of a high-energy nearly-collinear radiative
     contribution to the Wilson loop.
  }
\end {center}
\end {figure}

We need to review the origin of the remaining, explicit
$x$ and $E$ dependence
in (\ref{eq:LOeff0}) so that we can discuss how to resum it.
We will use the combined analysis of IR double and single logarithms
presented in ref.\ \cite{logs2}.
There, the usual, leading-order BDMPS-Z rate calculation (in $\hat q$
approximation) was
modified by replacing $\hat q$ by the effective transverse momentum
broadening
parameter $\hat q_\eff(\Delta b)$ originally calculated by
Liou, Mueller and Wu (LMW) \cite{LMW}, which incorporates the effect
of soft radiation carrying away transverse momentum.
The $\Delta b$ in $\hat q_\eff(\Delta b)$ represents transverse
separation.
Formally, $\hat q_\eff(\Delta b)$ is extracted from the thermal expectation
of a Wilson loop with long, light-like sides separated by transverse
distance $\Delta b$, as depicted in fig.\ \ref{fig:Wilson}a.
The bare $\qhat_{(0)}$ corresponds to the contribution from thermal-scale
correlations in the medium; the double and single logarithms come
from the exchange of a nearly collinear, high-energy gluon
($\omega \gg T$) as in fig.\ \ref{fig:Wilson}b.
In our application, those logarithms are cut-off at high energy by
the factorization scale $\Lambda_\fac$, so that
$T \ll \omega \le \Lambda_\fac$.
We should really write $\hat q_\eff(\Delta b\,; \Lambda_\fac)$ instead
of just $\hat q_\eff(\Delta b)$, but we will stick with the shorter
notation $\hat q_\eff(\Delta b)$ for now, with the $\Lambda_\fac$
dependence implicit.%
\footnote{
  In the original work of LMW \cite{LMW} on momentum broadening,
  the role of our ``$\Lambda_\fac$'' is played by the largest
  ``soft'' bremsstrahlung energy $\omega$
  that has a formation time that fits inside the length $L$ of the
  medium, which corresponds to $\Lambda_\fac \sim \qhat L^2$.
  Our canonical choice (\ref{eq:Lambdafac}) of $\Lambda_\fac$ in this
  paper corresponds to replacing that $L$ by the formation time of the
  underlying hard single-splitting process $E \to xE + (1{-}x)E$
  that one is computing soft radiative corrections to.
}

\begin {figure}[t]
\begin {center}
  \includegraphics[scale=0.6]{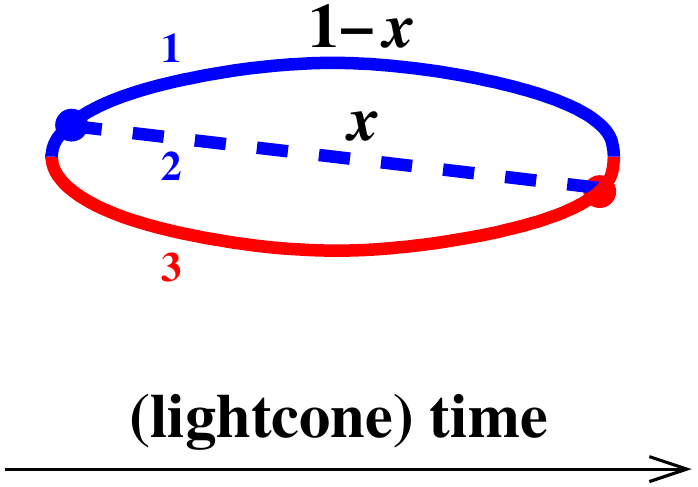}
  \caption{
     \label{fig:LO2}
     This is fig.\ \ref{fig:LO}b for LO splitting $g \to gg$, but here
     with the three lines labeled (1,2,3).
  }
\end {center}
\end {figure}

As reviewed in our notation in ref.\ \cite{logs2}, the Zakharov
picture of the usual BDMPS-Z calculation for $g{\to}gg$ involves
solving for the propagator of 3-particle quantum mechanics in the
two-dimensional transverse plane with Hamiltonian
\begin {equation}
  H = \frac{p_{\perp1}^2}{2|p_{z_1}|} + \frac{p_{\perp2}^2}{2|p_{z_2}|}
     - \frac{p_{\perp3}^2}{2|p_{z_3}|}
     - \frac{i \qhatA}{8} \, ( b_{12}^2 + b_{23}^2 + b_{31}^2 ) ,
\label {eq:H0}
\end {equation}
where $\b_{ij} \equiv \b_i - \b_j$ are the transverse separations between
the three ``particles'' in fig.\ \ref{fig:LO2} and
$(p_{z1},p_{z2},p_{z3}) = (1{-}x,x,-1)E$ are the corresponding
longitudinal momenta of those particles.
Symmetries are used to reduce this to a 1-particle quantum mechanics
problem in a single transverse position variable $\B$ related by
\begin {equation}
  \b_{12} = \B,
  \qquad
  \b_{23} = -(1{-}x)\B,
  \qquad
  \b_{31} = -x\B ,
\end {equation}
which reduces (\ref{eq:H0}) to
\begin {equation}
  H = \frac{P^2}{2x(1{-}x)E}
     - \frac{i \qhatA}{8} \, ( 1+(1{-}x)^2+x^2 ) B^2 ,
\label {eq:HBDMPS}
\end {equation}
where ${\bm P}$ is conjugate to $\B$.
In the LO splitting process of fig.\ \ref{fig:LO2}, transverse
separations vary with time, but the typical value
$\Bbar$ of $B$ during
the splitting is parametrically
\begin {equation}
   \Bbar \sim [ x(1{-}x)E \qhat]^{-1/4} .
\end {equation}

Ref.\ \cite{logs2} argued that, in the large-$\Nc$ limit, the modification
of (\ref{eq:H0}) that would correctly reproduce the IR double and single
logs from soft radiative corrections to the hard, underlying
$g{\to}gg$ process was, with one caveat,
\begin {equation}
  H = \frac{p_{\perp1}^2}{2|p_{z_1}|} + \frac{p_{\perp2}^2}{2|p_{z_2}|}
     - \frac{p_{\perp3}^2}{2|p_{z_3}|}
     - \frac{i}{8} \Bigl[
           \qhatAeff(b_{12}) \, b_{12}^2
           + \qhatAeff(b_{23}) \, b_{23}^2
           + \qhatAeff(b_{31}) \, b_{31}^2
      \Bigr] .
\label {eq:H1}
\end {equation}
The caveat is that the momentum broadening analysis of LMW \cite{LMW}
gives the $\qhat_\eff$ between an amplitude (blue) line and a
conjugate amplitude (red) line in fig.\ \ref{fig:LO2}.
The $\qhat_\eff$ between two amplitude (blue) lines is slightly
different.  In the analysis of ref.\ \cite{logs2}, this difference
was equivalent to replacing
\begin {equation}
  \qhatAeff(b_{12}) \longrightarrow
  \qhatAeff(e^{-i\pi/8} b_{12}) .
\end {equation}
in (\ref{eq:H1}).
The modified (\ref{eq:H1}) then reduces to
\begin {equation}
  H = \frac{P^2}{2x(1{-}x)E}
     - \frac{i}{8} \Bigl[
           \qhatAeff\bigl(e^{-i\pi/8} B\bigr)
           + (1{-}x)^2 \, \qhatAeff\bigl((1{-}x)B\bigr)
           + x^2 \, \qhatAeff\bigl(x B\bigr)
      \Bigr] B^2 .
\label {eq:Heff}
\end {equation}
Ref.\ \cite{logs2} used this Hamiltonian instead of
(\ref{eq:HBDMPS}) for the BDMPS-Z calculation and reproduced the
soft radiative corrections (\ref{eq:IRlogs0}) to the usual leading-order
BDMPS-Z rate (\ref{eq:LOrate0}).  The result may be summarized in the form%
\footnote{
   Though some broader claims were made at the end, ref.\ \cite{logs2}
   only did explicit calculations for the part of the double log region
   to the right of the corner marked $\beta$ in our fig.\ \ref{fig:LMWregion}.
   However, that region contains all of the $\Delta b$ dependence of the
   logarithms, which is our ultimate interest here.
}
\begin {multline}
   \left[ \frac{d\Gamma}{dx} \right]^\LO_\eff
   =
   \left[ \frac{d\Gamma}{dx} \right]^\LO
   \Re
   \Biggl\{
     \sqrt{2} \, e^{-i\pi/4}
     \Biggl[
     w_{12} \sqrt{ \frac{\qhatAeff(\Bbar)}{\qhat_{\rm A(0)}} }
     + w_{23} \sqrt{
        \frac{\qhatAeff\bigl(e^{-i\pi/8}(1{-}x)\Bbar\bigr)}{\qhat_{\rm A(0)}}
       }
\\
     + w_{31} \sqrt{
        \frac{\qhatAeff\bigl(e^{-i\pi/8}x\Bbar\bigr)}{\qhat_{\rm A(0)}}
       }
     \,
     \Biggr]
   \Biggr\} ,
\label {eq:LOeff}
\end {multline}
where here%
\footnote{
  Our $\Bbar$ defined in (\ref{eq:Bbar}) differs from the
  $\bar B$ defined in ref.\ \cite{logs2} by a factor of
  $i^{1/4} = e^{i\pi/8}$.
}
\begin {equation}
   \Bbar \equiv
   e^{-\gammaE/2} \bigl[ \tfrac12 x(1{-}x)(1{-}x{+}x^2) \qhatA E \bigr]^{-1/4} ,
\label {eq:Bbar}
\end {equation}
and the weights $(w_{12},w_{23},w_{31})$ are defined by
\begin {equation}
   w_{12} = \frac{1}{1+(1{-}x)^2+x^2} ,
   \quad
   w_{23} = \frac{(1{-}x)^2}{1+(1{-}x)^2+x^2} ,
   \quad
   w_{31} = \frac{x^2}{1+(1{-}x)^2+x^2}
\end {equation}
with
\begin {equation}
   w_{12} + w_{23} + w_{31} = 1.
\end {equation}
The intricate details of these formulas will not matter for our argument,
but we thought it useful to have something concrete to reference.
There are two aspects of (\ref{eq:LOeff}) that will matter.

The first is that, for our application, the arguments $\Delta b$ of
the three $\qhatAeff(\Delta b)$'s in (\ref{eq:LOeff}) are all of order
\begin {equation}
   \Delta b \sim \Bzero \equiv (\qhatA E_0)^{-1/4} .
\label {eq:Bzero}
\end {equation}
That's because,
as previously discussed, processes with parametrically (i) $E \ll E_0$ or (ii)
$x \ll 1$ or $1{-}x \ll 1$ are not important to determining the shape
function $S(Z)$.

The second important aspect is that, if one were
to replace all three of the different 
$\qhatAeff(\Delta b)$'s in (\ref{eq:LOeff})
by the fixed ($x$ and $E$ independent)%
\footnote{
  $\qhatAeff(\Delta b) = \qhatAeff(\Delta b; \Lambda_\fac)$
  also depends on $\Lambda_\fac$.
  Remember the earlier argument that the difference between using
  $\Lambda_\fac = rE_0$ and $\Lambda_\fac = rE$ or
  $\Lambda_\fac = \kappa x(1{-}x)E$ does not involve large logarithms in
  our application, and so, for simplicity,
  we would carry out our discussion of resumming large logarithms using
  the fixed scale choice $\Lambda_\fac = rE_0$.  That simplifies the discussion
  here because the only $x$ and $E$ dependence inside the braces $\{\cdots\}$
  in (\ref{eq:LOeff}) is that of the arguments
  $\Delta b$ of $\qhatAeff(\Delta b;\Lambda_\fac)$;
  we need not be distracted by the possibility of $x$ or $E$
  dependence of $\Lambda_\fac$ in this analysis.
}
value
$\qhatAeff(\Bzero)$, then the effective LO rate $[d\Gamma/dx]^{\rm LO}_\eff$
would be a fixed multiple of the original $[d\Gamma/dx]^{\rm LO}$
(i.e.\ something that
could be absorbed by a constant shift of $\hat q$), and so the shape of
the energy deposition distribution would be unchanged:
$S^\LO_\eff(Z) = S^\LO(Z)$.
That means that the actual difference between
$S^\LO_\eff(Z)$ and $S^\LO(Z)$ depends specifically
on how $\qhatAeff(\Delta b)$ varies when one varies $\Delta b$.

% --------------------------------------------------------------------------

\subsection{The dependence of resummed \boldmath$\qhatAeff(\Delta b)$
            on $\Delta b$}
\label {sec:dbDependence}

\begin {figure}[t]
\begin {center}
  \includegraphics[scale=1]{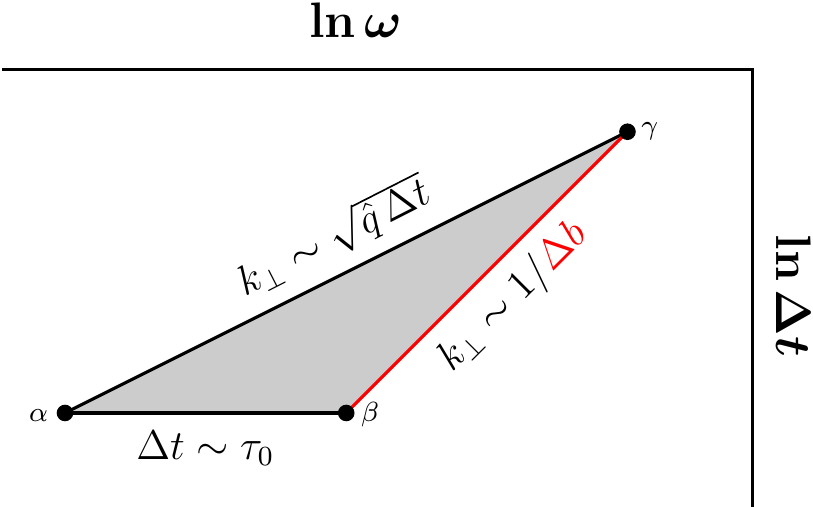}
  \caption{
     \label{fig:LMWregion}
     The integration region giving rise to the double logs of LMW \cite{LMW}.
     Here $\omega$ is the energy of the soft radiated gluon (which we called
     $yE$ earlier), and $\Delta t$ is the time over which it is radiated
     (the difference of the emission time in the amplitude and the emission
     time in the conjugate amplitude).  The transverse momentum of the
     soft radiated gluon is $k_\perp \sim \sqrt{\omega/\Delta t}$.
     The only boundary that is sensitive to $\Delta b$ is the red one.
     For a quark-gluon plasma, the three vertices $(\alpha,\beta,\gamma)$
     above respectively correspond to
     $(\omega,\Delta t)$ of order $(T,\tau_0)$,
     $\bigl(\tau_0/(\Delta b)^2,\tau_0)$, and
     $\bigl( 1/\qhat(\Delta b)^4 , 1/\qhat(\Delta b)^2 \bigr)$.
     The last one is also parametrically
     $\sim \bigl( \Lambda_\fac , t_{\rm form}(\Lambda_\fac) \bigr)$
     for our application.
     We have not shown any vertical snip off the $\gamma$ corner
     corresponding to constraining $\omega \le \Lambda_\fac$ because
     it is unimportant as far as large logarithms are concerned and
     so, for this purpose, is a detail hidden inside the circle
     marking that corner.
  }
\end {center}
\end {figure}

The dependence of the original LMW $\qhat_\eff(\Delta b)$ on $\Delta b$
is easy to extract from parametric arguments for the {\it double} log
in ref.\ \cite{LMW}, provided we rewrite their parametric formulas in
terms of variables more relevant here.
Fig.\ \ref{fig:LMWregion} shows the double log region, where
$\tau_0$ is the scale of the mean free path
for elastic scattering of high-energy particles from the medium.
The difference with similar discussion in LMW is that they were
interested specifically in the problem of transverse momentum broadening after
passing through a large length $L$ of medium, and in that context
they eventually set the transverse separation to be
$\Delta b \sim (\qhat L)^{-1/2}$.  We want to keep everything in
terms of $\Delta b$, which can be achieved by substituting back
$L \sim 1/\qhat(\Delta b)^2$ in their general discussion.
With this translation, they found
\begin {equation}
  \qhat_\eff(\Delta b) = \qhat_{(0)} + \delta \qhat(\Delta b)
  \approx \qhat_{(0)}
    \left[
      1 +
      \frac{\CA\alphas}{2\pi}
         \ln^2\left( \frac{1}{\qhat \tau_0(\Delta b)^2} \right)
    \right]
\label {eq:qhatLMW}
\end {equation}
at leading log order, to first order in $\alphas(\mu)$.
In fact,
the $\Delta b$ dependence of the double log above contains all of
the $\Delta b$ dependence including the single log as well \cite{LMW}.
We can therefore use LMW's results for leading-log order resummation to
all orders in $\alphas(\mu)$ to also obtain the results for the
$\Delta b$ {\it dependence} of a NLLO resummation.  (We outline a more
detailed argument of this claim in appendix \ref{app:NLLO}.)

Eq.\ (\ref{eq:qhatLMW}) was derived by LMW for the case where one ignores
running of $\alphas(k_\perp)$.  In that case, they obtained an analytic
result for the leading-log resummation.  We will continue with their
fixed-coupling analysis, but later argue that a running coupling will not
change our conclusion that $\NLO/\LO_\eff \simeq \NLO/\LO$ as in
(\ref{eq:ratio}).
Their resummed result, when translated from their $L$ back
to $\Delta b$, is
\begin {equation}
  \qhat_\eff(\Delta b)
  \approx \qhat_{(0)} \,
  \frac{
    I_1 \left(
      2\bigl(\frac{\CA\alphas}{\pi}\bigr)^{1/2}
      \ln\bigl( \frac{1}{\qhat \tau_0(\Delta b)^2} \bigr)
    \right)
  }{
      \bigl(\frac{\CA\alphas}{\pi}\bigr)^{1/2}
      \ln\bigl( \frac{1}{\qhat \tau_0(\Delta b)^2} \bigr)
  }
,
\label {eq:LMWresum}
\end {equation}
where $I_1$ is the modified Bessel function.
Remember that in our problem $\Delta b \sim \Bzero = (\qhatA E_0)^{-1/4}$,
and so%
\footnote{
  In the case of a weakly-coupled QGP with gauge coupling coupling $g$,
  we've used $\qhat_{(0)} \sim g^4 T^3$ and
  $\tau_0 \sim 1/g^2 T$ and so $\qhat_{(0)} \tau_0^2 \sim T$
  in (\ref{eq:logarg}).
  For a strongly-coupled QGP, the only relevant scale here is $T$.
  One can worry that one should self-consistently use
  $\qhat_\eff$ instead of $\qhat_{(0)}$ for $\qhat$ in (\ref{eq:logarg}), but
  the difference would only generate a sub-leading $O(\alphas)$
  correction to the $O(\sqrt{\alphas})$ exponents in
  (\ref{eq:qeff1}) and (\ref{eq:qeff2}) and will not affect
  the conclusion (\ref{eq:sqrtExpansion}).
}
\begin {equation}
   \frac{1}{\qhat \tau_0(\Delta b)^2} \sim \sqrt{\frac{E_0}{T}} \,.
\label {eq:logarg}
\end {equation}
In the high-energy limit of large logarithms, (\ref{eq:LMWresum})
becomes
\begin {equation}
  \qhat_\eff(\Delta b)
  \approx \qhat_{(0)}
  \left( \frac{1}{\qhat \tau_0(\Delta b)^2} \right)^{\!2\sqrt{\CA\alphas/\pi}}
  ,
\label {eq:qeff1}
\end {equation}
where we have suppressed a prefactor proportional to
$1/(\sqrt\alphas \,\log)^{3/2}$ that will not
affect the argument (see appendix \ref{app:prefactor} for details).
Since $\Delta b \sim \Bzero$,
this can be expanded as
\begin {equation}
  \qhat_\eff(\Delta b)
  \approx \qhat_{(0)}
  \left( \frac{1}{\qhat \tau_0\Bzero^2} \right)^{\!2\sqrt{\CA\alphas/\pi}}
  \left[
     1
     - 2 \Bigl(\frac{\CA\alphas}{\pi}\Bigr)^{\!1/2}
         \ln\left( \frac{(\Delta b)^2}{\Bzero^2} \right)
  \right]
\label {eq:qeff2}
\end {equation}
and so
\begin {equation}
  \qhat_\eff(\Delta b)
  =
  \qhat_\eff(\Bzero) \, \bigl[ 1 + O(\sqrt{\alphas}\,) \bigr]
  =
  \mbox{(fixed constant)} \times \bigl[ 1 + O(\sqrt{\alphas}\,) \bigr] .
\label {eq:sqrtExpansion}
\end {equation}
The expansion in $\sqrt{\alphas}$ made here
is valid because $\ln(\Delta b/\Bzero)$ is not a large
logarithm in our application.
Eq.\ (\ref{eq:sqrtExpansion}) is the justification for our earlier
claim (\ref{eq:ratio}) that we could ignore the difference between
$S^\LO(Z)$ and $S^\LO_\eff$(Z) when computing the relative size of
NLO corrections to $S^\LO_\eff(Z)$.

% --------------------------------------------------------------------------

\subsection{Running of \boldmath$\alphas(k_\perp)$}
\label{sec:running}

In the preceding, we used an explicit resummation formula
(\ref{eq:LMWresum}) that ignored running of $\alphas(k_\perp)$.
At leading-log order, one may find more sophisticated discussions
in refs.\ \cite{run1,run2a,run2b}.
However, that analysis is not needed for our argument.

First note that the red boundary $k_\perp \sim 1/\Delta b$ in
fig.\ \ref{fig:LMWregion} is the part of the double log region
where $k_\perp$ is the largest and so $\alphas(k_\perp)$ is the smallest.
In our previous argument, we were trying to show that
\begin {equation}
   \left|
     \frac{ \qhat_\eff(\Delta b) - \qhat_\eff(\Bzero) }
          { \qhat_\eff(\Bzero) }
   \right|
   \ll 1
\label {eq:runrat}
\end {equation}
for $\Delta b \sim \Bzero$,
so that $\qhat_\eff(\Delta b)$ could be replace by $\qhat_\eff(\Bzero)$.
For fixed coupling, we argued that this ratio was $O(\sqrt{\alphas}\,)$.
Imagine that the fixed coupling we had taken was the coupling associated
with the red boundary, $\alphas(1/\Delta b)$.  Note that
$1/\Delta b \sim 1/\Bzero \sim (\qhat E_0)^{1/4} \sim \mu$
in our application,
and so, up to higher-order corrections,
$\alphas(1/\Delta b)$ is just the $\alphas=\alphas(\mu)$ that we've
been using throughout this entire paper.
Now imagine replacing fixed $\alphas = \alphas(1/\Delta b)$ by a running
$\alphas(k_\perp)$.  The numerator in (\ref{eq:runrat}) does not
change, because it only involves the physics of $k_\perp \sim 1/\Delta b$.
But the denominator gets bigger because, in the {\it rest}\/ of the double-log
region, $\alphas(k_\perp)$ is bigger than before.
So, the parametric inequality (\ref{eq:runrat}) remains valid
for small $\alphas(\mu)$.

% --------------------------------------------------------------------------

\subsection{Notation: LO vs.\ bare}
\label{sec:notation}

Going forward, it will be helpful to somewhat streamline our notation.
From now on, we will use ``LO'' 
to refer to calculations based on
the leading-order splitting rates (\ref{eq:LOrate0}) with $\hat q$ taken
to be $\qhat_\eff(\Bzero)$, as opposed to the bare $\hat q_{(0)}$.
With this nomenclature, we now formally have
\begin {equation}
   \LO_\eff = \LO \times [1+O(\sqrt{\alphas}\,)]
\end {equation}
for any quantity we will discuss in the context of energy deposition,
including ones that are (unlike the shape function)
sensitive to constant shifts in $\qhat$.

% =========================================================================

\section{Energy deposition equation}
\label{sec:epsEquation}

In this section, we derive the basic equation satisfied by the
energy deposition distribution $\eps(z)$.
We will build on the methods of refs.\ \cite{stop,qedNfstop}.%
\footnote{
  See in particular appendix A.1 of ref.\ \cite{qedNfstop}, but specialize
  throughout to the case of a single type of particle (namely gluons).
  %%%%Oops: I don't think the following supposed difference is real.%%%%
  %%Warning: Our $\eps(E,z)$ in this paper is normalized as in
  %%(\ref{eq:epsnormalize}), whereas the
  %%$\varepsilon(E,z)$ of ref.\ \cite{qedNfstop} is normalized so
  %%that $\int_0^\infty dz \> \varepsilon(E,z) = 1$.  The conversion
  %%is simply that our $\eps(E,z) = E\,\varepsilon(E,z)$.
}
One might be able to directly figure out the final formula
in terms of the net rate $[d\Gamma/dx]_\net$, but we think it's
clearer to first review earlier results written in terms of
$[d\Gamma/dx]_{1\to2}$ and $[d\Gamma/dx]_{1\to3}$.

For simplicity, start by considering a shower composed of only
$1{\to}2$ splittings.
Let $\eps(E,z)$ represent the distribution of deposited energy as a function
of position $z$ for a shower initiated by a particle of energy $E$, with
\begin {equation}
  \int_0^\infty dz \> \eps(E,z) = E .
\label {eq:epsnormalize}
\end {equation}
The starting equation is
\begin {multline}
  \eps(E,z+\Delta z)
  \simeq
  [1 - \Gamma(E)\,\Delta z] \, \eps(E,z)
\\
  +
  \frac12
  \int_0^1 dx \> \left[\frac{d\Gamma}{dx}(E,x)\right]_{1\to2} \Delta z\,
  \bigl\{
    \eps(x E,z)
     + \eps\bigl((1{-}x)E,z\bigr)
  \bigr\}
\label {eq:epsEvolve0}
\end {multline}
for small $\Delta z$.
To see this,
think of traveling the distance $z+\Delta z$ indicated on the left-hand side
as first traveling $\Delta z$ followed by traveling distance $z$.
In the first $\Delta z$ of distance,
the particle has a chance $1-\Gamma(E)\,\Delta z$ of
not splitting at all, and then
the energy density deposited after
traveling the remaining distance $z$ will just be
$\eps(E,z)$.  This possibility is represented by the first term on the
right-hand side of (\ref{eq:epsEvolve0}).
Alternatively, there is a chance that the particle {\it does} split in
the first $\Delta z$.  In this case, we will have
two particles with energies $x E$ and $(1{-}x)E$, which will deposit
energy density $\eps(xE,z)$ and $\eps\bigl((1{-}x)E,z\bigr)$ respectively
after traveling the remaining distance $z$.
Both daughter's eventual contribution to the deposited energy
are added together in the second term of (\ref{eq:epsEvolve0}).
The factor of $\frac12$ in the second term is the identical final-state
particle factor for the two daughter gluons:
\begin {equation}
  \Gamma(E) = \frac12
  \int_0^1 dx \> \left[\frac{d\Gamma}{dx}(E,x)\right]_{1\to2} .
\label {eq:gamma1}
\end {equation}
Rearranging the terms in (\ref{eq:epsEvolve0}) and taking the limit
$\Delta z \to 0$ yields the integro-differential equation
\begin {equation}
  \frac{\partial \eps(E,z)}{\partial z}
  =
  - \Gamma(E) \, \eps(E,z)
  +
  \frac12
  \int_0^1 dx \> \left[\frac{d\Gamma}{dx}(E,x) \right]_{1\to2}
  \bigl\{
    \eps(x E,z)
     + \eps\bigl((1{-}x)E,z\bigr)
  \bigr\}
  .
\label {eq:epseq1}
\end {equation}
Now use the symmetry of $[d\Gamma/dx]_{1\to2}$ under exchange of the
final-state daughters $x$ and $1{-}x$ to rewrite this as
\begin {equation}
  \frac{\partial \eps(E,z)}{\partial z}
  =
  - \Gamma(E) \, \eps(E,z)
  +
  \int_0^1 dx \> \left[\frac{d\Gamma}{dx}(E,x) \right]_{1\to2}
    \eps(x E,z)
  .
\label {eq:epseq1b}
\end {equation}

$1{\to}3$ splittings may be included by following the same steps.
First, add a $1{\to}3$ term
\begin {equation}
  +
  \frac{1}{3!}
  \int_0^1 dx \int_0^{1-x} dy \>
  \left[ \frac{d\Gamma}{dx\,dy}(E,x,y) \right]_{1\to3}
  \bigl\{
    \eps(x E,z)
     + \eps(y E,z)
     + \eps\bigl((1{-}x{-}y)E,z\bigr)
  \bigr\}
\end {equation}
to the right-hand side of (\ref{eq:epseq1}).  Using the symmetry of
the three daughters, this generalizes (\ref{eq:epseq1b}) to
\begin {align}
  \frac{\partial \eps(E,z)}{\partial z}
  &=
  - \Gamma(E) \, \eps(E,z)
  +
  \int_0^1 dx \> \left[\frac{d\Gamma}{dx}(E,x) \right]_{1\to2}
    \eps(x E,z)
\nonumber\\
  & \hspace{6em}
  +
  \frac12
  \int_0^1 dx \int_0^{1-x} dy \>
    \left[\frac{d\Gamma}{dx\,dy}(E,x,y) \right]_{1\to3}
    \eps(x E,z)
\nonumber\\
  &=
  - \Gamma(E) \, \eps(E,z)
  +
  \int_0^1 dx \> \left[\frac{d\Gamma}{dx}(E,x) \right]_\net
    \eps(x E,z)
  ,
\end {align}
where the last equality uses (\ref{eq:dGnet}).
We may now express everything in terms of $[d\Gamma/dx]_\net$ by (i) using
(\ref{eq:GammaAlt}) to rewrite $\Gamma$ as $\int dx \> x [d\Gamma/dx]_\net$
and (ii) combining the $x$ integrals:
\begin {equation}
  \frac{\partial \eps(E,z)}{\partial z}
  =
  \int_0^1 dx \> \left[\frac{d\Gamma}{dx}(E,x)\right]_\net \,
  \bigl\{
     \eps(x E,z)
     - x\,\eps(E,z)
  \bigr\}
  .
\label {eq:epseq3}
\end {equation}

Provided $[d\Gamma/dx]_\net$ scales with parent energy as $E^{-1/2}$,
e.g.\ like the leading-order rate (\ref{eq:LOrate0}) does, we may
define an energy-independent, rescaled rate $[d\tilde\Gamma/dx]_\net$ by%
\footnote{
  It might be more elegant to scale out a factor of
  $\CA\alphas\sqrt{\qhatA/E}$
  in (\ref{eq:raterescale}) instead of just $E^{-1/2}$, so that the
  rescaled rate $[d\tilde\Gamma/dx]_\net$ (and also eventually the coordinate
  $\tilde z)$ would be dimensionless.
  We will find it convenient to do this later, in section \ref{sec:shape}.
  We don't do it now because it would slightly clutter our equations
  and de-emphasize the most essential
  point, the $E^{-1/2}$ dependence.
}
\begin {equation}
  \left[ \frac{d\Gamma}{dx}(E,x) \right]_\net
    = E^{-1/2} \left[ \frac{d\tilde\Gamma}{dx}(x) \right]_\net .
\label {eq:raterescale}
\end {equation}
If rates scale like $E^{-1/2}$, then the distances $z$ characteristic of
shower development will scale like $E^{1/2}$, so
the energy deposition distribution should scale as
\begin {equation}
  \eps(E,z) \propto \tilde\eps(E^{-1/2} z) .
\label {eq:epsrescale}
\end {equation}
We want the rescaled function $\tilde\eps(s)$ to be independent of $E$
and so have a normalization independent of $E$.  We choose to
normalize it so that
\begin {equation}
  \int_0^\infty d{s} \> \tilde\eps(s) = 1,
\end {equation}
which, together with (\ref{eq:epsnormalize}),
fixes the proportionality constant in (\ref{eq:epsrescale}):
\begin {equation}
  \eps(E,z) = E^{1/2} \, \tilde\eps(E^{-1/2} z) .
\end {equation}
For a shower initiated by a particle of energy $E_0$,
(\ref{eq:epseq3}) becomes
\begin {equation}
  \frac{\partial \tilde\eps(\tilde z)}{\partial\tilde z}
  =
  \int_0^1 dx \> x \biggl[\frac{d\tilde\Gamma}{dx}(x) \biggr]_\net
  \bigl\{ x^{-1/2}\,\tilde\eps(x^{-1/2} \tilde z)
          - \tilde\eps(\tilde z) \bigr\} ,
\label {eq:epseq0}
\end {equation}
where
\begin {subequations}
\label {eq:zepsrescale}
\begin {equation}
  \tilde z \equiv E_0^{-1/2} z ,
\label {eq:tildez}
\end {equation}
and the original energy deposition distribution $\eps(z)$ that we were
looking for is
\begin {equation}
   \eps(z) \equiv \eps(E_0,z) = E_0^{1/2} \tilde\eps(\tilde z) .
\end {equation}
\end {subequations}

Now that the variable $\tilde z$ has served its purpose, we may
use (\ref{eq:raterescale}) with $E=E_0$, along with (\ref{eq:zepsrescale}),
to rewrite (\ref{eq:epseq0}) in terms of the original, unscaled variables
as
\begin {equation}
  \frac{\partial \eps(z)}{\partial z}
  =
  \int_0^1 dx \> x \biggl[\frac{d\Gamma}{dx}(E_0,x) \biggr]_\net
  \bigl\{ x^{-1/2}\,\eps(x^{-1/2} z)
          - \eps(z) \bigr\} ,
\label {eq:epseq}
\end {equation}
Just remember that this formula is only valid if $[d\Gamma/dx]_\net$
scales with energy as exactly $E^{-1/2}$.

Eq.\ (\ref{eq:epseq}) will be the basic equation underlying the analysis
in the rest of this paper.
Like $[d\Gamma/dx]^\LO$ of (\ref{eq:LOrate0}),
$[d\Gamma/dx]_\net$ diverges
$\propto [x(1{-}x)]^{-3/2}$ for $x\to 0$ and $x\to 1$.
It's useful to note that, nonetheless, the $x$ integration in
(\ref{eq:epseq}) is convergent as $x{\to}1$
because the two terms inside the braces then cancel, and it is also
convergent as $x\to0$ because of (i) the overall factor of $x$ in the integrand
and (ii) the fact that the energy deposition distribution
$\eps(z')$ must fall rapidly (at least exponentially)
to zero as $z' \to \infty$.

% =========================================================================

\section{Moments of the shape \boldmath$S(Z)$}
\label {sec:moments}

The simplest aspects to calculate,
of the energy deposition distribution $\eps(z)$ and its shape $S(Z)$,
are their moments.

Before we start, we give a clarification about numerical accuracy.
In this section, we give a variety of numerical results for moments
in tables \ref{tab:moments}-\ref{tab:shapeF}, where we will implicitly
pretend that the fit (\ref{eq:fit}) to our NLO/LO rate ratio
$f(x)$ is exactly correct.
In reality, though our fit is good, it is only an approximation to
$f(x)$.  We have not attempted to make systematic estimates of the
error arising from this approximation.
However, from our experience in (i)
varying the number of terms in our fits and (ii)
improvement over time of the accuracy of the values that culminated
in our table \ref{tab:dGnet}, we estimate that the final results
for the relative size of overlap effects on moments of $S(Z)$
should be accurate to roughly two significant figures.

% --------------------------------------------------------------------------

\subsection{Recursion formula for moments of \boldmath$\eps(z)$}

To find a formula for the moments, multiply both sides of (\ref{eq:epseq})
by $z^n$ and integrate over $z$.
After integrating by parts on the left-hand side of the equation,
one finds the recursion relation
\begin {equation}
  -n \langle z^{n-1}\rangle
  =
  \int_0^1 dx \> x \biggl[\frac{d\Gamma}{dx}(E_0,x)\biggr]_\net
  \bigl\{ x^{n/2} \langle z^n \rangle
         - \langle z^n \rangle \bigr\} ,
\end {equation}
giving
\begin {subequations}
\label {eq:znMaster}
\begin {equation}
   \langle z^n \rangle =
   \frac{ n \langle z^{\,n-1} \rangle }
        { \Avg[ x(1-x^{n/2}) ] }
   \,,
\label {eq:zn}
\end {equation}
where we find it convenient to introduce the notation
\begin {equation}
   \Avg[g(x)] \equiv
   \int_0^1 dx \> \biggl[\frac{d\Gamma}{dx}(E_0,x)\biggr]_\net g(x) .
\end {equation}
\end {subequations}
The moments $\langle Z^n \rangle$
of the shape $S(Z)$ [defined by (\ref{eq:shape})]
are given in terms
of the moments (\ref{eq:zn}) as simply
\begin {equation}
  \langle Z^n \rangle =
  \frac{\langle z^n \rangle}{\langle z \rangle^n} \,.
\end {equation}

As examples, the stopping distance is
\begin {equation}
  \ell_\stop \equiv \langle z \rangle
  =
   \frac{ 1 }
        { \Avg[x(1-\sqrt{x}\,)] }
   \,,
\end {equation}
and the width of the energy deposition distribution is
$\sigma = \bigl( \langle z^2\rangle - \langle z \rangle^2 \bigr)^{1/2}$
with
\begin {equation}
  \langle z^2 \rangle
  =
   \frac{ 2 \ell_\stop }
        { \Avg[x(1-x)] }
   \,.
\end {equation}
The width of the shape $S(Z)$ is then
\begin {equation}
  \sigma_S = \frac{\sigma}{\lstop}
  = \left(
      \frac{ 2 \Avg[x(1-\sqrt{x}\,)] }{ \Avg[x(1-x)] }
      - 1
    \right)^{1/2} .
\end {equation}

% --------------------------------------------------------------------------

\subsection{Expansion in \boldmath$\alphas$ and results}

We now want to expand results to NLO in
$\alphas = \alphas(\mu)$ to compute the relative size of the
changes to
the moments due to overlapping formation times effects.
We imagine splitting the rate into
\begin {equation}
   \Bigl[ \frac{d\Gamma}{dx} \Bigr]_\net =
   \Bigl[ \frac{d\Gamma}{dx} \Bigr]^\LO_\eff
   + \Bigl[ \frac{d\Gamma}{dx} \Bigr]^{\NLO,\fac}_\net
\end {equation}
as discussed in section \ref{sec:fac}.  We expand the
moments as
\begin {subequations}
\label {eq:znexpand}
\begin {equation}
   \langle z^n \rangle \simeq
   \langle z^n \rangle_\LO^\eff + \delta \langle z^n \rangle \,,
\label {eq:znexpand0}
\end {equation}
where $\langle z^n \rangle_\LO^\eff$ represents the result obtained
using $[d\Gamma/dx]_\eff^\LO$ instead of $[d\Gamma/dx]_\net$ in
(\ref{eq:znMaster}), and $\delta \langle z^n \rangle$ represents the
factorized NLO correction to $\langle z^n \rangle_\LO^\eff$ at first order
in $[d\Gamma/dx]^{\NLO,\fac}_\net$.
Remember that, adopting the nomenclature of section \ref{sec:notation},
\begin {equation}
   \langle z^n \rangle_\LO^\eff
   =
   \langle z^n \rangle_\LO \, [1 + O(\sqrt{\alphas}\,)] .
\end {equation}
\end {subequations}
Expanding the recursion relation (\ref{eq:zn}) gives
\begin {equation}
  \delta \langle z^n\rangle
  = \langle z^n \rangle_\LO
    \left[
       \frac{ \delta\langle z^{n-1} \rangle }
            { \langle z^{n-1} \rangle_\LO }
       -
       \frac{ \dAvg[ x(1-x^{n/2}) ] }
            { \Avg[ x(1-x^{n/2}) ]_\LO }
    \right] ,
\label {eq:dzn}
\end {equation}
where
\begin {subequations}
\label {eq:Avg}
\begin {align}
   \Avg[g(x)]_\LO &\equiv
   \int_0^1 dx \> \biggl[\frac{d\Gamma}{dx}(E_0,x)\biggr]^\LO g(x) ,
\label {eq:AvgLO}
\\
   \dAvg[g(x)]\, &\equiv
   \int_0^1 dx \> \biggl[\frac{d\Gamma}{dx}(E_0,x)\biggr]^{\NLO,\fac}_\net g(x) ,
\end {align}
\end {subequations}
and $\delta\langle z^0 \rangle \equiv 0$.
The LO moments are determined recursively by the analog of
(\ref{eq:zn}),
\begin {equation}
   \langle z^n \rangle_\LO =
   \frac{ n \langle z^{\,n-1} \rangle_\LO }
        { \Avg[ x(1-x^{n/2}) ]_\LO }
   \,.
\label {eq:znLO}
\end {equation}

Though it's not our ultimate goal, we give results for the first few
moments $\langle z^n \rangle$ in table \ref{tab:moments}.
These were calculated using (\ref{eq:LOrate0}) for the LO rate and
using
\begin {equation}
  \left[ \frac{d\Gamma}{dx} \right]^{\NLO,\fac}_\net
  =
  \CA\alphas \left[ \frac{d\Gamma}{dx} \right]^\LO f(x)
\end {equation}
with fit function (\ref{eq:fit}) and
$\Lambda_\fac = x(1{-}x) E$
for the NLO rate.
The parametric scale for the stopping distance is
\begin {equation}
   \ell_\stop \sim \frac{1}{\CA\alphas} \sqrt{ \frac{E_0}{\qhatA} } ,
\end {equation}
and so we've expressed the moments in table \ref{tab:moments} in
appropriate units of
\begin {equation}
   \ell_0 \equiv \frac{1}{\CA\alphas} \sqrt{ \frac{E_0}{\qhatA} } \,.
\label{eq:ell0}
\end {equation}

\begin {table}[t]

\setlength{\tabcolsep}{7pt}
\begin{tabular}{lccccc}
\toprule
  \multicolumn{1}{c}{$z^n$}
  & \multicolumn{1}{c}{$\langle z^n\rangle_\LO$}
  & \multicolumn{1}{c}{$\delta\langle z^n \rangle$}
  &
  & \multicolumn{1}{c}{$\langle z^n\rangle_\LO^{1/n}$}
  & \multicolumn{1}{c}{$\delta[\langle z^n \rangle^{1/n}]$}
 \\
\cline{2-3}\cline{5-6}
  & \multicolumn{2}{c}{in units of $\ell_0^{\kern1pt n}$}
  &
  & \multicolumn{2}{c}{in units of $\ell_0$} \\
\hline
$z$   & 2.1143 & 2.2338\,$\CA\alphas$ && 2.1143 & 2.2338\,$\CA\alphas$ \\ 
$z^2$ & 5.7937 & 12.191\,$\CA\alphas$ && 2.4070 & 2.5324\,$\CA\alphas$ \\
$z^3$ & 18.758 & 59.214\,$\CA\alphas$ && 2.6570 & 2.7959\,$\CA\alphas$ \\
$z^4$ & 68.534 & 289.00\,$\CA\alphas$ && 2.8772 & 3.0332\,$\CA\alphas$ \\
\botrule
\end{tabular}
\caption{
   \label{tab:moments}
   Expansions (\ref{eq:znexpand}) of the moments $\langle z^n \rangle$
   of the energy deposition distribution $\eps(z)$
   for $\Lambda_\fac = x(1{-}x)E$ [(\ref{eq:Lambdafac}) with $\kappa=1$].
   The last two columns show similar expansions of
   $\langle z^n \rangle^{1/n}$, for which
   $\delta[\langle z^n \rangle^{1/n}] =
    \frac{1}{n} \langle z^n \rangle^{(1/n)-1}_\LO \, \delta\langle z^n\rangle$.
   The unit $\ell_0$ is defined by (\ref{eq:ell0}).
}
\end{table}

Because different moments $\langle z^n \rangle$
have different dimensions, comparing those
moments would
be comparing apples and oranges.
So we've also converted all the moments
into lengths by presenting the expansions of
$\langle z^n\rangle^{1/n}$ in the last two columns.
In that comparison, the overlap corrections are
roughly $O(100\%) \times \CA\alphas$
relative to the LO results.
This is similar in size to the NLO
corrections that we saw for $[d\Gamma/dx]_\net$ in section
\ref{sec:dGfacNums}.

Now look instead at the analog of $\langle z^n \rangle^{1/n}$
for moments of the shape function $S(Z)$:
\begin {equation}
   \langle Z^n \rangle^{1/n} = \frac{\langle z^n \rangle^{1/n}}{\langle z \rangle}
   \,.
\end {equation}
Their expansions to NLO are given in table \ref{tab:shape},
now using the adjustable factorization scale
$\Lambda_\fac = \kappa x(1{-}x)$ and explicitly showing the $\kappa$
dependence of the results.%
\footnote{
   If we had shown $\kappa$ dependence
   for the moments of table \ref{tab:moments}, they
   would have double log dependence on $\kappa$.  For example,
   $\langle z \rangle =
    2.1143 + (2.2338 + 0.3084 \ln\kappa - 0.0841 \ln^2\kappa)$
   in units of $\ell_0$.
   We didn't show this for everything since
   we are focused on the shape function, which is
   not affected by constant changes in $\qhat$.
}
In all these entries, $\chi\alphas$ is our name for the {\it relative}
size of NLO corrections:
\begin {equation}
   \chi\alphas \equiv \frac{\delta Q}{Q_\LO}
\end {equation}
for any quantity $Q$.

Table \ref{tab:shape} similarly show results for
$(\mu_{n,S})^{1/n}$,
where the {\it reduced} moment $\mu_{n,S}$ of the shape $S(Z)$ is
\begin {equation}
   \mu_{n,S} \equiv \bigl\langle (Z-\langle Z\rangle)^n \bigr\rangle .
\end {equation}
Our motivational example of such a moment \cite{finale} is
\begin {equation}
  \sigma_S = \frac{\sigma}{\ell_\stop} = \mu_{2,S}^{1/2} ,
\end {equation}
for which the relative size $\chi\alphas$ of NLO corrections is
roughly $-2\% \times \CA\alphas$ for $\kappa=1$ and which remains small
for $\kappa$ varied over any reasonable range.
All the other $\langle Z^n \rangle^{1/n}$ and $(\mu_{n,{\rm S}})^{1/n}$
entries in table \ref{tab:shape} have similarly small NLO corrections.

\begin {table}[t]

\setlength{\tabcolsep}{7pt}
\begin{tabular}{lccc}
\toprule
  \multicolumn{1}{c}{quantity $Q$}
  & \multicolumn{1}{c}{$Q_\LO$}
  & \multicolumn{1}{c}{$\delta Q$}
  & \multicolumn{1}{c}{$\chi\alphas$}
 \\
\hline
$\langle Z \rangle$
   & 1 \\
$\langle Z^2 \rangle^{1/2}$
   & 1.1384
   & $(-0.0050 + 0.0004 \ln\kappa)\,\CA\alphas$
   & $(-0.0044 + 0.0003 \ln\kappa)\,\CA\alphas$ \\ 
$\langle Z^3 \rangle^{1/3}$
   & 1.2567
   & $(-0.0053 + 0.0006 \ln\kappa)\,\CA\alphas$
   & $(-0.0042 + 0.0005 \ln\kappa)\,\CA\alphas$ \\ 
$\langle Z^4 \rangle^{1/4}$
   & 1.3608
   & $(-0.0031 + 0.0007 \ln\kappa)\,\CA\alphas$
   & $(-0.0023 + 0.0005 \ln\kappa)\,\CA\alphas$ \\[1pt]
\hline
$\mu_{2,S}^{1/2} = k_{2,{\rm S}}^{1/2} = \sigma_S$
   & 0.5441
   & $(-0.0104 + 0.0008 \ln\kappa)\,\CA\alphas$
   & $(-0.0191 + 0.0014 \ln\kappa)\,\CA\alphas {}^{\strut}$\\ 
$\mu_{3,S}^{1/3} = k_{3,{\rm S}}^{1/3}$
   & 0.4587
   & $(\phantom{+}0.0139 + 0.0004 \ln\kappa)\,\CA\alphas$
   & $(\phantom{+}0.0303 + 0.0010 \ln\kappa)\,\CA\alphas$ \\ 
$\mu_{4,S}^{1/4}$
   & 0.7189
   & $(\phantom{+}0.0011 + 0.0006 \ln\kappa)\,\CA\alphas$
   & $(\phantom{+}0.0016 + 0.0009 \ln\kappa)\,\CA\alphas$ \\[2pt]
\hline
$k_{4,S}^{1/4}$
   & 0.2561
   & $(\phantom{+}0.3242 - 0.0086 \ln\kappa)\,\CA\alphas$
   & $(\phantom{+}1.2662 - 0.0338 \ln\kappa)\,\CA\alphas {}^{\strut}$ \\[2pt] 
\botrule
\end{tabular}
\caption{
   \label{tab:shape}
   Expansions involving moments $\langle Z^n \rangle$, reduced moments
   $\mu_{n,S}$, and cumulants $k_{n,S}$ of the shape function $S(Z)$.
   Here we take $\Lambda_\fac = \kappa x(1{-}x)$ and show the $\kappa$
   dependence of the results.
   There are no NLO entries for $\langle Z \rangle$ because
   $\langle Z \rangle = 1$ and $\langle Z \rangle_\LO = 1$
   by definition of $Z \equiv z/\langle z\rangle$.
   See the caveat about significant figures given at the beginning of
   section \ref{sec:moments}; we estimate that our results for
   $\chi\alphas$ are valid to roughly two significant digits, once
   one accounts for approximation error to the NLO/LO rate ratio
   $f(x)$.
}
\end{table}

Not content to leave well enough alone, we also considered similar
expansions involving the cumulants
$k_{n,S}$ of $S(Z)$ up through $n=4$.  For $n<4$, cumulants are the
same as reduced moments, but
\begin {equation}
   k_{4,S} \equiv \mu_{4,S} - 3 \mu_{2,S}^2 .
\label {eq:k4}
\end {equation}
As can be seen in table \ref{tab:shape}, the NLO correction for
$k_{4,S}^{1/4}$ is large --- more than $100\% \times \CA\alphas$!
This is because the LO values on the right-hand side of (\ref{eq:k4})
cancel to within 2\%, and so the relatively small NLO corrections
to $\mu_{4,S}$ and $3 \mu_{2,S}^2$ become a large relative correction
to what's left over.

One can worry if the large correction to $k_{4,S}$ is an
important effect, or whether something important may
happen for moments beyond $n{=}4$.  A simple way to settle
this is to calculate the corrections to the shape function $S(Z)$
itself rather than merely its moments.
It's trickier to get accurate numerics for $S(Z)$, but we will be
be able to see that the NLO corrections to $S(Z)$ are all very small,
the fourth cumulant $k_{4,S}$ not withstanding.

% --------------------------------------------------------------------------

\subsection{A formula for later}

We gave recursive expressions for
$\delta\langle z^n\rangle$ and $\langle z^n\rangle_\LO$ in
(\ref{eq:dzn}) and (\ref{eq:znLO}),
but we have not bothered to explicitly write formulas for each
$\delta Q$ in table \ref{tab:moments} in terms of
$\delta\langle z^n\rangle$ and $\langle z^n\rangle_\LO$
and thence in terms of integrals.
For later reference, it will be helpful to have one explicit
example:  $\chi\alphas = (\delta Q)/(Q_\LO)$ in the
case of $Q = \sigma_S = \sigma/\lstop$.
Starting from
$\sigma = \bigl( \langle z^2 \rangle - \langle z \rangle^2 \bigr)^{1/2}$
and $\lstop = \langle z \rangle$, we have
\begin {equation}
  \delta\sigma_S =
  \delta\left( \frac{\sigma}{\lstop} \right)
  =
  \sigma_{S,\LO} \left(
     \frac{\delta(\sigma^2)}{2\sigma^2_\LO}
     - \frac{\delta\langle z\rangle}{\langle z\rangle_\LO}
  \right)
  =
  \sigma_{S,\LO} \left(
     \frac{
       \delta\langle z^2 \rangle
       - 2 \langle z\rangle_\LO \, \delta\langle z \rangle
     }{
       2(\langle z^2 \rangle_\LO - \langle z\rangle_\LO^2)
     }
     - \frac{\delta\langle z\rangle}{\langle z\rangle_\LO}
  \right) ,
\end {equation}
and so
\begin {equation}
  [\chi\alphas]_{\sigma_S} =
     \frac{
       \delta\langle z^2 \rangle
       - 2 \langle z\rangle_\LO \, \delta\langle z \rangle
     }{
       2(\langle z^2 \rangle_\LO - \langle z\rangle_\LO^2)
     }
     - \frac{\delta\langle z\rangle}{\langle z\rangle_\LO}
  .
\label {eq:chi2a}
\end {equation}
Combined with (\ref{eq:dzn}) and (\ref{eq:znLO}), that's good enough for
numerics.  If desired, one may simplify this formula to%
\footnote{
  The averages in the first term of (\ref{eq:chi2}) are related to
  the averages of $x(1-x^{n/2})$ that arise in an evaluation of
  (\ref{eq:chi2a}) by
  the linearity of the definitions
  (\ref{eq:Avg}) of $\dAvg$ and $\Avg_\LO$ in their
  argument, which gives
  $\dAvg[x(1-\sqrt{x})^2] = 2\dAvg[x(1-\sqrt{x})] - \dAvg[x(1-x)]$
  and similarly for $\Avg_\LO$.
}  
\begin {equation}
  [\chi\alphas]_{\sigma_S} =
   \frac{ \dAvg[x(1-\sqrt{x})^2] }
        { 2 \Avg[x(1-\sqrt{x})^2]_\LO }
   -
   \frac{ \dAvg[x(1-x)] }
        { 2 \Avg[x(1-x)]_\LO } \,.
\label {eq:chi2}
\end {equation}

% --------------------------------------------------------------------------

\subsection{An alternate choice: \boldmath$\Lambda_\fac = r E$}

Before moving on, there is another check that can be made of the
robustness of our qualitative conclusion that NLO corrections to moments
(other than the fourth cumulant) are tiny relative to LO results.
In section \ref{sec:rE}, we argued that the choice $\Lambda_\fac = rE$,
where $r$ is an $O(1)$ constant, is a poor choice of factorization
scale for small $x(1{-}x)$ but should be adequate for defining
the factorization of
the shower's energy deposition distribution $\eps(z)$, and hence
shape $S(Z)$, into
$\LO_\eff$ and NLO pieces.
Our $[d\Gamma/dx]^{\NLO,\fac}_\net$ can be converted from our original choice
$\Lambda_\fac = x(1{-}x)E$ to $\Lambda_\fac = rE$ using (\ref{eq:convertr})
and then used to compute moments.
Table \ref{tab:shaper} shows the result of converting the last column
$\chi\alphas$ of table \ref{tab:shape} to $\Lambda_\fac = rE$.%
\footnote{
   $\kappa=1$ was our canonical choice for
   $\Lambda_\fac = \kappa x(1{-}x)E$.
   In table \ref{tab:shaper}, we implicitly
   made $r=\frac14$ our ``canonical'' choice
   for $\Lambda_\fac = rE$, just because it matches
   $\Lambda_\fac = x(1{-}x)E$ for perfectly
   democratic splittings $x=0.5$.
   This is the reason we write the logs in table \ref{tab:shaper} as
   $\ln(4r)$, so that the logs vanish for $r=\tfrac14$.
}

\begin {table}[t]

\setlength{\tabcolsep}{7pt}
\begin{tabular}{lc}
\toprule
  \multicolumn{1}{c}{quantity $Q$}
  & \multicolumn{1}{c}{$\chi\alphas$ ($\Lambda_\fac{=}rE$)}
 \\
\hline
$\langle Z \rangle$
   & \\
$\langle Z^2 \rangle^{1/2}$
   & $(0.0023 + 0.0058 \ln(4r))\,\CA\alphas$ \\ 
$\langle Z^3 \rangle^{1/3}$
   & $(0.0051 + 0.0082 \ln(4r))\,\CA\alphas$ \\ 
$\langle Z^4 \rangle^{1/4}$
   & $(0.0081 + 0.0090 \ln(4r))\,\CA\alphas$ \\[1pt]
\hline
$\mu_{2,S}^{1/2} = k_{2,{\rm S}}^{1/2} = \sigma_S$
   & $(0.0102 + 0.0252 \ln(4r))\,\CA\alphas {}^{\strut}$\\ 
$\mu_{3,S}^{1/3} = k_{3,{\rm S}}^{1/3}$
   & $(0.0429 + 0.0140 \ln(4r))\,\CA\alphas$ \\ 
$\mu_{4,S}^{1/4}$
   & $(0.0236 + 0.0169 \ln(4r))\,\CA\alphas$ \\[2pt]
\hline
$k_{4,S}^{1/4}$
   & $(0.8415 - 0.4878 \ln(4r))\,\CA\alphas {}^{\strut}$ \\[2pt] 
\botrule
\end{tabular}
\caption{
   \label{tab:shaper}
   Like the last column of table \ref{tab:shape} (the relative size
   of NLO corrections) but computed here for factorization scale
   $\Lambda_\fac = r E$.
}
\end{table}

Like table \ref{tab:shape}, the relative sizes of NLO corrections
remain small, except for $k_{4,S}^{1/4}$.
Note that results for $\Lambda_\fac = rE$
are more sensitive to the exact choice of $r$ than results
for $\Lambda_\fac = \kappa x(1{-}x)E$ were to the choice of $\kappa$.

% --------------------------------------------------------------------------

\subsection{The relative importance of F diagrams}

Table \ref{tab:dGnet}, or a comparison of figs.\ \ref{fig:dGnet}
and \ref{fig:dGnetF}, shows that F=4+I diagrams
(like those of fig.\ \ref{fig:Fexamples}) make a relatively small
contribution to $[d\Gamma/dx]^{\NLO,\fac}_\net$ for
$\Lambda_\fac=x(1{-}x)E$.
Was it (with hindsight) important to include them in our analysis?
It's interesting to examine their contribution to the shape $S(Z)$ of
energy deposition,
which is insensitive to changes that can be absorbed into $\hat q$.
How much do F diagrams affect the relative size $\chi\alphas$ of NLO
corrections, like those given in table \ref{tab:shape}?
Table \ref{tab:shapeF} shows the relative contribution of
F diagrams to $\chi\alphas$ compared to the total of all NLO diagrams.
Their effect is small for
our favorite characteristic $\mu_{2,S}^{1/2} = \sigma/\lstop$ of the shape.
However, their relative effect is larger for higher moments like
$\mu_{4,S}^{1/4}$.

\begin {table}[t]

\setlength{\tabcolsep}{7pt}
\begin{tabular}{lr}
\toprule
  \multicolumn{1}{c}{quantity $Q$}
  & \multicolumn{1}{c}{
    $\qquad\frac{\displaystyle{\chi\alphas~(\mbox{F diags only}) \strut}}
          {\displaystyle{\chi\alphas~(\mbox{total})}}$
    }
 \\[5pt]
\hline
$\langle Z \rangle$
   & \\
$\langle Z^2 \rangle^{1/2}$
   & $-14\%$ \\ 
$\langle Z^3 \rangle^{1/3}$
   & $-25\%$ \\
$\langle Z^4 \rangle^{1/4}$
   & $-63\%$ \\[1pt]
\hline
$\mu_{2,S}^{1/2} = k_{2,{\rm S}}^{1/2} = \sigma_S$
   & $-14\% {}^{\strut}$\\ 
$\mu_{3,S}^{1/3} = k_{3,{\rm S}}^{1/3}$
   & $18\%$ \\ 
$\mu_{4,S}^{1/4}$
   & $225\%$ \\[2pt]
\hline
$k_{4,S}^{1/4}$
   & $4\% {}^{\strut}$ \\[2pt] 
\botrule
\end{tabular}
\caption{
   \label{tab:shapeF}
   The relative contribution of F=4+I diagrams to the $\chi\alphas$
   values listed in table \ref{tab:shape} for $\kappa=1$.
}
\end{table}

The take-away is that calculation of the F diagrams \cite{qcdI} was
important for getting good estimates of
some of the shape moments in a particular factorization scheme,
but their inclusion or exclusion did
not affect the answer to the qualitative question of whether NLO corrections
are large.

% =========================================================================

\section{The full shape \boldmath$S(Z)$}
\label{sec:shape}

We now turn to finding the full shape function $S(Z)$ expanded to first
order in $[d\Gamma/dx]^{\NLO,\fac}_\net$.

% -------------------------------------------------------------------------

\subsection{Method}

First, return to the basic equation (\ref{eq:epseq}) for $\eps(z)$.
It will be useful for numerics and the following discussion to switch to
dimensionless variables
\begin {equation}
   \hat z \equiv \frac{z}{\ell_0} ,
   \qquad
   \hat \eps(\hat z) \equiv
       \frac{\ell_0}{E_0} \, \eps( \ell_0 \hat z ) ,
   \qquad
   \frac{d\hat\Gamma}{dx} =
       \ell_0 \, \frac{d\Gamma}{dx}
   \,,
\label {eq:hatvars}
\end {equation}
with $\ell_0$ defined by (\ref{eq:ell0}).
Then
\begin {equation}
  \frac{\partial\hat\eps(\hat z)}{\partial\hat z}
  =
  \int_0^1 dx \> x \biggl[\frac{d\hat\Gamma}{dx} \biggr]_\net
  \bigl\{ x^{-1/2}\,\hat\eps(x^{-1/2}\hat z)
          - \hat\eps(\hat z) \bigr\} .
\label {eq:epseq2}
\end {equation}
The leading-order version is just
\begin {equation}
  \frac{\partial\hat\eps_\LO(\hat z)}{\partial\hat z}
  =
  \int_0^1 dx \> x \biggl[\frac{d\hat\Gamma}{dx}\biggr]^\LO
  \bigl\{ x^{-1/2}\,\hat\eps_\LO(x^{-1/2}\hat z)
          - \hat\eps_\LO(\hat z) \bigr\} .
\label {eq:epseqLO}
\end {equation}

To solve (\ref{eq:epseqLO}) numerically, we follow a procedure similar to
ref.\ \cite{qedNfstop}.%
\footnote{
  Specifically, see appendix B of ref.\ \cite{qedNfstop}.
}
First, we start with an approximate asymptotic solution for large $\hat z$,
\begin {equation}
   \hat\eps_\LO(\hat z) \sim e^{-\hat z^2/\pi} ,
\label {eq:zasymp}
\end {equation}
which is derived in appendix \ref{app:asymp}.
[This leading exponential dependence is also the same as that for
the Blaizot/Iancu/Mehtar-Tani (BIM) model for showers, discussed in
appendix \ref{app:BIM}.]
We choose a large value $\hat z_{\rm max} \gg 1$
and use (\ref{eq:zasymp}) for
$\hat z > \hat z_{\rm max}$.
Since (\ref{eq:epseqLO}) is a linear equation, it does not care about the
overall normalization of $\hat\eps_\LO$, and so we initially take
$\hat\eps_\LO(\hat z) = e^{-\hat z^2/\pi}$ for $\hat z > \hat z_{\rm max}$
and postpone
normalizing $\hat\eps_\LO$ until later.

Next, we choose a small increment $\Delta\hat z \ll 1$ and approximate
(\ref{eq:epseqLO}) by
\begin {equation}
  \hat\eps_\LO(\hat z - \Delta z)
  \simeq
  \hat\eps_\LO(\hat z)
  - \Delta z
  \int_0^1 dx \> x \biggl[\frac{d\hat\Gamma}{dx}\biggr]^\LO
  \bigl\{ x^{-1/2}\,\hat\eps_\LO(x^{-1/2}\hat z)
          - \hat\eps_\LO(\hat z) \bigr\} .
\label {eq:epseqLO2}
\end {equation}
Note that, for any value of $\hat z$, the arguments of the function
$\hat\eps_\LO$ on the right-hand side of (\ref{eq:epseqLO2}) are never smaller
than $\hat z$ itself.
So, starting with $\hat z=\hat z_\max$, we use (\ref{eq:epseqLO2})
repeatedly, step by step,
to calculate $\hat\eps_\LO(\hat z)$ for smaller and smaller values of $\hat z$,
until we get to $\hat z=0$.
When we are done, we then normalize $\hat\eps_\LO(\hat z)$ so that
\begin {equation}
   \int_0^\infty d\hat z \> \hat\eps_\LO(\hat z) = 1 .
\label {eq:epsLOnorm}
\end {equation}
A few more details about numerical implementation are given in
appendix \ref{app:epsNumerics}.

Next, we substitute
\begin {equation}
   \hat\eps(\hat z) \simeq \hat\eps_\LO(\hat z) + \delta\hat\eps(\hat z)
\end {equation}
into (\ref{eq:epseq2}) and expand to first order in NLO quantities, giving
\begin {multline}
  \frac{\partial\,\delta\hat\eps(\hat z)}{\partial\hat z}
  =
  \int_0^1 dx \> x \biggl[\frac{d\hat\Gamma}{dx} \biggr]^\LO
  \bigl\{ x^{-1/2}\,\delta\hat\eps(x^{-1/2}\hat z)
          - \delta\hat\eps(\hat z) \bigr\}
\\
  +
  \int_0^1 dx \> x \biggl[\frac{d\hat\Gamma}{dx} \biggr]^{\NLO,\fac}_\net
  \bigl\{ x^{-1/2}\,\hat\eps_\LO(x^{-1/2}\hat z)
          - \hat\eps_\LO(\hat z) \bigr\} .
\label {eq:epseqNLO}
\end {multline}
If not for the last term, this would have the same form as the
LO equation (\ref{eq:epseqLO}).  The last term, however, acts as
a driving term generated by the previously computed $\hat\eps_\LO(\hat z)$.
To solve (\ref{eq:epseqNLO}), we discretize it similar to
(\ref{eq:epseqLO2}) and start with $\delta\hat\eps(\hat z)=0$ for
$\hat z > \hat z_\max$.  Let $\delta\hat\eps_1(\hat z)$ be the solution obtained
through this procedure.

If $\delta\hat\eps_1(\hat z)$ is
a solution to (\ref{eq:epseqNLO}), then so is
\begin {equation}
   \delta\hat\eps(\hat z) = \delta\hat\eps_1(\hat z) + c\, \hat\eps_\LO(\hat z)
\label {eq:depsGeneral}
\end {equation}
for any constant $c$.
The solution we need is one consistent with normalizing
$\hat\eps = \hat\eps_\LO + \delta\hat\eps$ so that
$\int d\hat z \> \hat\eps(\hat z) = 1$ through first order.
That normalization requires
\begin {equation}
   \int_0^\infty d\hat z \> \delta\hat\eps(\hat z) = 0 .
\end {equation}
The properly normalized solution (\ref{eq:depsGeneral}) can be obtained
from any particular solution $\delta\hat\eps_1$ by
\begin {equation}
   \delta\hat\eps(\hat z) =
   \delta\hat\eps_1(\hat z)
   - \hat\eps_\LO(\hat z)
     \int_0^\infty d\hat z \> \delta\hat\eps_1(\hat z) ,
\end {equation}
provided we have normalized $\hat\eps_\LO$ as in (\ref{eq:epsLOnorm}).

Finally, the expansion
\begin {equation}
   S(Z) \simeq S_\LO(Z) + \delta S(Z)
\end {equation}
of the shape function (\ref{eq:shape}) to first order in
$[d\Gamma/dx]^{\NLO,\fac}_\net$ can be written in the form
\begin {equation}
  S_\LO(Z) =
  \langle\hat z\rangle_\LO \,
  \hat\eps_\LO\bigl( Z \langle\hat z\rangle_\LO \bigr) ,
\end {equation}
\begin {equation}
  \delta S(Z) =
  \left[
     \langle\hat z\rangle_\LO \, \delta\hat\eps_\LO(\hat\zeta)
     + \delta\langle\hat z\rangle \,
       \frac{d}{d\hat\zeta}\bigl( \hat\zeta\, \hat\eps_\LO(\hat\zeta) \bigr)       
  \right]_{\hat\zeta=Z\langle\hat z\rangle_\LO} ,
\end {equation}
where $\langle\hat z\rangle_\LO$ is evaluated
using $\hat\eps_\LO$, and
$\delta\langle \hat z\rangle$ is
\begin {equation}
   \delta\langle\hat z\rangle =
   \int_0^\infty d\hat z \> \hat z \,\delta\hat\eps(\hat z) .
\end {equation}

% -------------------------------------------------------------------------

\subsection{Results and Checks}

\begin {figure}[t]
\begin {center}
  \includegraphics[scale=0.3]{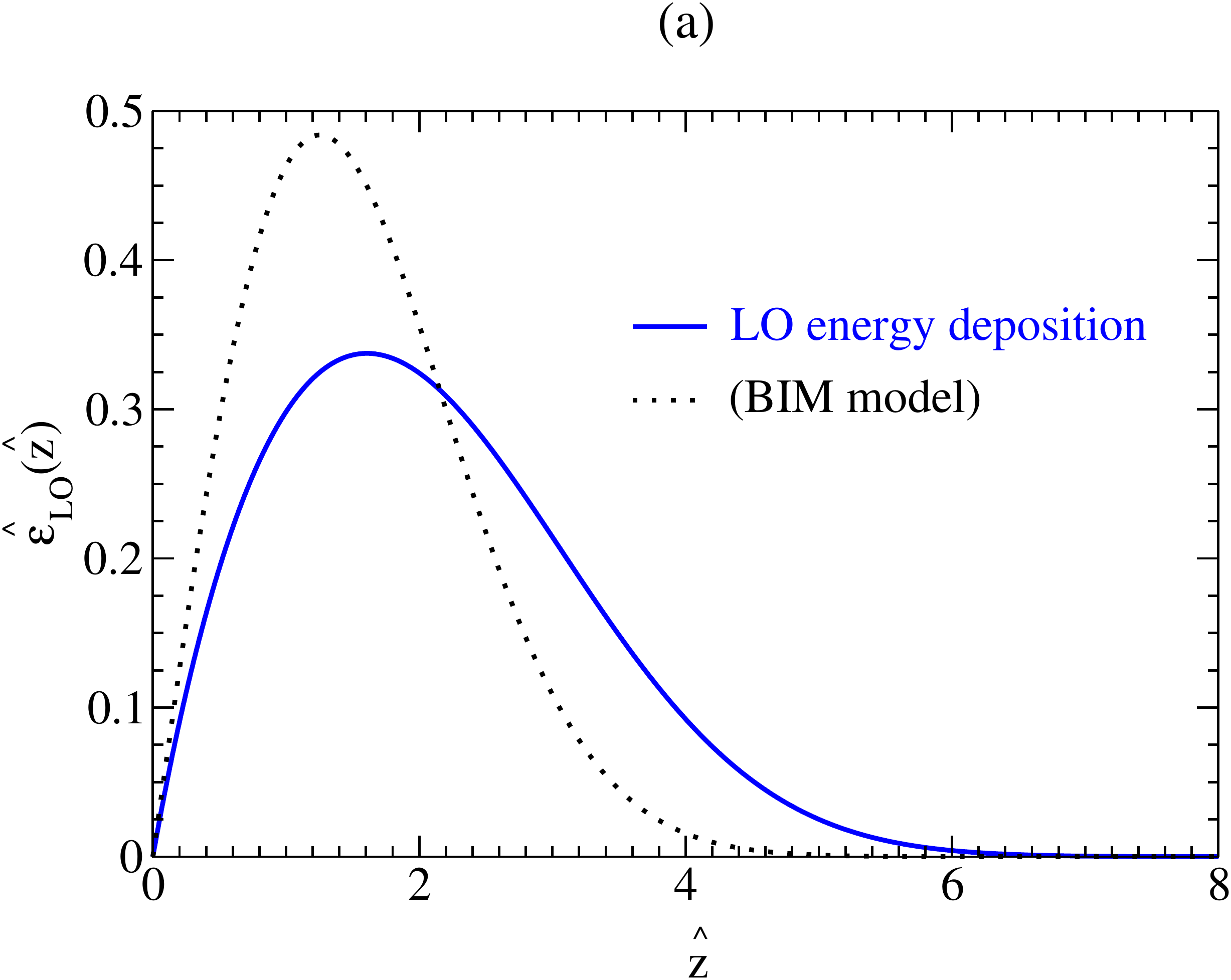}
  \hspace{2em}
  \includegraphics[scale=0.3]{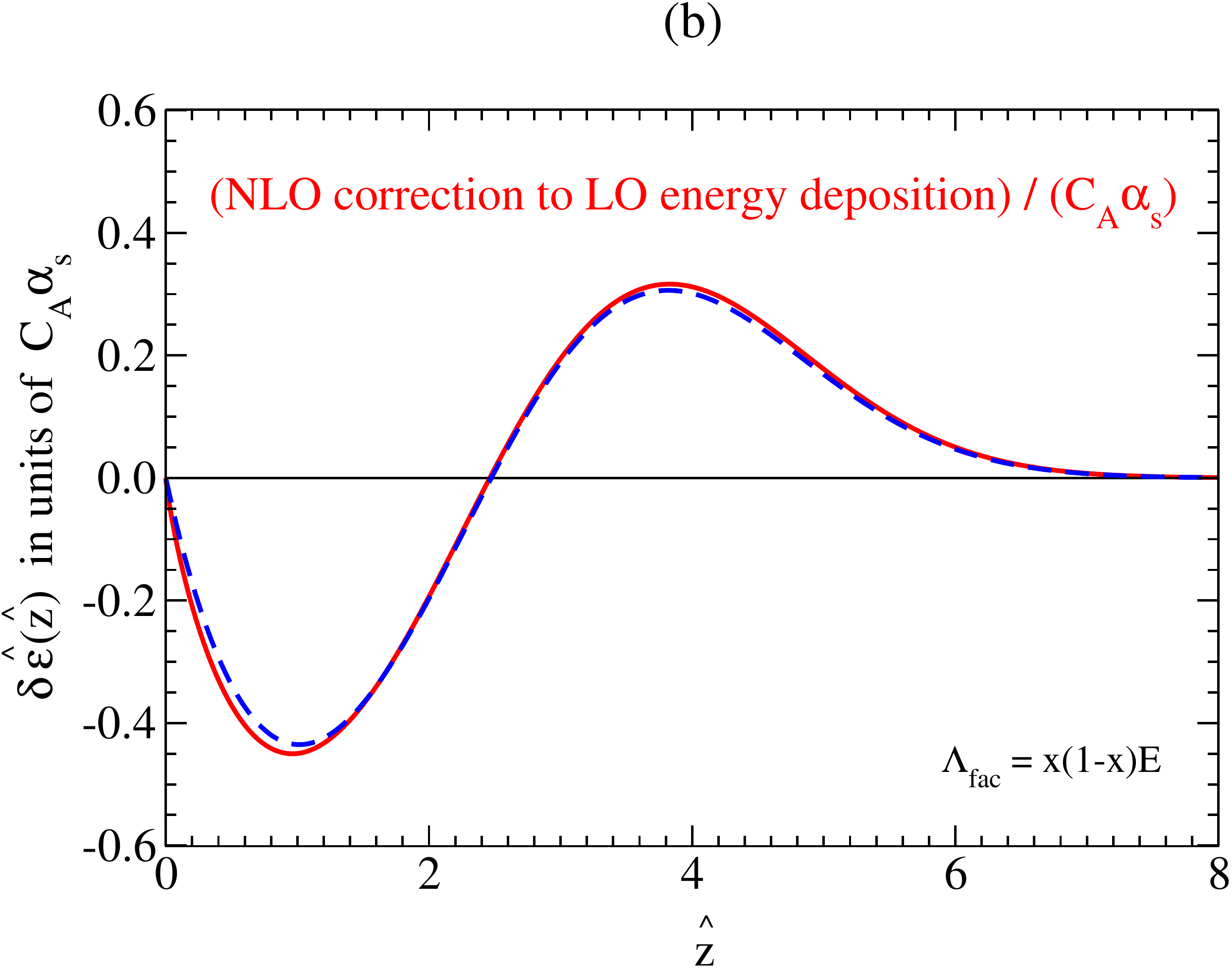}
  \caption{
     \label{fig:eps}
     (a) The solid curve shows the energy deposition distribution
     $\hat\eps_\LO(\hat z)$
     vs.\ $\hat z \equiv z/\ell_0$,
     where the unit $\ell_0$ is defined in (\ref{eq:ell0}).
     [For comparison, the dotted curve shows an analytic result
     (\ref{eq:BIMeps}) derived from the BIM model.]
     (b) A similar plot of $\delta\hat\eps_\LO(\hat z)/\CA\alphas$
     for our canonical choice $\Lambda_\fac=x(1{-}x)E$ of
     factorization scale.  For comparison, the dashed curve shows
     the first-order change (\ref{eq:dilate}) that would
     be induced in $\hat\eps_\LO(\hat z)$ by rescaling the
     $\hat z$ axis in fig.\ (a).
  }
\end {center}
\end {figure}

Fig.\ \ref{fig:eps} shows our numerical results for
$\hat\eps_\LO(\hat z)$ and $\delta\hat\eps(\hat z)/\CA\alphas$.
From the latter, we see that NLO corrections to the leading-order
energy deposition distribution are large unless $\CA\alphas$
is indeed small.
Similar to our earlier discussion of the table \ref{tab:moments}
results for the moments of $\eps(z)$, this is not surprising:
Back in fig.\ \ref{fig:dGnet}, we saw that NLO corrections
for the net rate $[d\Gamma/dx]_\net$ decreased the rate by
$O(100\%)\times \CA\alphas$.  A large decrease to the
rate will mean a large change to how soon the shower stops, and
so a large change to where the energy is deposited.
%%In fact, note that our expansion
%%$\hat\eps(\hat z) \simeq \hat\eps_\LO(\hat z) + \delta\hat\eps(\hat z)$
%%would not make physical sense
%%unless $\CA\alphas$ were smaller than roughly 0.25 because
%%the energy distribution $\hat\eps(\hat z)$ must be everywhere positive and
%%$\hat\eps_\LO(\hat z) + \delta\hat\eps(\hat z)$ would not be.

To understand the shape of $\delta\hat\eps(\hat z)$ in fig.\ \ref{fig:eps}b,
consider any change to $\hat\eps_\LO(\hat z)$ that
simply rescales the $\hat z$ axis:
\begin {equation}
   \hat\eps_\LO(\hat z) \rightarrow \lambda\, \hat\eps_\LO(\lambda \hat z) .
\end {equation}
If we increase the stopping distance by choosing $\lambda = 1-\xi$
and then formally expand to first order in $\xi$ (just as we formally
expand our overlap results to first order in $\alphas$), then
the change in $\hat\eps_\LO$ would be proportional to
\begin {equation}
  - \bigl[
      \hat\eps_\LO(\hat z)
      + \hat z\, \hat\eps_\LO^{\,\prime}(\hat z)
    \bigr] .
\label {eq:dilate}
\end {equation}
The dashed line in fig.\ \ref{fig:eps}b is a plot of
(\ref{eq:dilate}) which, to excellent approximation, is proportional
to the solid curve for $\delta\hat\eps(\hat z)/\CA\alphas$.
That is, the corrections that we see in fig.\ \ref{fig:eps}b
can mostly be absorbed into a change in the stopping distance and so
into the value of $\hat q$.

\begin {figure}[t]
\begin {center}
  \includegraphics[scale=0.3]{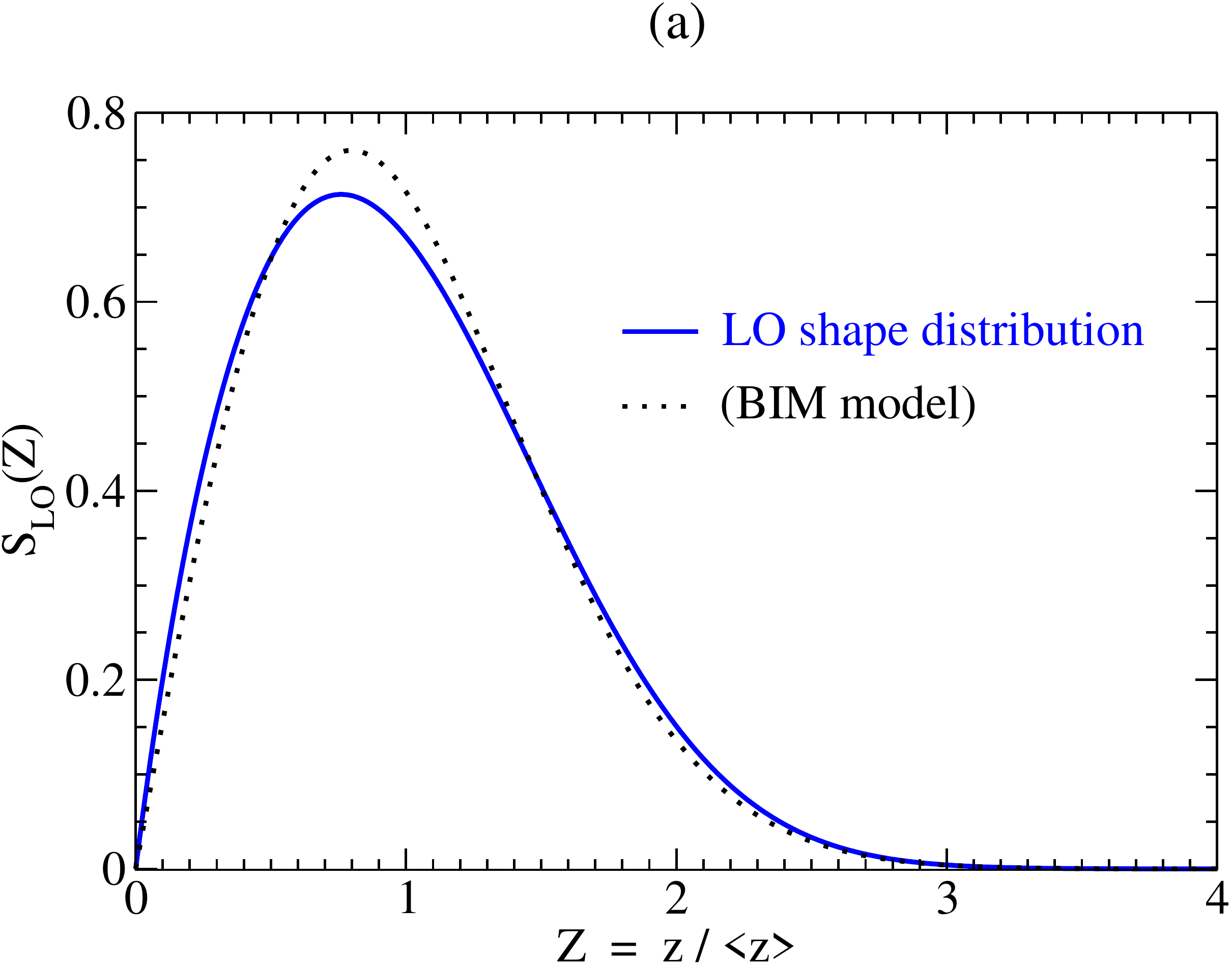}
  \hspace{2em}
  \includegraphics[scale=0.3]{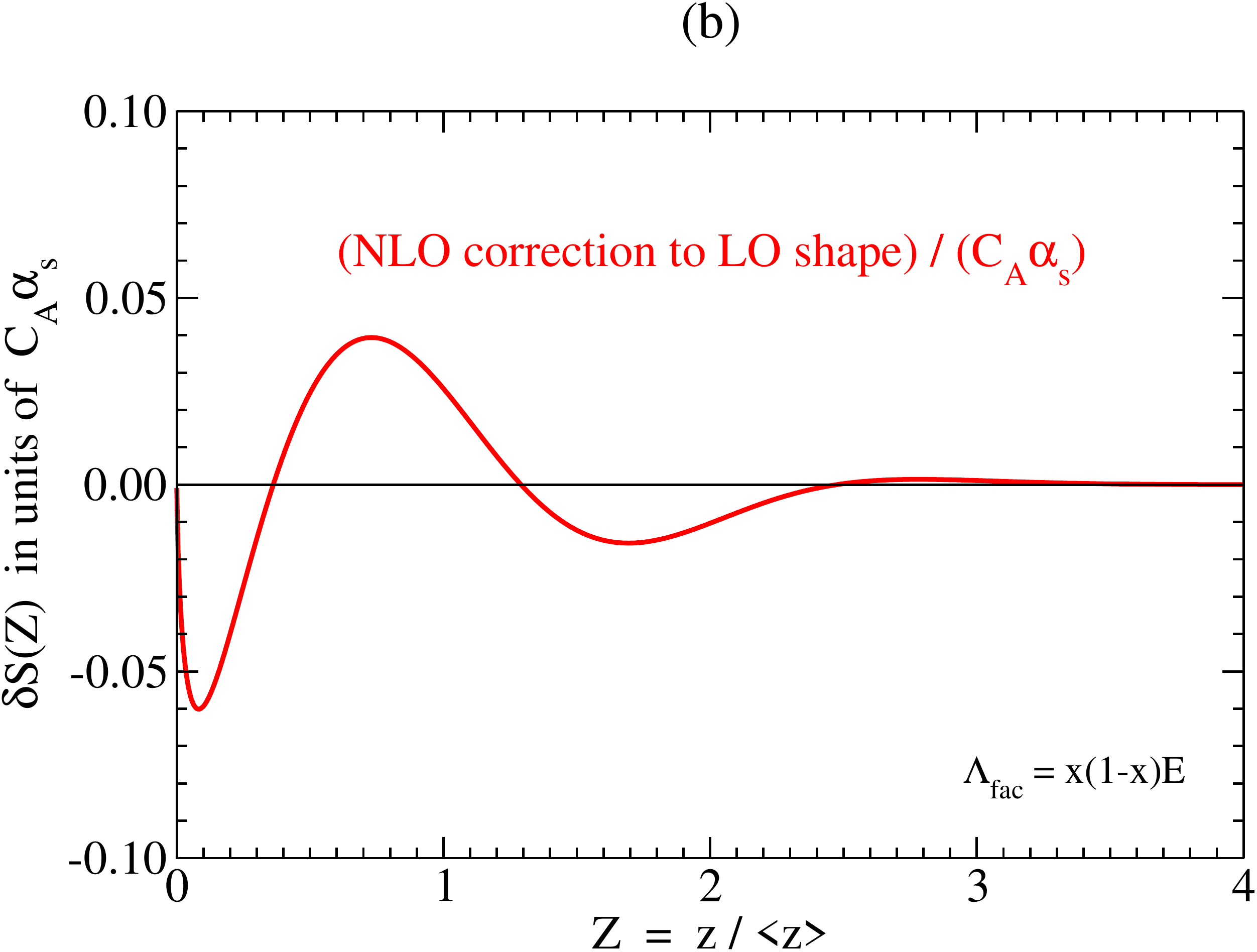}
  \caption{
     \label{fig:shape}
     (a) The solid curve shows $S_\LO(Z)$
     vs.\ $Z \equiv z/\langle z\rangle_\LO$.
     [For comparison, the dotted curve shows the analytic
     result
     (\ref{eq:SBIM}) from the BIM model.]
     (b) A plot of $\delta S(Z)/\CA\alphas$
     for
     our canonical choice $\Lambda_\fac = x(1{-}x)E$ of factorization scale.
     Note the different scale of the vertical axis compared to (a).
  }
\end {center}
\end {figure}

\begin{figure}
\includegraphics[scale=0.3]{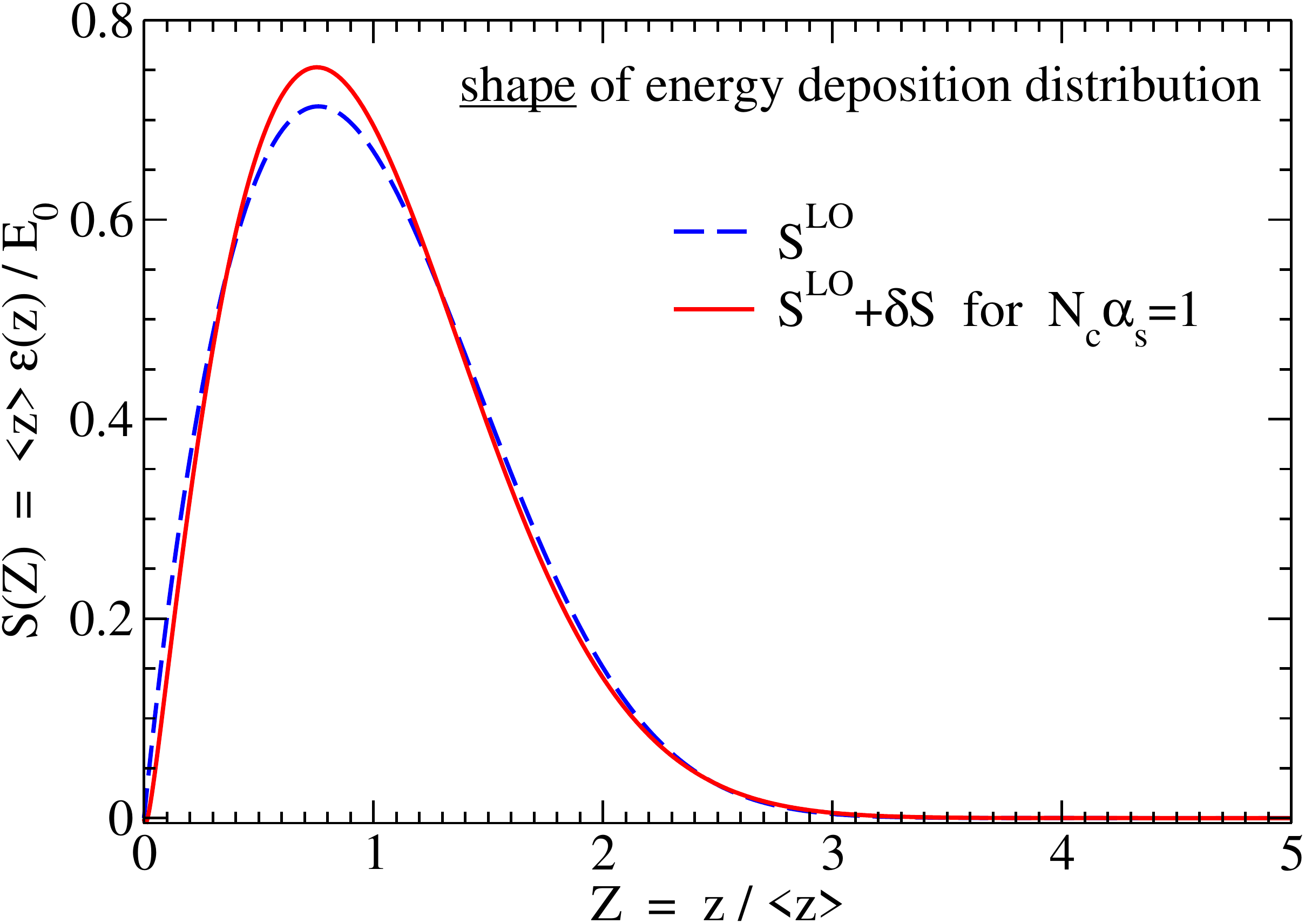}
\caption{
   \label{fig:S}
   Energy deposition shape with and without
   first-order overlapping formation time effects $\delta S$,
   for $\CA\alphas=1$.
}
\end{figure}

Now turn to the shape function $S(Z) \simeq S_\LO(Z) + \delta S(Z)$,
which is insensitive to constant changes that can be absorbed into $\qhat$.
Fig.\ \ref{fig:shape} shows plots of $S_\LO(Z)$ and $\delta S(Z)$.
Here, NLO corrections to $S_\LO(Z)$ are small even for
$\CA\alphas=1$, qualitatively consistent with our results for
the moments of the shape function in table \ref{tab:shape}, but now with
the clarification that the relatively large correction to the
delicate 4th cumulant does not correspond to a significant effect
on the shape distribution $S(Z)$.  To emphasize this point, we
reproduce in fig.\ \ref{fig:S} the comparison presented in
our summary paper \cite{finale} of $S_\LO(Z)$ vs.\
$S_\LO(Z)+\delta S(Z)$ for $\CA\alphas=1$.%
\footnote{
  We've been careful to say $S_\LO(Z)+\delta S(Z)$ instead of simply
  $S(Z)$.  That's because $S(Z)$ at this order is really
  $S_\LO^\eff(Z) + \delta S(Z)$.  Section \ref{sec:LOvEff} explained
  that $S_\LO$ and $S_\LO^\eff$ can be expected to differ already at
  $O(\sqrt{\alphas})$, and we have not calculated $S_\LO^\eff$.
  However, the comparison of $S_\LO$ and $S_\LO + \delta S$ made in
  fig.\ \ref{fig:S} is enough to investigate the relative
  importance of overlap effects $\delta S$.
}

\begin {figure}[t]
\begin {center}
  \includegraphics[scale=0.3]{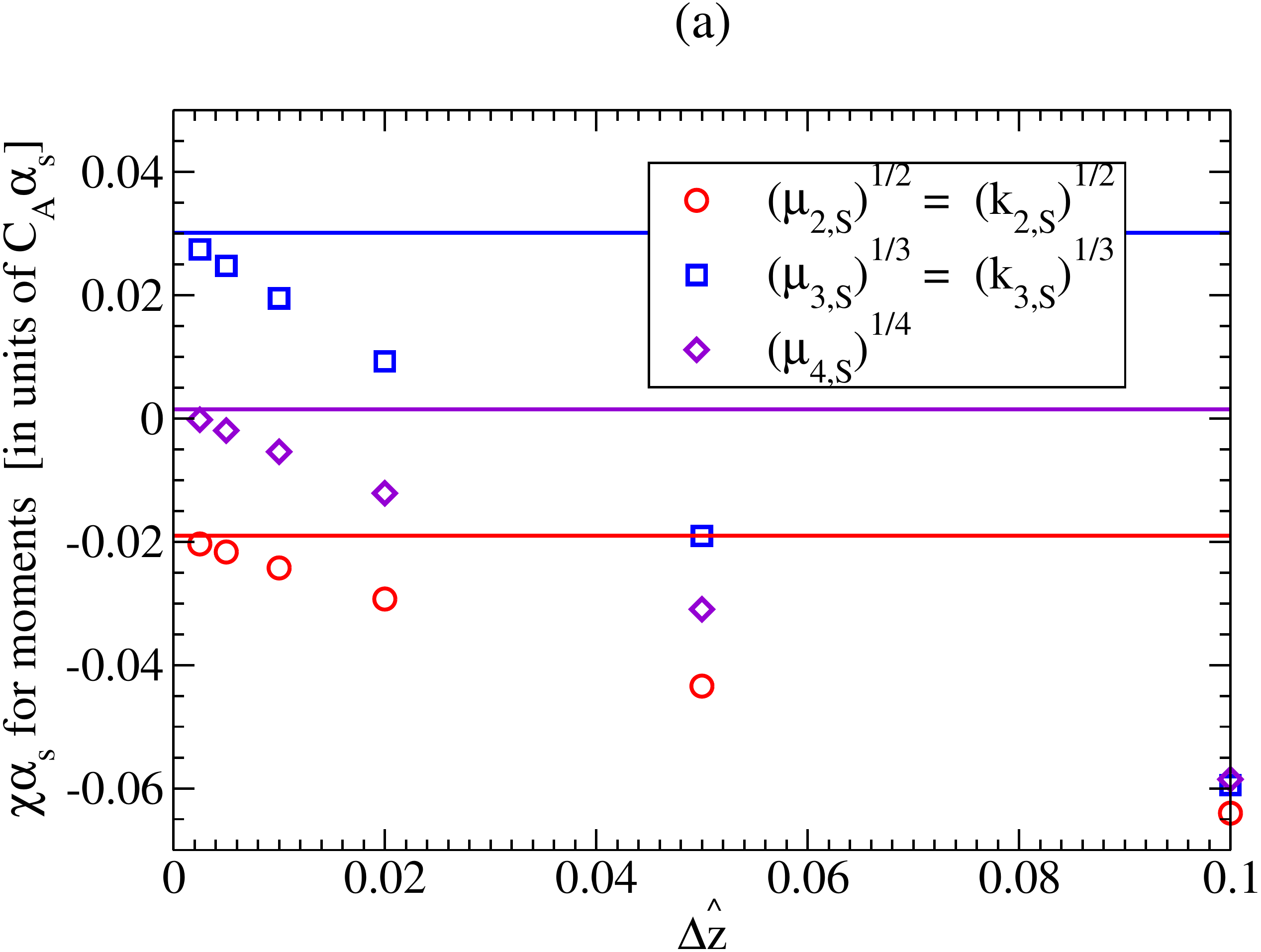}
  \hspace{2em}
  \includegraphics[scale=0.3]{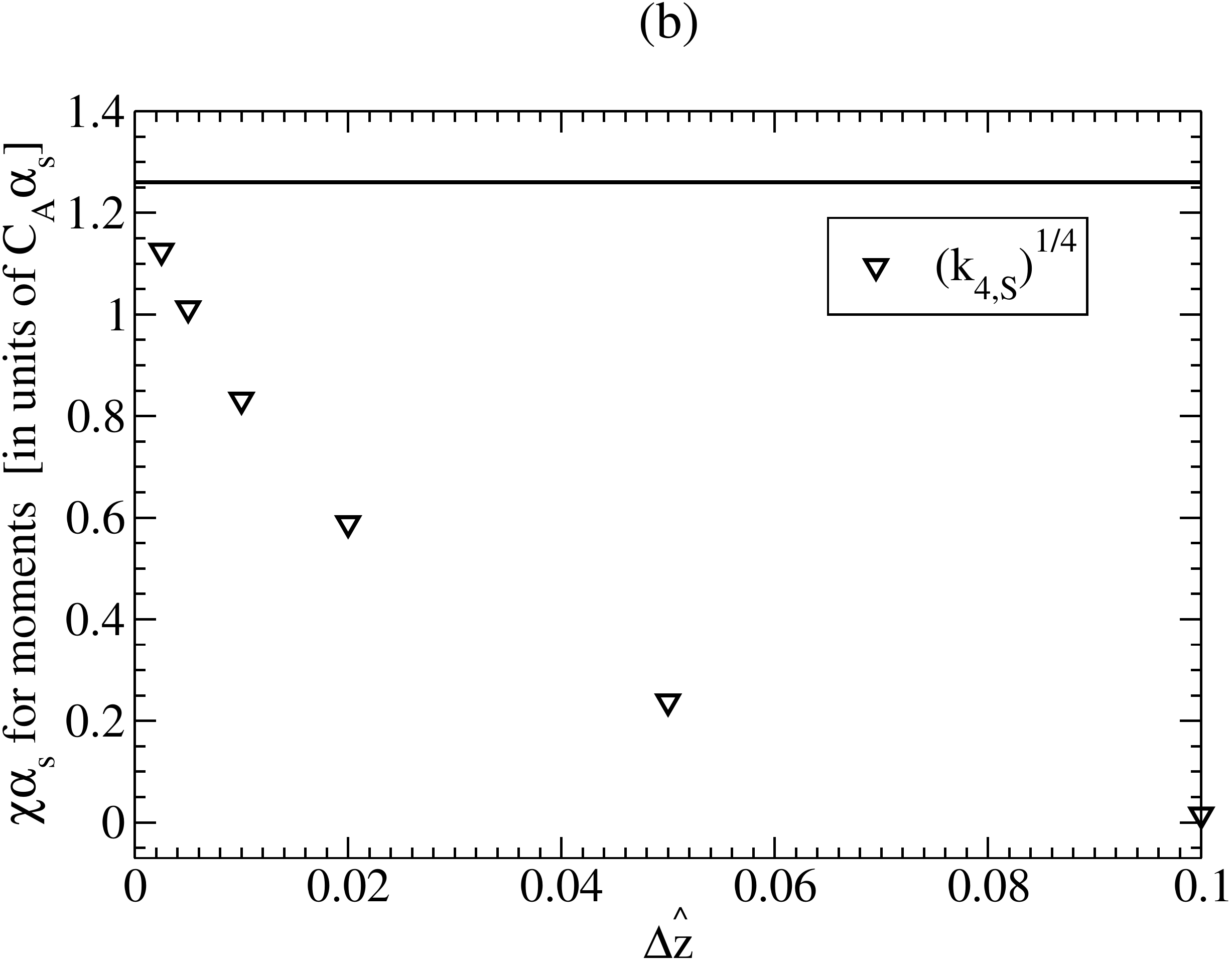}
  \caption{
     \label{fig:dz}
     The horizontal lines show the $\chi\alphas$ results of table
     \ref{tab:shape} for the relative size of NLO corrections to
     reduced moments and cumulants, as computed using
     the direct integration method of section \ref{sec:moments} for
     $\Lambda_\fac = x(1{-}x)E$, i.e.\ $\kappa=1$.
     The data points show, as a function of step size $\Delta\hat z$
     for $\hat z_\max = 20$,
     the same moments computed instead from the $S_\LO(Z)$ and
     $\delta S(Z)$ functions found by the numerical
     methods of section \ref{sec:shape}.
  }
\end {center}
\end {figure}

The shape functions shown in fig.\ \ref{fig:shape} were linearly
extrapolated to the continuum limit $\Delta\hat z = 0$ from simulations
at $(\Delta\hat z,\hat z_\max)=(0.0025,20)$ and $(0.005,20)$.
To check that this is adequate, we compute
moments from our numerical results for $S_\LO(Z)$ and $\delta S(Z)$
and compare them to our earlier moment calculations in
table \ref{tab:shape}.  Specifically, fig.\ \ref{fig:dz} shows
the approach to the continuum limit of the relative size $\chi\alphas$
of NLO corrections to the reduced moments and cumulants.
As one can see from the figure, a linear extrapolation from our
two smallest $\Delta\hat z$ values will do fairly well at reproducing
our earlier (and more precise) moment results.%
\footnote{
  See appendix \ref{app:epsNumerics} for a demonstration that
  errors associated with out choice of $\hat z_\max$ were negligible.
}
The precise numbers do not matter: The point of this exercise is simply
to feel confident enough in the accuracy of figs.\ \ref{fig:shape}
and \ref{fig:S} to support our
qualitative conclusion that the NLO corrections to the shape function
are small for $\CA\alphas \le 1$.

% .........................................................................

\subsubsection*{An aside: BIM model for LO results}

Our focus in this paper is on NLO corrections, which we have compared
to the size of LO results.
Like our NLO corrections, the LO energy deposition
$\eps_\LO(z)$ and shape function $S_\LO(Z)$ have been
computed numerically in figs.\ \ref{fig:eps}a and \ref{fig:shape}a.
It's interesting to compare those numerical results to a
model of LO shower development investigated by
Blaizot, Iancu, and Mehtar-Tani (BIM) \cite{BIM1,BIM2}, which replaces the
LO splitting rate (\ref{eq:LOrate0}) by something simpler that allows
for analytic solutions.
The BIM model of LO shower development gives the dotted curves
in figs.\ \ref{fig:eps}a and \ref{fig:shape}a.
(See our appendix \ref{app:BIM} for details.)
The BIM model result is notably different for the energy
deposition $\eps_\LO(z)$ but is close to the exact LO
result for the shape function $S_\LO(Z)$.
Since our conclusion is that NLO effects for the shape function are
small, the BIM model appears to give a reasonably good approximation to
the shape $S(Z)$ of energy deposition (for the purely gluonic
showers studied here).%
\footnote{
  If one compares the BIM model curve in fig.\ \ref{fig:shape}a
  to the total LO+NLO curve in fig.\ \ref{fig:S}, then the BIM curve
  looks like it matches the total curve even better than it matches
  the LO curve.  But this is accidental and represents a
  somewhat faulty comparison:
  The BIM curve in fig.\ \ref{fig:shape}a is independent of the
  value of $\CA\alphas$, but the difference between the LO and LO+NLO
  curves in fig.\ \ref{fig:S} is proportional to $\CA\alphas$,
  which was somewhat arbitrarily chosen to be $\CA\alphas(\mu)=1$
  for the purpose of fig.\ \ref{fig:S}.
}
That is, its more significant
deviation in the case of $\eps(z)$ could be absorbed into
the value of $\hat q$.

% =========================================================================

\section{Time evolution of gluon distribution}
\label{sec:time}

In this paper, we have focused on characteristics of the energy
deposition distribution $\eps(z)$, for which the basic equation was
(\ref{eq:epseq3}).
One might also be interested, more fundamentally, in the
time evolution of the
distribution of all shower gluon energies as a function of time.
Though we will not make use of it in this paper,
we present here
the basic evolution equation as another example that
all the necessary information about splitting rates is encoded
in the net rate
$[d\Gamma/dx]_\net$.

Ref.\ \cite{qcd} packaged the basic evolution equation as%
\footnote{
  See section 3.1.1 of ref.\ \cite{qcd}, where
  our $n(\zeta,E_0,t)$ here is called $N(\zeta,E_0,t)$ there.
  For a sanity check of why
  $[d\Gamma/dz]_\net$ is appropriate in (\ref{eq:Nevolve0}), see
  footnote 27 of ref.\ \cite{qcd}.
}
\begin {equation}
  \frac{\partial}{\partial t}\, n(\zeta,E_0,t)
  =
  - \Gamma(\zeta E_0) \, n(\zeta,E_0,t)
  + \int_\zeta^1 \frac{dx}{x} \> 
    \left[ \frac{d\Gamma}{dx} \bigl(\tfrac{\zeta E_0}{x},x\bigr)
           \right]_{\rm net}
    n\bigl( \tfrac{\zeta}{x}, E_0, t \bigr) ,
\label {eq:Nevolve0}
\end {equation}
where $n(\zeta,E_0,t)\,d\zeta$ represents the number of gluons with
energy between $\zeta E_0$ and $(\zeta+d\zeta) E_0$ at time $t$.
Our new observation about this equation is simply that (\ref{eq:GammaAlt})
can be used to rewrite (\ref{eq:Nevolve0}) completely in terms of
$[d\Gamma/dx]_\net$:
\begin {multline}
  \frac{\partial}{\partial t} \, n(\zeta,E_0,t)
  =
  \int_0^1 dx \>
  \biggl\{
    \frac{\theta(x>\zeta)}{x}
    \left[
       \frac{d\Gamma}{dx} \bigl(\tfrac{\zeta E_0}{x},x\bigr)
     \right]_{\rm net}
     n\bigl( \tfrac{\zeta}{x}, E_0, t \bigr)
\\
    -
    x \left[ \frac{d\Gamma}{dx} (\zeta E_0,x) \right]_{\rm net}
    n(\zeta,E_0,t)
  \biggr\} .
\label {eq:Nevolve}
\end {multline}

When discussing energy deposition, it's a little easier
to describe the shower (following \cite{BIM1}) in terms of
gluon energy density in $\zeta$,
\begin {equation}
  D(\zeta,E_0,t) \equiv \zeta E_0 \, n(\zeta,E_0,t) ,
\end {equation}
instead of $n(\zeta,E_0,t)$.
The corresponding version of (\ref{eq:Nevolve}) is
\begin {multline}
  \frac{\partial}{\partial t} \, D(\zeta,E_0,t)
  =
  \int_0^1 dx \>
  \biggl\{
    \theta(x>\zeta)
    \left[
       \frac{d\Gamma}{dx} \bigl(\tfrac{\zeta E_0}{x},x\bigr)
     \right]_{\rm net}
     D\bigl( \tfrac{\zeta}{x}, E_0, t \bigr)
\\
    -
    x \left[ \frac{d\Gamma}{dx} (\zeta E_0,x) \right]_{\rm net}
    D(\zeta,E_0,t)
  \biggr\} .
\label {eq:Devolve}
\end {multline}
As time progresses, $D(\zeta,E_0,t)$ develops a $\delta$-function
piece representing the amount of stopped energy:%
\begin {equation}
  D(\zeta,E_0,t) =
  E_{\rm stopped}(E_0,t)\,\delta(\zeta)
  + D_{\rm moving}(\zeta,E_0,t) .
\end {equation}
For a sanity check,
we verify in appendix \ref{app:energy}
that the evolution equation (\ref{eq:Devolve}) conserves total energy.

In applications where the relevant rates scale with energy exactly as
$E^{-1/2}$, one may rescale variables as
\begin {subequations}
\begin {equation}
  t = E_0^{1/2} \tilde t,
  \qquad
  n(\zeta,E_0,t) = \tilde n(\zeta,\tilde t\,) ,
  \qquad
  D(\zeta,E_0,t) = E_0 \tilde D(\zeta,\tilde t\,)
\end {equation}
\begin {equation}
  \biggl[ \frac{d\Gamma}{dx}(E,x) \biggr]_\net
    = E^{-1/2} \biggl[ \frac{d\tilde\Gamma}{dx}(x) \biggr]_\net ,
\end {equation}
\end {subequations}
to simplify (\ref{eq:Nevolve}) to
\begin {equation}
  \frac{\partial}{\partial \tilde t} \, \tilde n(\zeta,\tilde t\,)
  =
  \frac{1}{\zeta^{1/2}}
  \int_0^1 dx \>
  \biggl[ \frac{d\tilde\Gamma}{dx} \biggr]_{\rm net}
  \left\{
    \frac{\theta(x>\zeta)}{x^{1/2}} \,
     \tilde n\bigl( \tfrac{\zeta}{x}, \tilde t\, \bigr) \,
    -
    x \,
    \tilde n(\zeta,\tilde t\,)
  \right\}
\label {eq:NevolveE}
\end {equation}
or equivalently
\begin {equation}
  \frac{\partial}{\partial \tilde t} \, \tilde D(\zeta,\tilde t\,)
  =
  \frac{1}{\zeta^{1/2}}
  \int_0^1 dx \>
  \biggl[ \frac{d\tilde\Gamma}{dx} \biggr]_{\rm net}
  \left\{
    \theta(x>\zeta)\, x^{1/2} \,
     \tilde D\bigl( \tfrac{\zeta}{x}, \tilde t\, \bigr) \,
    -
    x \,
    \tilde D(\zeta,\tilde t\,)
  \right\} .
\label {eq:DevolveE}
\end {equation}
At leading order, where there are only $1{\to}2$ splitting processes,
(\ref{eq:DevolveE}) is equivalent to an evolution equation
used previously by refs.\ \cite{BIM1,BIM2} to study leading-order shower
development in the BIM model.%
\footnote{
  See eq.\ (4) of ref.\ \cite{BIM1}, where their $(x,z)$ are our $(\zeta,x)$.
  Their ${\cal K}(x)$
  (before they make
  the BIM model approximation of replacing ${\cal K}$ by ${\cal K}_0$)
  is our $[d\Gamma/dx]^\LO$,
  up to a trivial overall normalization difference associated with
  their definition of rescaled time $\tau$ vs.\ our $\tilde t$.
}
Through the use of $[d\Gamma/dx]_\net$, our
(\ref{eq:DevolveE}) extends their equation to situations where
there are more than just $1{\to}2$ splitting processes.

Note that $E^{-1/2}$ energy scaling is subtle at NLO, even
when one chooses a factorization scale $\Lambda_\fac \propto E$ such that
$[d\Gamma/dx]^{\NLO,\fac}_\net$ scales as $E^{-1/2}$.
The subtlety is that $[d\Gamma/dx]^\LO_\eff$ then has $E^{-1/2}\ln^2E$
instead of $E^{-1/2}$
dependence on energy.
We have managed to ignore this difficulty
in our analysis only because we have been specifically
interested in the size of $\NLO/\LO_\eff$ ratios, as discussed in
section \ref{sec:LOvEff}.

One reason that we have not attempted to simulate (\ref{eq:DevolveE}) for
this paper is that we expect it would be more numerically challenging
to accurately reproduce the tiny NLO effects of table \ref{tab:shape}.

% =========================================================================

\section{Why are NLO effects so small?}
\label {sec:why}

Why are our results for overlap effects on the shape of energy deposition
so very small?  The simplest characteristic of the shape function, for
example, is its width $\sigma_S = \sigma/\lstop$,
for which the relative size of NLO
corrections listed in table \ref{tab:shape} was
\begin {equation}
   [\chi\alphas]_{\sigma/\lstop}^{\rm energy} =
   (-0.0191 + 0.0014\ln\kappa) \CA\alphas
\label {eq:chig}
\end {equation}
Seemingly, overlap effects which cannot be absorbed into $\qhat$
are almost negligible even for $\CA\alphas(\mu) = 1$ in
large-$\Nc$ Yang-Mills theory.
As noted in the summary paper \cite{finale}, this conclusion
is vastly different than an earlier analysis \cite{qedNfstop} of
overlap effects in large-$\Nf$ QED for
{\it charge} (rather than energy) deposition of a shower initiated
by an electron.  There, the result was
\begin {equation}
   [\chi\alphaqed]_{\sigma/\lstop}^{\rm charge} =
   -0.87\, \Nf\alphaqed ,
\label {eq:chiQED}
\end {equation}
which would be an $O(100\%)$ effect for $\Nf\alphaqed(\mu) = 1$.
When we set out performing the calculations in this paper, we
were expecting gluon shower results somewhat similar in size
to (\ref{eq:chiQED}).
We were very surprised by the tiny result (\ref{eq:chig}).

One could wonder if there might be some miraculous reason why
(\ref{eq:chig}) should be exactly zero for a purely gluonic shower.
Perhaps we were not careful enough with the precision of our numerics,
or perhaps there was some tiny mistake in the rate formulas of
refs.\ \cite{2brem,seq,dimreg,qcd}?  But $\kappa$ parametrizes
our choice of factorization scale $\Lambda=\kappa x(1{-}x)$, and
the $\kappa$ dependence of (\ref{eq:chig}) originates solely from the
double and single IR logarithms subtracted by the definitions
(\ref{eq:factorization}) and (\ref{eq:dGfac}).  The double logarithms
have long been known \cite{Blaizot,Iancu,Wu} and are well studied.
The full single logarithms have been derived by two completely different
methods \cite{logs,logs2} which give the same result.
The steps that lead from there to the $\kappa$ dependence
(\ref{eq:convert}) of the net rate, and then to the $\ln\kappa$ term
in (\ref{eq:chig}), are pretty straightforward.%
\footnote{
   It's worth noting that the $x$-independent terms of the
   $\kappa$ dependence shown in  (\ref{eq:convert}) can be absorbed
   into a constant shift in $\qhat$ and so do not affect the shape
   distribution and so give no NLO corrections $\chi\alphas$ to moments
   of the shape distribution.
   The only term in (\ref{eq:convert}) that does affect $\chi\alphas$
   is the $\hat s(x) \ln\kappa$ term associated with IR single logs.
}   
Since one
$O(1)$ value of $\kappa$ is a good as another, we do not see
how (\ref{eq:chig}) could be a mistaken value for something that
is actually exactly zero for all choices of $\kappa$.

Can we get any insight as to why (\ref{eq:chig}) is so small compared
to the analogous (\ref{eq:chiQED})?  Though we do not have an explanation
of why (\ref{eq:chig}) is as very small as it is, it is possible to
investigate some aspects of the suppression in more detail.

To study this, we will separate how the result (\ref{eq:chig})
depends on $[d\Gamma/dx]^{\NLO,\fac}_\net$ from how it depends on
everything else.  Eq.\ (\ref{eq:chi2}) for
(\ref{eq:chig}) can be rewritten as
\begin {subequations}
\label {eq:chi2W}
\begin {equation}
   [\chi\alphas]_{\sigma/\lstop}^{\rm energy} =
   \int_0^1 dx \> W(x) \, \left[ \frac{d\Gamma}{dx} \right]^{\NLO,\fac}_\net
\label {eq:chi2Wint}
\end {equation}
with weight function $W$ defined by%
\footnote{
  Note that, in (\ref{eq:W}), the variables $x$ appearing in the
  $\Avg[\cdots]_\LO$'s are dummy variables associated with the definition
  (\ref{eq:AvgLO}), unrelated to the integration
  variable $x$ in (\ref{eq:chi2Wint}).
}
\begin {equation}
  W(x') =
     \frac{ x'(1-\sqrt{x'})^2 }
          { 2 \Avg[x(1-\sqrt{x})^2]_\LO }
     -
     \frac{ x'(1-x') }
          { 2 \Avg[x(1-x)]_\LO }
  \,.
\label {eq:W}
\end {equation}
\end {subequations}
Now rewrite the above in
terms of the NLO/LO rate ratio $f(x)$ defined by (\ref{eq:dGnetRatio}):
\begin {subequations}
\label {eq:chi2w}
\begin {equation}
   [\chi\alphas]_{\sigma/\lstop}^{\rm energy} =
   \CA\alphas \int_0^1 dx \> w(x) \, f(x) ,
\label {eq:chi2wint}
\end {equation}
\begin {equation}
  w(x') =
   \left[ \frac{d\Gamma}{dx} (x') \right]^\LO
   \left\{
     \frac{ x'(1-\sqrt{x'})^2 }
          { 2 \Avg[x(1-\sqrt{x})^2]_\LO }
     -
     \frac{ x'(1-x') }
          { 2 \Avg[x(1-x)]_\LO }
   \right\} .
\end {equation}
\end {subequations}
Note that the definition (\ref{eq:AvgLO}) of $\Avg[\cdots]_\LO$ means that
\begin {equation}
  \int_0^1 dx' \> w(x') = 0 .
\label {eq:wnorm}
\end {equation}
This had to be: If $f(x)$ had been an $x$-independent constant, so that
$[d\Gamma/dx]^{\NLO,\fac}_\net \propto [d\Gamma/dx]^\LO$, then the
NLO effects could be completely absorbed into a constant shift in
$\qhat$, and the whole point of looking at shape characteristics such
as $\sigma_S$ is that the shape is insensitive to constant shifts in
$\qhat$.  So the integral (\ref{eq:chi2wint}) must vanish for
constant $f$.

\begin {figure}[t]
\begin {center}
  \includegraphics[scale=0.3]{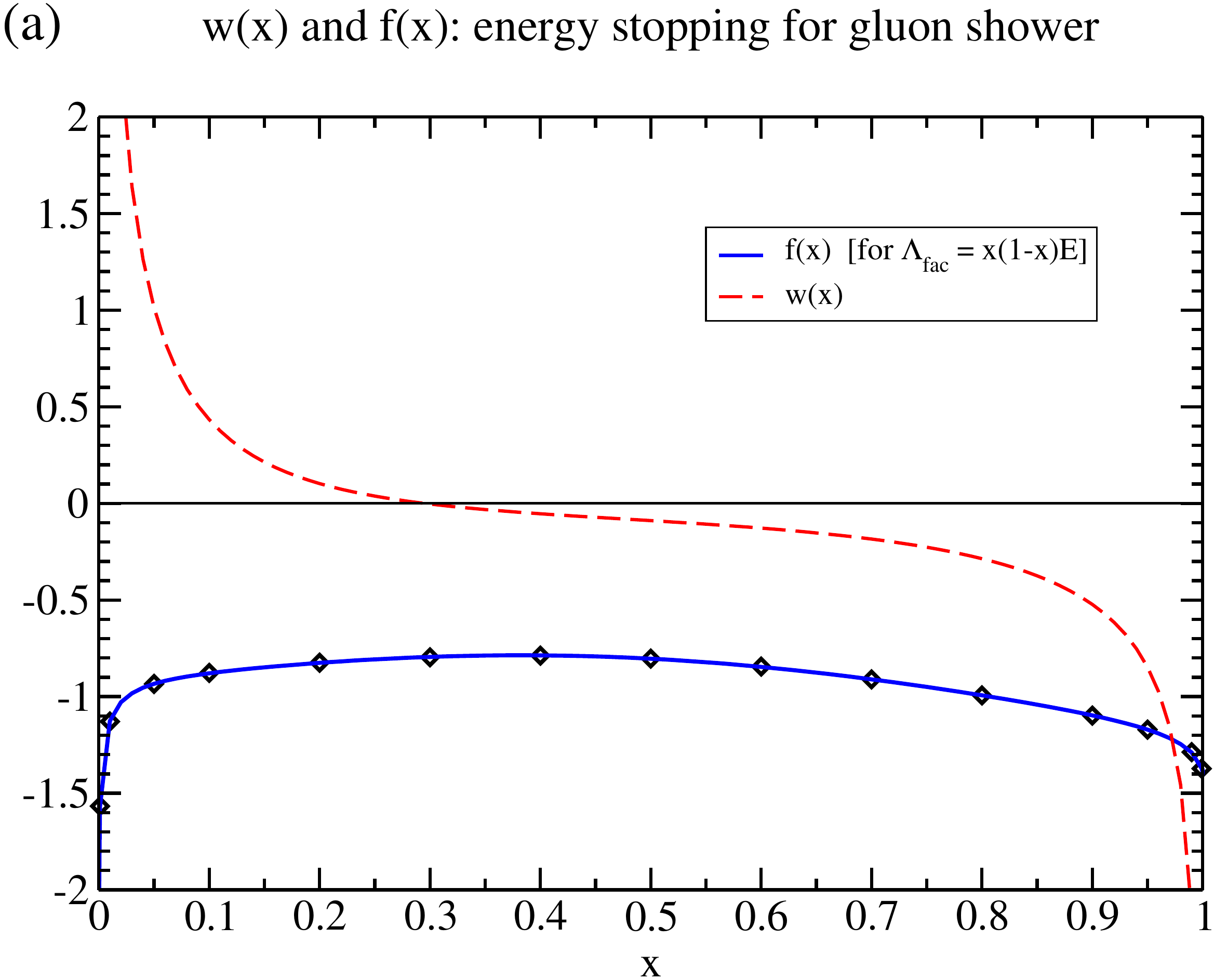}
  \hspace{2em}
  \includegraphics[scale=0.3]{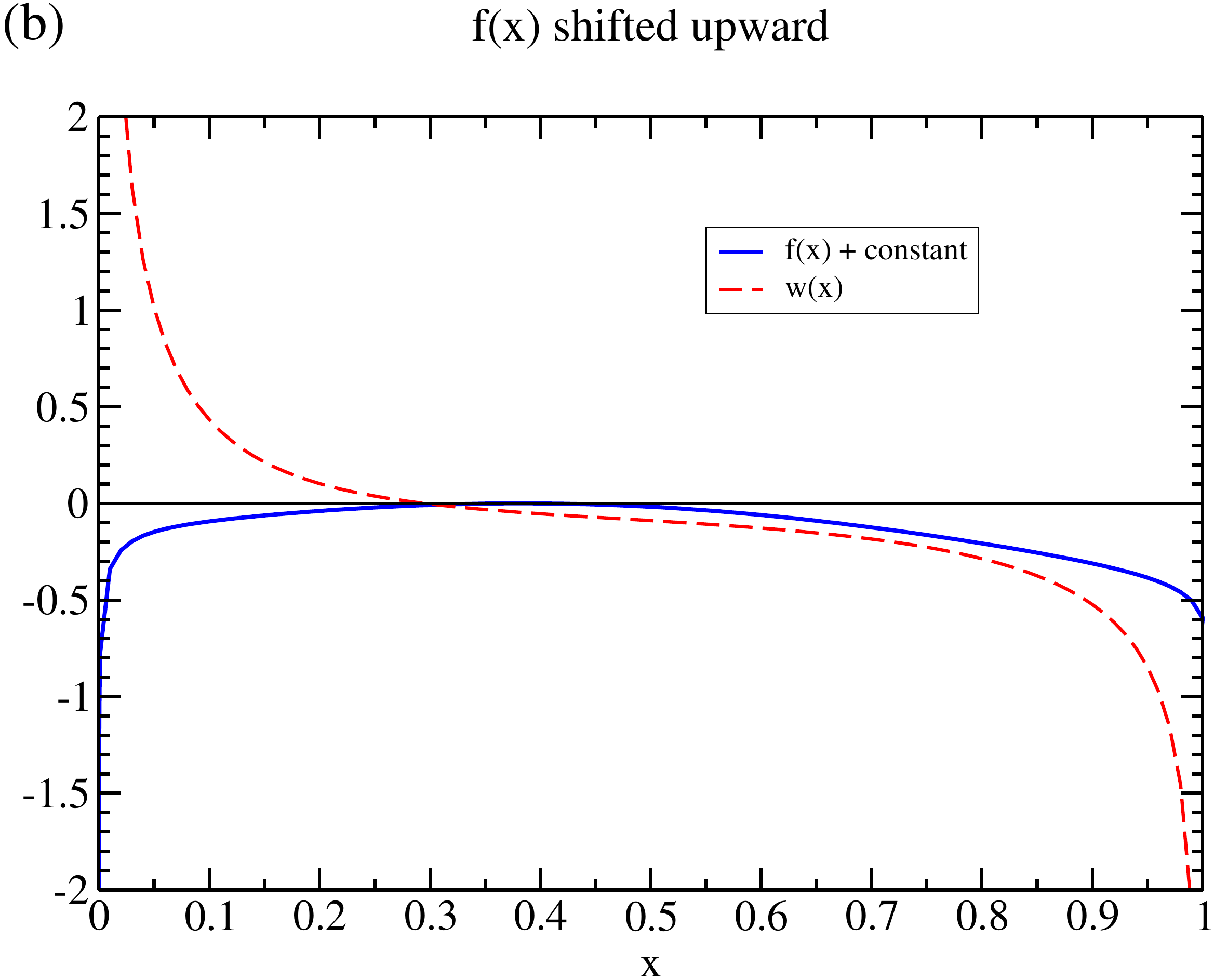}
  \caption{
     \label{fig:w}
     (a) Plot of the NLO/LO net rate ratio $f(x)$ (solid curve) and
     the weight function $w(x)$ in the integral (\ref{eq:chi2wint})
     that gives (\ref{eq:chig}).
     (b) The same, but $f(x)$ is shifted upward by a constant,
     as described in the text.
  }
\end {center}
\end {figure}

Fig.\ \ref{fig:w}a shows a plot of $w(x)$ and $f(x)$.  Because of
(\ref{eq:wnorm}), the $w$ function has to be positive in some places and
negative in others, but note how that manifests: it's positive on the
left of the plot and negative on the right.  It's not really
anti-symmetric in $x \to 1{-}x$, but qualitatively it's a crude
distortion of something ``anti-symmetric.''  In contrast, $f(x)$ has
the same sign on both sides of the plot; it is not really symmetric
in $x \to 1{-}x$, but qualitatively it's a crude distortion of something
symmetric.  Note that the NLO $g{\to}gg$ contribution to $f(x)$ must be exactly
symmetric because the daughter gluons are identical particles, but this
symmetry is not respected by the $g{\to}ggg$ contribution.%
\footnote{
  It wouldn't make sense to plot the NLO $g{\to}gg$ and $g{\to}ggg$
  contributions separately because they have canceling {\it power}-law
  IR divergences \cite{qcd}, which are not handled by our factorization
  scheme (\ref{eq:dGfac}).  One might in principle imagine enhancing our
  factorization scheme to subtract power-law divergences for the
  separate contributions, but it doesn't seem worth the effort (and we do not
  currently have complete analytic results for all of the power-law divergences
  \cite{qcd}).
}
These properties of $f(x)$ and $w(x)$ explain
a partial cancellation when we compute the integral (\ref{eq:chi2wint})
of their product $w(x)\,f(x)$.

\begin {figure}[t]
\begin {center}
  \includegraphics[scale=0.5]{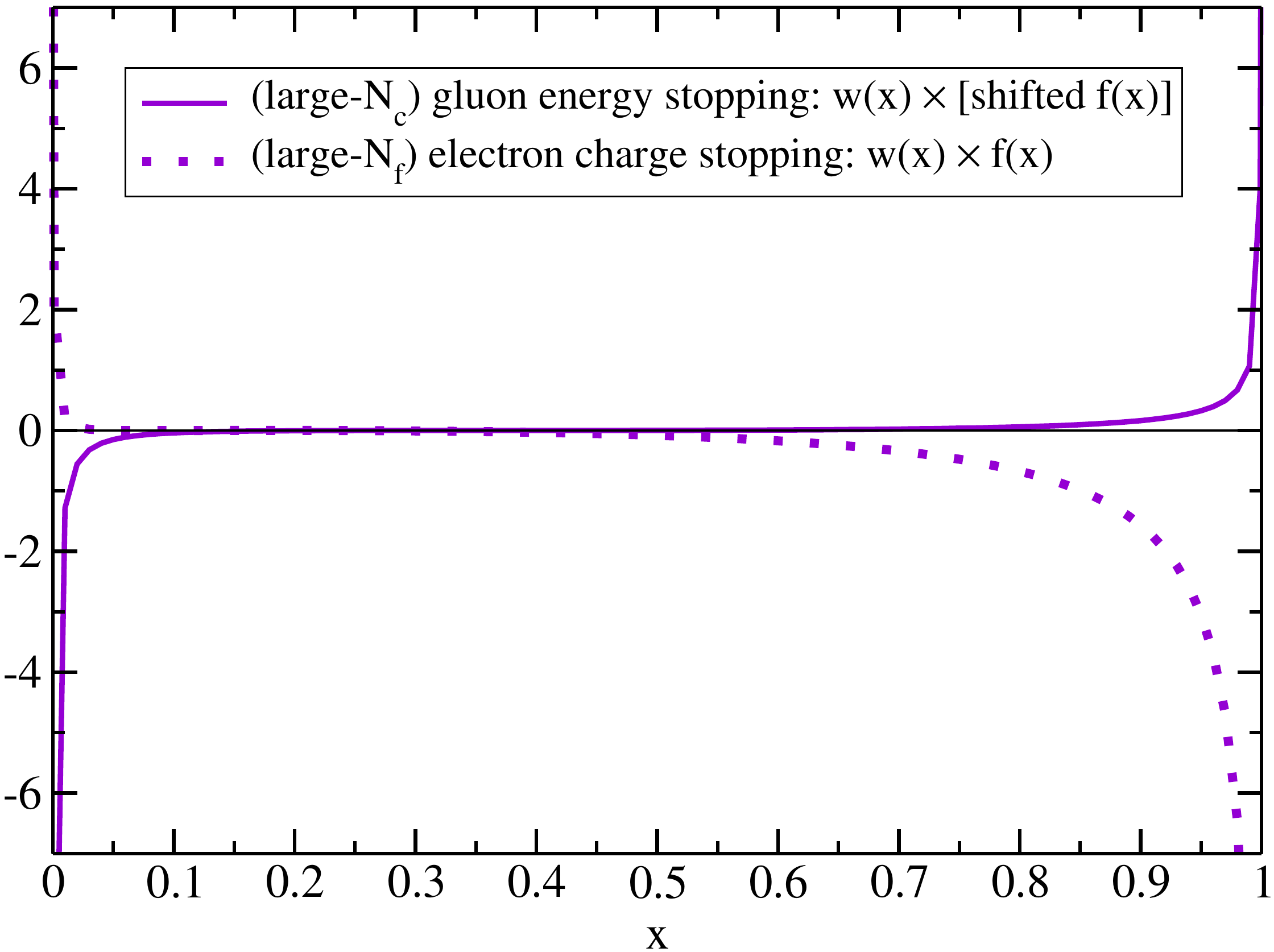}
  \caption{
     \label{fig:wf}
     The solid curve is the product of the $w(x)$ and shifted $f(x)$
     functions of fig.\ \ref{fig:w}, and its integral gives
     (\ref{eq:chig}).
     For comparison, the dotted curve shows a similar product for
     large-$\Nf$ QED (\ref{eq:chiQED}).
  }
\end {center}
\end {figure}

We will make the last statement more concrete by plotting $w(x)\,f(x)$,
but we find it more visually advantageous to first eliminate one
piece that does not contribute to $\chi\alphas$.
Note that, because of (\ref{eq:wnorm}), the integral (\ref{eq:chi2wint})
for $\chi\alphas$ will be unchanged
if we replace $f(x)$ by $f(x) + c$, for any constant $c$.
We choose to
replace fig.\ \ref{fig:w}a by fig.\ \ref{fig:w}b, where we've
chosen $c$ to make $f(x)+c$ small for the middle range of $x$
values, while still maintaining that $f(x)+c$, like $f(x)$, has
the same sign everywhere.  Now we plot the product $w(x)\,[f(x)+c]$
as the solid curve in fig.\ \ref{fig:wf}.
The value of $\chi\alphas$ is the area under that curve.
One sees a positive contribution from the far right of the plot,
partly canceled by
a negative contribution from the far left, though
it's hard to judge visually how precisely they cancel.

\begin {figure}[t]
\begin {center}
  \includegraphics[scale=0.3]{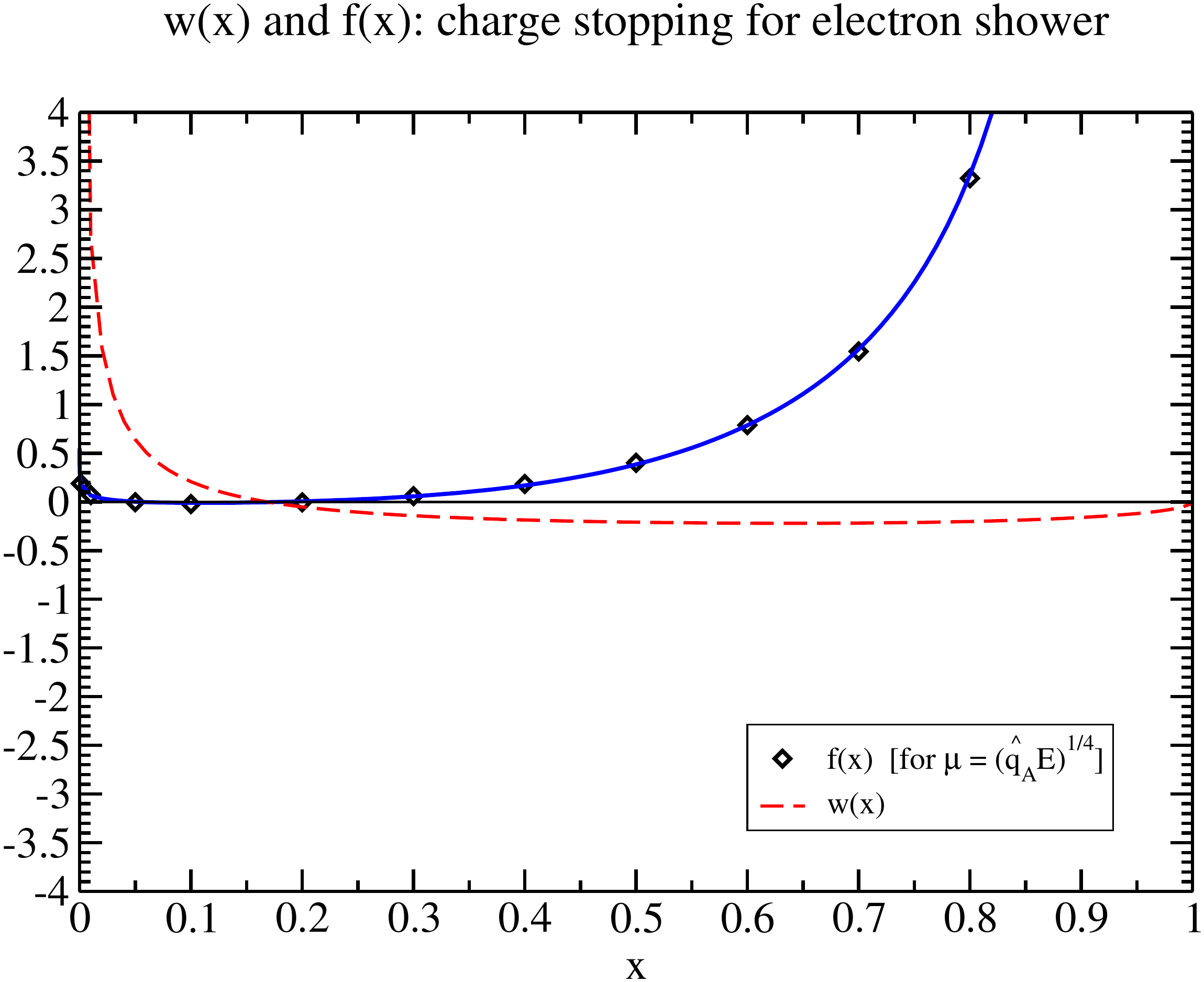}
  \caption{
     \label{fig:wqed}
     Like fig.\ \ref{fig:w}a, but here for charge stopping of electron
     showers in large-$\Nf$ QED.
  }
\end {center}
\end {figure}

Now let's look at a similar analysis for the analogous, charge stopping
calculation for an electron-initiated
shower in large-$\Nf$ QED.
In the large $\Nf$ limit, it is possible to distinguish the original
electron throughout the evolution of the shower, and the overall charge
deposition of the shower is simply given by where the original electron
finally stops and deposits its charge.%
\footnote{
  See the discussion in section 2.2 of ref.\ \cite{qedNfstop}.
}
The relevant splitting rate for computing charge deposition is then
the electron splitting rate $[d\Gamma/dx]_e$, where $x$ represents
the energy fraction of the original electron after the splitting compared
to before the splitting.
In ref.\ \cite{qedNfstop}, the formula
analogous to (\ref{eq:chi2W}) was (with minor adjustment%
\footnote{
  Specifically, see eq.\ (2.17) of ref.\ \cite{qedNfstop}.  The analysis
  of that paper later used a more complicated version,
  eq.\ (2.26) of ref.\ \cite{qedNfstop}, which
  accounted for a piece of the rate that scaled with energy as
  $\beta_0 E^{-1/2}\ln E$, arising
  from a {\it fixed}\/ choice of renormalization scale $\mu$.
  One will get the simpler equation we have used by
  instead choosing $\mu \propto (\qhat r E)^{1/4}$ with constant $r$,
  similar to our (\ref{eq:Lambdafacr}).
  The difference with the fixed-$\mu$ result
  turns out to be small and does not significantly affect
  (\ref{eq:chiQED}).
  [The change is less than 3\% and does not depend on the choice of $r$.]
  We have not shown other reasonable
  choices, such as $\mu = (\qhat\kappa x E)^{1/4}$ analogous to
  our (\ref{eq:Lambdafac}).
}%
)
\begin {equation}
  [\chi\alphaqed]_{\sigma/\lstop}^{\rm charge} =
   \frac{ \dAvg[(1-\sqrt{x})^2] }
        { 2 \Avg[(1-\sqrt{x})^2]_\LO }
   -
   \frac{ \dAvg[(1-x)] }
        { 2 \Avg[(1-x)]_\LO } \,,
\label {eq:chi2e}
\end {equation}
where here $\dAvg$ is computed using $[d\Gamma/dx]_{e\to e}^\NLO$ instead
of $[d\Gamma/dx]_\net^{\NLO,\fac}$.
IR factorization is not necessary (there are no log IR divergences),
and so there is no IR factorization scale $\Lambda_\fac$.
Eq.\ (\ref{eq:chi2e}) can now be rewritten as
\begin {subequations}
\label {eq:chi2We}
\begin {equation}
   [\chi\alphaqed]_{\sigma/\lstop}^{\rm charge} =
   \int_0^1 dx \> W_e(x) \, \left[ \frac{d\Gamma}{dx} \right]^{\NLO}_{e\to e}
\label {eq:chi2Weint}
\end {equation}
with weight function
\begin {equation}
  W_e(x') =
     \frac{ (1-\sqrt{x'})^2 }
          { 2 \Avg[(1-\sqrt{x})^2]_\LO }
     -
     \frac{ (1-x') }
          { 2 \Avg[1-x]_\LO }
  \,.
\label {eq:We}
\end {equation}
\end {subequations}
To put it in a form similar to (\ref{eq:chi2w}),
\begin {subequations}
\label {eq:chi2we}
\begin {equation}
   [\chi\alphaqed]_{\sigma/\lstop}^{\rm charge} =
   \Nf\alphaqed \int_0^1 dx \> w_e(x) \, f_e(x) ,
\label {eq:chi2weint}
\end {equation}
\begin {equation}
  w_e(x') =
   \left[ \frac{d\Gamma}{dx} (x') \right]^\LO_{e\to e}
   \left\{
     \frac{ (1-\sqrt{x'})^2 }
          { 2 \Avg[(1-\sqrt{x})^2]_\LO }
     -
     \frac{ (1-x') }
          { 2 \Avg[(1-x)]_\LO }
   \right\} ,
\end {equation}
\begin {equation}
   f_e(x) \equiv
   \frac{ \left[\frac{d\Gamma}{dx}\right]_{e\to e}^{\NLO} }
        { \Nf\alphaqed \left[\frac{d\Gamma}{dx}\right]^\LO_{e\to e} } \,.
\end {equation}
\end {subequations}
Fig.\ \ref{fig:wqed} shows plots of $w_e(x)$ and $f_e(x)$ analogous to
the plots of $w(x)$ and $f(x)$ in fig.\ \ref{fig:w}.

There is no crude symmetry or anti-symmetry here.  Note in particular
that even LO and NLO single splitting rates for $e \to e\gamma$ will
not be symmetric in $x \to 1{-}x$ because the two daughters are not
identical particles.  (Unlike the discussion of fig.\ \ref{fig:w}, we
will not shift $f_e(x)$ by a constant because it already, like
fig.\ \ref{fig:w}b, is almost as close as it can get to $f_e=0$ while having
the same sign of $f_e(x)$ for all $x$.)
The product of $w_e(x)$ and $f_e(x)$ is shown
by the dotted curve in fig.\ \ref{fig:wf}.  One can see the qualitative
difference with the gluonic case: the area under the dotted curve does not
have any significant cancellation between positive and negative contributions.
But also, the area associated with the right-hand side of the dotted curve
is already bigger than that associated with the right-hand side of
the solid curve.%
\footnote{
   We find numerically that (up to logarithms) both curves blow up
   as $(1-x)^{-1/2}$ as $x {\to 1}$, which is an integrable divergence.
}

It is natural to wonder how much of the huge difference between
the small vs.\ large $\chi\alphas$'s of (\ref{eq:chig}) and (\ref{eq:chiQED})
are due to
having fermions in large-$\Nf$ QED (e.g. $e \to e\gamma$
and $\gamma \to e\bar e$ processes), and so how much different
our QCD results might be if we included quarks in addition to gluons
(e.g.\ $q \to qg$ and $g \to q\bar q$).  Formally, quark processes
are suppressed in the large-$\Nc$ limit if one takes $\Nc$ large while
keeping $\Nf$ fixed, and so can be ignored for large-$\Nc$ gluon-initiated
showers.  However, since $\Nc=3$ and $\Nf\ge3$ [depending
on the size of $\mu \sim (\qhat E)^{1/4}$] in QCD,
a more relevant large-$\Nc$ limit would be to
include quarks and treat $\Nf$ as also potentially large.

It is also natural to wonder whether, even for electron-initiated
showers in large-$\Nf$ QED,
there might be a significant difference between the size of
overlap corrections for (i) the shape of the energy deposition
distribution and (ii) the shape of the charge deposition distribution.
And similarly for quark-initiated showers in QCD.  We leave all
of these questions for future study.

% =========================================================================

\section{Theorist Error}
\label{sec:error}

We should comment on the possibility
of error in our calculation.  There is, of course, theoretical error
associated with the unknown size of yet-higher-order corrections and,
in our case, the choice of factorization and renormalization scales.
But one may be more concerned with what we instead
refer to as theorist error.  The calculation of overlapping splitting
rates \cite{2brem,seq,dimreg,4point,QEDnf,qcd,qcdI}
was very long and very complicated.
Though we and our previous collaborators have tried very hard to be
meticulously careful, to independently check the details of all calculations,
and to devise cross-checks, we can't completely
rule out the possibility of error.
Ref.\ \cite{qcd} lists a number of non-trivial
sanity checks on our rate calculations,%
\footnote{
  Specifically, see section 5 of ref.\ \cite{qcd}.
}
though we later found one error in the calculation
after the first publication of ref.\ \cite{qcd}.%
\footnote{
   See appendix A of ref.\ \cite{logs}.
}
More recently, our best cross-check has been to show
that the IR contribution to our very complicated,
full expression for $[d\Gamma/dx]_\net$
gives the correct result for single (and not just double)
IR logarithms.  This was shown by (i) extracting \cite{logs}
the single log coefficient
(\ref{eq:sbar}) from the IR limit of our full rate calculation and
comparing to (ii) a {\it much}\/ simpler and completely independent
derivation of the IR single logarithm \cite{logs2},
found by substituting the known single-log result
\cite{LMW} for soft radiative corrections to in-medium transverse momentum
broadening into a BDMPS-Z-like calculation of the leading-order
rate for a hard $g{\to}gg$ splitting.

In principle,
the best way to have full confidence in our full result for
$[d\Gamma/dx]_\net$
would be for an independent
group to repeat the calculation, preferably using
an independent method.  A less arduous check might be
to independently calculate $[\Delta\,d\Gamma/dx\,dy]_{g\to ggg}$ in the IR limit
$y{\to}0$ (for fixed $x$) and extract the non-logarithm piece of that limit.
Or to somehow independently compute $[d\Gamma/dx]_{\rm net}$ in the limits
$x{\to}0$ and/or $x{\to}1$.  But we are unsure how complicated such
calculations might be.

All that said, we feel fairly confident in our final conclusion.

% =========================================================================

\section{Concluding Remarks}
\label{sec:conclusion}

Our specific conclusion is that the effects of overlapping gluon
splittings are numerically very small and inconsequential for
the {\it shape} of the energy deposition of a purely-gluonic in-medium
shower, at least with the simplifying assumptions used in
our thought experiment.
Put another way, the effects of overlapping formation times
on the energy deposition distribution $\eps(z)$ itself are
small {\it provided}\/ one
allows $\hat q$ to be an energy-dependent
phenomenological jet quenching parameter for
this purpose.
The energy-dependence of $\hat q_\eff(\omega)$ was investigated at
leading-log order by the early work of refs.\ \cite{Blaizot, Iancu, Wu},
and expanded on in refs.\ \cite{run1,run2a,run2b}.  It would be interesting
if those analyses could be extended to next-to-leading-log order
(for which our very limited NLLO analysis of section \ref{sec:LOvEff}
would be inadequate).

The results of this paper and its companion \cite{finale} represent a
first exploratory investigation into these topics.  In particular,
motivated by section \ref{sec:why},
it remains to be seen whether overlap corrections become more important
when quarks are incorporated into our gluonic showers.

% =========================================================================
% =========================================================================

\begin{acknowledgments}

The work of Arnold and Elgedawy was supported, in part, by the U.S. Department
of Energy under Grant No.~DE-SC0007974.
We are deeply indebted to Han-Chih Chang and Tyler Gorda for their
collaboration in the long chain of previous work that
made our current results possible.
We also thank Zifeng Liu who, several years ago, checked the argument
for the asymptotic behavior (\ref{eq:zasymp2}) of the energy
deposition distribution and also derived (unpublished) corrections to that
behavior.

\end{acknowledgments}

% =========================================================================
\appendix
% =========================================================================

\section{NLO rates in terms of the \boldmath$\NLObar$ formulas
         of ref.\ \cite{qcd}}
\label{app:NLObar}

The NLO rates used in this paper are given in refs.\ \cite{qcd,qcdI}
(and in particular appendix A of each).
But most of the rate formulas in those references are given for what they
call $\NLObar$ rates.  The purpose of this appendix is
to be clear how the various NLO rates needed for this paper can be written in
terms of the $\NLObar$ rate formulas given in refs.\ \cite{qcd,qcdI}.

The difference between NLO and $\NLObar$ is that ref.\ \cite{qcd}
found it convenient to separate the renormalization
scale dependence $\mu$ from the rest of the NLO $g{\to}gg$ rate,
writing
\begin {equation}
   \left[ \Delta \frac{d\Gamma}{dx} \right]^{\rm NLO}_{g\to gg}
   =
   \left[ \Delta \frac{d\Gamma}{dx} \right]^{\NLObar}_{g\to gg}
   +
   \left[ \frac{d\Gamma}{dx} \right]_\renlog
\label {eq:NLOvNLObar}
\end {equation}
with%
\footnote{
  Above, eqs.\ (\ref{eq:NLOvNLObar}) and (\ref{eq:Om0}) correspond to
  eqs.\ (A.49) and (A.4) of ref.\ \cite{qcd}.
  Eq.\ (\ref{eq:renlog2}) above is a slight
  rewriting of eq.\ (A.50) of ref.\ \cite{qcd}.
  For that, we've used eqs.\ (A.6) and (A.7) of ref.\ \cite{qcd},
  and we've also used the fact that
  $\Omega_0 = e^{-i\pi/4}|\Omega_0|$ to rewrite
  $\Re\bigl(i\Omega_0 \ln(1/\Omega_0)\bigr) =
   \Re\bigl(i\Omega_0) \bigl[ \ln(1/|\Omega_0|) - \frac{\pi}{4} \bigr]$.
}
\begin {equation}
   \left[ \frac{d\Gamma}{dx} \right]_\renlog
   \equiv
   - \frac{\beta_0\alphas}{2}
       \left[ \frac{d\Gamma}{dx} \right]^\LO
         \left[
            \ln \Bigl( \frac{\mu^2}{|\Omega_0| E} \Bigr)
            + \ln\Bigl( \frac{x(1{-}x)}{4} \Bigr)
            + \gammaE
            - \frac{\pi}{4}
         \right]
\label {eq:renlog2}
\end {equation}
and $\beta_0$ given by our (\ref{eq:beta0}).
Above, $\Omega_0$ is the complex frequency associated with the leading-order
BDMPS-Z $g{\to}gg$ splitting rate (\ref{eq:LOrate0}), given by
\begin {equation}
   \Omega_0
   = \sqrt{ \frac{-i\qhatA}{2 E}
            \left( -1 + \frac{1}{x} + \frac{1}{1-x} \right) }
   = \sqrt{ \frac{-i (1-x+x^2) \qhatA}{2x(1-x)E} } ,
\label {eq:Om0}
\end {equation}
and $\gammaE$ is the Euler-Mascheroni constant.
Note that the $\ln\mu$ dependence in (\ref{eq:renlog2}) matches
(\ref{eq:MuDependence}).  There is not necessarily anything significant
about the $x$ dependence and dimensionless constants in the rest of
(\ref{eq:renlog2}) --- they were just a combination that was
convenient
to algebraically separate from the $\NLObar$ rate in ref.\ \cite{qcd}
and to integrate over $y$.

When written in terms of the $\NLObar$ rates of refs.\ \cite{qcd,qcdI},
our eq.\ (\ref{eq:virt}) is then%
\footnote{
  Eq.\ (\ref{eq:NLOconversion}) above is just the combination of
  eqs.\ (A.47--49) and (A.52) of ref.\ \cite{qcd} for the case of
  renormalized rates.
}
\begin {multline}
   \left[ \Delta \frac{d\Gamma}{dx} \right]^\NLO_{g\to gg}
   =
     \biggl(
       \int_0^{1-x} dy \>
       \left[ \Delta \frac{d\Gamma}{dx\,dy} \right]_\virtI
     \biggr)
     + (x \to 1{-}x)
\\
   + \int_0^1 dy \> \left[ \Delta \frac{d\Gamma}{dx\,dy} \right]_\virtII
   ~+~ \left[ \frac{d\Gamma}{dx} \right]_\renlog
   ,
\label {eq:NLOconversion}
\end {multline}
where
$[\Delta\,d\Gamma/dx\,dy]_\virtI$ and $[\Delta\,d\Gamma/dx\,dy]_\virtII$
is the notation in those references for the $\NLObar$ versions of
what we call
$[\Delta\,d\Gamma/dx\,dy]^\NLO_\classI$ and $[\Delta\,d\Gamma/dx\,dy]^\NLO_\classII$
in this paper.
Correspondingly, eqs.\ (\ref{eq:dGnetNLO}), (\ref{eq:VRdef}),
and (\ref{eq:dGfac})
of this paper
can be rewritten, in terms of the rates presented in refs.\ \cite{qcd,qcdI},
as%
\footnote{
   The $\NLObar$ rate in (\ref{eq:dGnet2}) above is
   eq.\ (1.7) of ref.\ \cite{qcd}.
   $v(x,y)$ and $r(x,y)$ are defined as in eq.\ (1.8) of ref.\ \cite{qcd}.
}
\begin {align}
   \left[ \frac{d\Gamma}{dx} \right]_{\rm net}^{\NLO} &=
   \left[ \frac{d\Gamma}{dx} \right]_\renlog
   + \left[ \frac{d\Gamma}{dx} \right]_{\rm net}^{\NLObar}
\nonumber\\
   &=
   \left[ \frac{d\Gamma}{dx} \right]_\renlog
   +
   \int_0^{1/2} dy \>
   \Bigl\{
      v(x,y) \, \theta(y<\tfrac{1-x}{2})
      + v(1{-}x,y) \, \theta(y<\tfrac{x}{2})
\nonumber\\ &\hspace{20em}
      + r(x,y) \, \theta(y<\tfrac{1-x}{2})
   \Bigr\}
 ,
\label {eq:dGnet2}
\end {align}
\begin {subequations}
\begin {align}
  v(x,y) &\equiv 
  \left(
     \left[ \Delta \frac{d\Gamma}{dx\,dy} \right]_\virtI
     + \left[ \Delta \frac{d\Gamma}{dx\,dy} \right]_\virtII
  \right)
  + ( y \leftrightarrow 1{-}x{-}y ) ,
\\
  r(x,y) & \equiv
  \left[ \Delta \frac{d\Gamma}{dx\,dy} \right]_{g\to ggg} ,
\end {align}
\end {subequations}
and, most importantly,
\begin {multline}
   \left[ \frac{d\Gamma}{dx} \right]_{\rm net}^{\NLO,\fac} \equiv
   \left[ \frac{d\Gamma}{dx} \right]_\renlog
   +
   \int_0^\infty dy \>
   \biggl\{
      v(x,y) \, \theta(y<\tfrac{1-x}{2})
      + v(1{-}x,y) \, \theta(y<\tfrac{x}{2})
\\
      + r(x,y) \, \theta(y<\tfrac{1-x}{2})
      + \frac{\CA\alphas}{4\pi}
          \left[ \frac{d\Gamma}{dx} \right]^{\rm LO}
          \frac{\ln y + \bar s(x)}{y} \,
        \theta(yE < \Lambda_\fac)
   \biggr\} .
\label {eq:dGfacAlt}
\end {multline}

Take care when using these formulas to note that the definitions of
$[\Delta\,d\Gamma/dx\,dy]_{g\to ggg}$,
$[\Delta\,d\Gamma/dx\,dy]_\virtI$, and $[\Delta\,d\Gamma/dx\,dy]_\virtII$
in ref.\ \cite{qcd} have been updated to include F diagrams in
ref.\ \cite{qcdI}.%
\footnote{
  Specifically, see eqs.\ (A.1), (A.18), and (A.19) of ref.\ \cite{qcdI}.
}

% ===========================================================================

\section{Numerical methods}

\subsection{Computation of \boldmath$[d\Gamma/dx]^{\NLO,\fac}_\net$}
\label {app:dGfacNumerics}

In (\ref{eq:dGfacAlt}) for $[d\Gamma/dx]^{\NLO,\fac}_\net$,
there is a subtraction in the $y$ integrand that removed the
$y^{-1}\ln y$ and $y^{-1}$ behavior of the integrand at small $y$
which would otherwise have generated
IR double and single logarithmic divergences.
With that subtraction, the left-over behavior
of the integrand at small $y$ turns
out to be of order $y^{-1/2} \ln y$,
which is an integrable divergence.  However, as a practical matter
for numerical integration, it is more efficient to soften the
integrable divergence by changing integration variable from $y$ to
$u = y^{1/2}$, so that the behavior
of the $u$ integrand is merely $\ln u$ as $u \to 0$.

We use Mathematica \cite{Mathematica}
for the evaluations of the $y$ integrand, including
the necessary $\Delta t$ integrations in the formulas for
$[\Delta\,d\Gamma/dx\,dy]_\virtI$, $[\Delta\,d\Gamma/dx\,dy]_\virtII$,
and $[\Delta\,d\Gamma/dx\,dy]_{g\to ggg}$ presented in refs.\ \cite{qcd,qcdI}.
Our unsophisticated attempts to use Mathematica's built-in integrator to do the
$u=y^{1/2}$ integrals were inefficient, however.
Instead, we did the $u$ integration by brute force
using a simple
mid-point Riemann sum
covering the integration region
$u=0$ to $u_\max=[\max(x/2,(1{-}x)/2,\Lambda_\fac/E)]^{1/2}$ where the
integrand is non-zero.  For sufficiently smooth functions,
the error of a mid-point Riemann sum should scale as
$O( (\Delta u)^2 )$, where $\Delta u$ is the small step
size.  But there are two issues that spoil this rate
of convergence:
our integrand (i) has discontinuities
at the thresholds for the various $\theta$ functions in
(\ref{eq:dGfacAlt}), and (ii) diverges as $\ln u$ as $u \to 0$.
The simplest way to take care of issue (i) is to divide the integral
up into the three regions
where the integrand is continuous, and do each region separately
with a mid-point Riemann sum.%
\footnote{
   Alternatively, one can do a single integral over the total integration
   region and correct the mid-point rule in the steps where discontinuities
   occur, given that we know exactly where the points of discontinuity are.
}

For the second issue, we numerically extract the coefficient $c$
of the $c \ln u$ behavior as $u \to 0$, and then we
correct the midpoint Riemann sum approximation to
\begin {equation}
  \int_0^{u_\max} du \> f(u) =
  - \frac{\Delta u \, \ln 2}{2} \, c
  + \sum_{n=1}^N \Delta u \, f\bigl( (n-\tfrac12)\Delta u\bigr) ,
\end {equation}
where $\Delta u = u_\max/N$.  The factor of $\frac12 \, \Delta u \,\ln 2$
in the correction term comes from the identity
\begin {equation}
   \lim_{N\to\infty}
   \left[
     \int_0^{N\,\Delta u} du \> \ln u
     - \sum_{n=1}^N \Delta u \, \ln\bigl( (n-\tfrac12)\Delta u \bigr)
   \right]
   =
   -\tfrac12 \,\Delta u \, \ln 2 .
\end {equation}
There are, no doubt, much more sophisticated integration methods that could
have been used, but these were the simplest for us to quickly implement
without diagnosing how to fine-tune the performance
of general-purpose integrators.
Because our integration method is non-adaptive, however, one must
monitor the numerical convergence with increasing $N$.

% --------------------------------------------------------------------------

\subsection{More details on numerical evaluation of
         \boldmath$\hat\eps(\hat z)$}
\label {app:epsNumerics}

In the backward-evolution equation (\ref{eq:epseqLO2}) for
$\hat\eps_\LO(\hat z)$, the integral
\begin {equation}
  \int_0^1 dx \> x \biggl[\frac{d\hat\Gamma}{dx}\biggr]^\LO
  \Bigl\{ x^{-1/2}\,\hat\eps_\LO(x^{-1/2}\hat z)
          - \hat\eps_\LO(\hat z) \Bigr\}
\label {eq:epsIntegral}
\end {equation}
has integrable singularities at the endpoints.  Specifically, the
integrand scales like $x^{-1/2}$
as $x\to0$ and $(1-x)^{-1/2}$ as $x\to1$.
It is numerically more efficient to make a change of integration variable,
similar to the $u=y^{1/2}$ earlier in this appendix, to reduce the
singularity.  Changing variables to $u=x^{1/2}$ in (\ref{eq:epsIntegral})
will help $x\to0$ but won't do anything for $x\to 1$.  A simple
solution is to first split the integral up as
\begin {equation}
   \int_0^1 dx \> \cdots =
   \int_0^{1/2} dx \> \cdots +
   \int_{1/2}^1 dx \> \cdots ,
\end {equation}
and then change integration variable $x \to 1{-}x$ in the last integral.
Remembering that $[d\Gamma/dx]^\LO$ is symmetric under exchange of its
two daughters, (\ref{eq:epsIntegral}) then becomes
\begin {multline}
  \int_0^{1/2} dx \> 
  \biggl[\frac{d\hat\Gamma}{dx}\biggr]^\LO
  \Bigl(
  x \Bigl\{ x^{-1/2}\,\hat\eps_\LO(x^{-1/2}\hat z)
          - \hat\eps_\LO(\hat z) \Bigr\}
\\
  +
  (1{-}x) \Bigl\{ (1{-}x)^{-1/2}\,\hat\eps_\LO\bigl((1{-}x)^{-1/2}\hat z\bigr)
          - \hat\eps_\LO(\hat z) \Bigr\}
  \Bigr) .
\end {multline}
Now the change of integration variable to $u=x^{1/2}$ will remove all
$1/\sqrt{\phantom{u}}$ divergences.

To do the integral (\ref{eq:epsIntegral}) with the discretized representation
of $\hat\eps(\hat\zeta)$ that
we obtain for $\hat z \le \hat\zeta \le \hat z_\max$,
we used Mathematica to interpolate the function and then integrated
using that interpolation.

The integrals in (\ref{eq:epseqNLO}) that determine $\delta\hat\eps(\hat z)$
may be treated similarly, except that one must remember that
$[d\Gamma/dx]_\net$ is {\it not}\/ symmetric under $x \to 1{-}x$.
So the driving term
\begin {equation}
  \int_0^1 dx \> x \biggl[\frac{d\hat\Gamma}{dx} \biggr]^{\NLO,\fac}_\net
  \Bigl\{ x^{-1/2}\,\hat\eps_\LO(x^{-1/2}\hat z)
          - \hat\eps_\LO(\hat z) \Bigr\}
\end {equation}
for that equation should be replaced by
\begin {multline}
  \int_0^{1/2} dx \> 
  \Biggl(
  x \biggl[\frac{d\hat\Gamma}{dx}(x)\biggr]^{\NLO,\fac}_\net
  \Bigl\{ x^{-1/2}\,\hat\eps_\LO(x^{-1/2}\hat z)
          - \hat\eps_\LO(\hat z) \Bigr\}
\\
  +
  (1{-}x) \biggl[\frac{d\hat\Gamma}{dx}(1{-}x)\biggr]^{\NLO,\fac}_\net
  \Bigl\{ (1{-}x)^{-1/2}\,\hat\eps_\LO\bigl((1{-}x)^{-1/2}\hat z\bigr)
          - \hat\eps_\LO(\hat z) \Bigr\}
  \Biggr) ,
\end {multline}
followed by a change of variables to $u=x^{1/2}$.

\begin {figure}[t]
\begin {center}
  \includegraphics[scale=0.3]{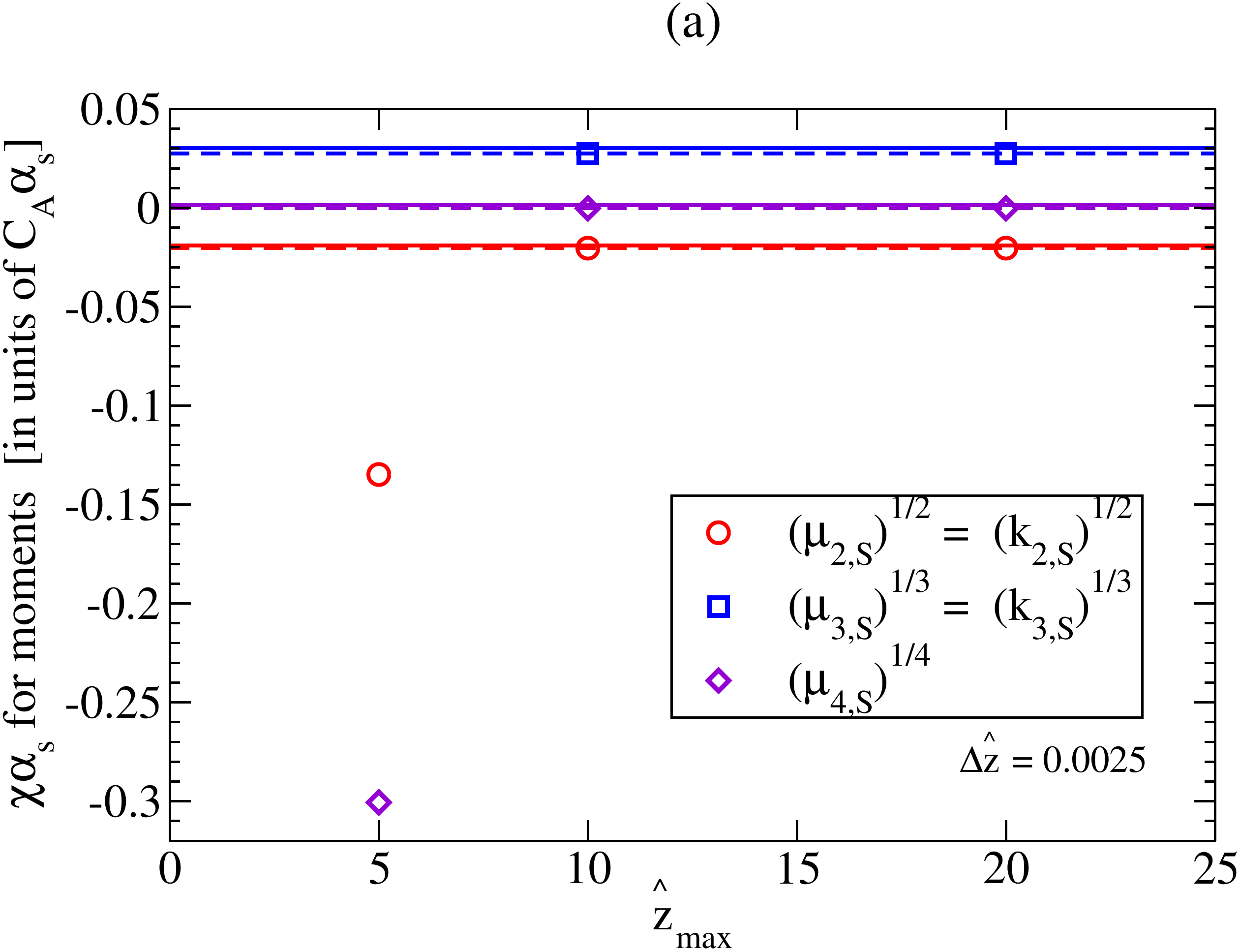}
  \hspace{2em}
  \includegraphics[scale=0.3]{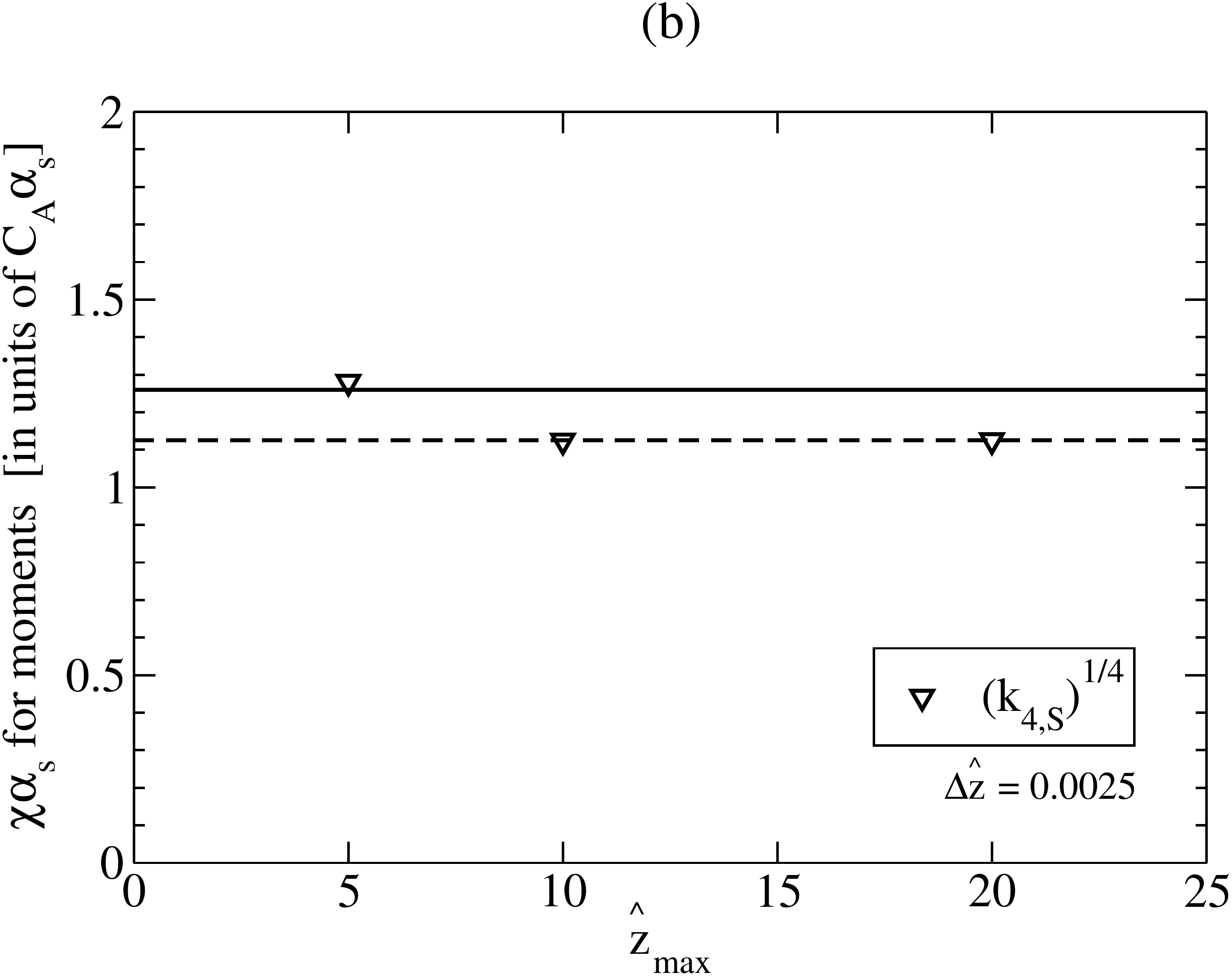}
  \caption{
     \label{fig:zmax}
     Like fig.\ \ref{fig:dz} but here the data points show the dependence
     on $\hat z_\max$ for $\Delta\hat z = 0.0025$.
     The solid horizontal lines again show the
     results of table \ref{tab:shape}, and their difference with
     the $(\Delta\hat z,\hat z_\max)=(0.0025,20)$ data points is the same
     as that in fig.\ \ref{fig:dz}, due to the non-zero value of
     $\Delta\hat z$.  We've drawn dashed horizontal lines corresponding
     to the $(\Delta\hat z,\hat z_\max)=(0.0025,20)$ value to
     instead emphasize the relevant point for approximating
     $\hat z_\max \to\infty$: there is no significant
     difference between $\hat z_\max = 10$ and $\hat z_\max = 20$.
  }
\end {center}
\end {figure}

In the main text, we demonstrated approach to the continuum limit
in fig.\ \ref{fig:dz}.
Fig.\ \ref{fig:zmax} shows our approach to the $\hat z_\max \to \infty$ limit
for the smallest $\Delta\hat z$ value of fig.\ \ref{fig:dz}.
There is no noticeable difference between the results for
$\hat z_\max = 10$ and $\hat z_\max = 20$, and so the
value $\hat z_\max = 20$ used in fig.\ \ref{fig:dz} was plenty large
enough.

% ===========================================================================

\section{More on \boldmath$\Delta b$ dependence of NLLO resummation}
\label {app:NLLO}

In this appendix, we argue that the resummation (\ref{eq:LMWresum})
is adequate to capture
the $\Delta b$ dependence of resummation at next-to-leading-log order (NLLO)
but would fail at the next order (NNLLO).
As in section \ref{sec:dbDependence}, we will ignore the running of
$\alphas(k_\perp)$, which was argued not to affect our conclusions in
section \ref{sec:running}.

% ----------------------------------------------------------------------------

\subsection{Review of LLO resummation}

We first review the leading-log order (LLO) resummation of
LMW \cite{LMW}.  In our notation, we find it convenient to express
the leading-log contribution to $\qhat_\eff$ from $n$-th order in
$\alphas(\mu)$ as
\begin {equation}
  \delta^n \hat q_\eff(\Delta b)
  \approx
  \baralphas^{\,n}
  \hat q_{(0)}
  \int_{\tau_0}^\infty \frac{dt_1}{t_1}
    \int_{\qhat t_1}^{1/(\Delta b)^2} \frac{dk_{\perp 1}^2}{k_{\perp 1}^2}
  \int_{\tau_0}^{t_1} \frac{dt_2}{t_2}
    \int_{\qhat t_2}^{k_{\perp 1}^2} \frac{dk_{\perp 2}^2}{k_{\perp 2}^2}
  \cdots
  \int_{\tau_0}^{t_{n-1}} \frac{dt_n}{t_n}
    \int_{\qhat t_n}^{k_{\perp,n-1}^2} \frac{dk_{\perp n}^2}{k_{\perp n}^2} \,,
\label {eq:resum1}
\end {equation}
where in this appendix we use the short-hand notation
\begin {equation}
  \baralphas \equiv \frac{\CA \alphas}{\pi} .
\end {equation}
In our convention, $(k_{\perp 1},t_1)$ are the transverse momentum and
emission duration%
\footnote{
  The emission duration
  $t_1$ is what we called $\Delta t$ in fig.\ \ref{fig:LMWregion}.
}
of the first soft gluon, $(k_{\perp 2},t_2)$ are
those of an even softer gluon emission, and so forth, with $k_\perp$ ordering
\begin {equation}
   \frac{1}{\Delta b} \gg k_{\perp1} \gg k_{\perp 2} \gg \cdots .
\label {eq:ktordering}
\end {equation}
The first inequality in (\ref{eq:ktordering})
can be understood as following a pattern ($k_{\perp0} \gg k_{\perp1}$) similar
to the others, because $1/\Delta b$ is the transverse momentum scale
($k_{\perp0}$)
corresponding to the lightlike Wilson loop of fig.\ \ref{fig:Wilson}
from which the first gluon ($k_{\perp 1}$) is emitted.
The other conditions for leading logs are that
softer emissions take place within the
duration of harder emissions, so that
\begin {equation}
   t_1 \gg t_2 \gg t_3 \gg \cdots \gg \tau_0.
\label {eq:tordering}
\end {equation}
The last inequality in (\ref{eq:tordering}),
implemented in the lower limits of all
the time integrals, reflects the breakdown of the $\qhat$ approximation
for emission times smaller than the mean free path $\tau_0$, which was
also a constraint in fig.\ \ref{fig:LMWregion}.
The lower limits on the $k_\perp$ integrals
correspond to the fact that the transverse momentum kicks
$\Delta p_\perp \sim \sqrt{\qhat t}$ accumulated over the duration of an
emission will disrupt the vacuum-like logarithms if $\Delta p_\perp$
is as large as the $k_\perp$ of that emission.
Each double logarithm relies on nearly-collinear emissions, and the kicks
from the medium disturb collinearity.

Mathematically, in order to implement the conditions just described,
the $k_\perp$
integrals in (\ref{eq:resum1}) should be understood as requiring that
each upper limit of integration be greater than the corresponding
lower limit.  That means in particular
that the $k_{\perp1}$ integration sets an upper limit
\begin {equation}
   t_1 < \frac{1}{\qhat(\Delta b)^2}
\label {eq:t1limit}
\end {equation}
on the $t_1$ integration.  We could have explicitly written that in
(\ref{eq:resum1}), but the motivation for the limits was easier to
explain by initially writing the $t_1$ integral as unbounded.

In LMW's application, the relevant scale for $\Delta b$ was
$(\qhat L)^{-1/2}$, where $L$ was the length of the medium traversed:
\begin {equation}
   \Delta b ~ \mbox{here}
   ~\longrightarrow~
   \frac{1}{Q_{\rm s}} \sim \frac{1}{\sqrt{\qhat L}} ~ \mbox{in LMW \cite{LMW}} .
\label{eq:dbtranslate}
\end {equation}
In our application, the scale analogous to $L$ is,
parametrically, the formation time for the underlying, hard splitting process.
However, for the sake of the discussion of section \ref{sec:LOvEff}, we
keep things here explicitly in terms of $\Delta b$.

There are many different ways to rewrite (\ref{eq:resum1}), and we will
provide several for the sake of reference when comparing to other papers.
LMW use the variables%
\footnote{
  LMW represent (\ref{eq:altx}) with the symbol $x$.
  We use $\altx$ here to avoid confusion with
  our use of $x$ elsewhere in this paper.
}
\begin {equation}
  \altx \sim \frac{\tau_0}{t}
\label {eq:altx}
\end {equation}
in place of our $t$'s.  After making this change of integration variable
in (\ref{eq:resum1}),
one may change the order of integrations to write a formula equivalent to
LMW's version:%
\footnote{
  Specifically, see eq.\ (50) of ref.\ \cite{LMW}, which only explicitly
  writes out the example $n{=}2$, and make use of the translation
  (\ref{eq:dbtranslate}).
  Our $\delta^2 \hat q_\eff$ corresponds to their eq.\ (50) divided by $L$,
  except that their numbering of the gluons is the reverse of ours, i.e.\
  their $(k_{\perp 1}, \cdots k_{\perp n})$ are our $(k_{\perp n},\cdots,k_{\perp 1})$
  and their $(x_1,\cdots,x_n)$ are our $(\altx_n,\cdots,\altx_1)$.
  Their $Q_0^2 = \qhat \tau_0$.
}
\begin {multline}
  \delta^n \hat q_\eff(\Delta b)
  \approx
  \baralphas^{\,n} \hat q_{(0)}
  \int_{\qhat\tau_0}^{1/(\Delta b)^2} \frac{dk_{\perp 1}^2}{k_{\perp 1}^2}
  \int_{\qhat\tau_0}^{k_{\perp 1}^2} \frac{dk_{\perp 2}^2}{k_{\perp 2}^2}
  \cdots
  \int_{\qhat\tau_0}^{k_{\perp,n-1}^2} \frac{dk_{\perp n}^2}{k_{\perp n}^2}
\\ \times
    \int_{\qhat\tau_0/k_{\perp n}^2}^{1} \frac{d\altx_n}{\altx_n}
    \cdots
    \int_{\qhat\tau_0/k_{\perp 2}^2}^{\altx_3} \frac{d\altx_2}{\altx_2}
    \int_{\qhat\tau_0/k_{\perp 1}^2}^{\altx_2} \frac{d\altx_1}{\altx_1}
  \,.
\end {multline}

Alternatively, to make contact with the variables $(t,\omega)$ used
in fig.\ \ref{fig:LMWregion}, change integration variables in (\ref{eq:resum1})
by using the parametric relation
$t \sim \omega/k_\perp^2$ for the duration of vacuum-like gluon fluctuations,%
\begin {multline}
  \delta^n \hat q_\eff(\Delta b)
  \approx
  \baralphas^{\,n} \hat q_{(0)}
  \int_{\tau_0}^\infty \frac{dt_1}{t_1}
    \int_{\qhat t_1^2}^{t_1/(\Delta b)^2} \frac{d\omega_1}{\omega_1}
  \int_{\tau_0}^{t_1} \frac{dt_2}{t_2}
    \int_{\qhat t_2^2}^{\omega_1 t_2/t_1} \frac{d\omega_2}{\omega_2}
\\
  \cdots
  \int_{\tau_0}^{t_{n-1}} \frac{dt_n}{t_n}
    \int_{\qhat t_n^2}^{\omega_{n-1} t_n/t_{n-1}} \frac{d\omega_n}{\omega_n} \,,
\label {eq:resum2}
\end {multline}
where the limits of the $\omega_1$ integration again implicitly set
the upper limit (\ref{eq:t1limit}) on $t_1$.

The analysis of ref.\ \cite{run1} (which reviews the fixed coupling case
as a warm-up) uses the logarithmic
variables
\begin {equation}
   Y \equiv \ln\left( \frac{t}{\tau_0} \right) ,
   \qquad
   \rho \equiv \ln\left( \frac{k_\perp^2}{\hat q \tau_0} \right) ,
\label {eq:Yrho}
\end {equation}
in terms of which (\ref{eq:resum1}) can be written
\begin {subequations}
\label {eq:resum3}
\begin {equation}
  \delta^n \qhat_\eff
  \approx
  \baralphas^{\,n} \hat q_{(0)} \,
  f_n\Bigl(
     \ln\bigl(\tfrac{1}{\qhat\tau_0(\Delta b)^2}\bigr),
     \ln\bigl(\tfrac{1}{\qhat\tau_0(\Delta b)^2}\bigr)
  \Bigr) ,
\label {eq:qhatfn}
\end {equation}
where (introducing our own notation ``$f_n$'')
\begin {equation}
  f_n(Y,\rho) \equiv
  \int_0^Y dY_1
    \int_{Y_1}^\rho d\rho_1
  \int_0^{Y_1} dY_2
    \int_{Y_2}^{\rho_1} d\rho_2
  \cdots
  \int_0^{Y_{n-1}} dY_n
    \int_{Y_n}^{\rho_{n-1}} d\rho_n
  .
\label {eq:resum3f}
\end {equation}
\end {subequations}
Eqs.\ (\ref{eq:resum3}) tell us that the
leading-log result at $n$-th order is just
just $\baralphas^{\,n}\qhat_{(0)}$ times
the hyper-volume of the integration region in (\ref{eq:resum3}).

LMW's summation of all the leading-log $\delta^n\qhat_\eff$ gives the
formula (\ref{eq:LMWresum}) presented in the main text.
Iancu and Triantafyllopoulos \cite{run1} give a little more detail,
showing that
\begin {equation}
  f_n(Y,\rho) =
  \frac{Y^n\rho^n}{(n!)^2} - 
  \frac{Y^{n+1}\rho^{n-1}}{(n{+}1)!\,(n{-}1)!}
  \qquad (n>0)
\end {equation}
(which can be proven by induction).
Summing all orders of $\alphas$ gives
\begin {equation}
   1 + \sum_{n=1}^\infty \baralphas^{\,n} f_n(Y,\rho)
   =
   I_0\!\left( 2\sqrt{\baralphas Y \rho} \right)
   -
   \frac{Y}{\rho} \,
   I_2\!\left( 2\sqrt{\baralphas Y \rho} \right) ,
\label {eq:fnsum}
\end {equation}
and setting $Y = \rho = \ln\bigl(\frac{1}{\qhat\tau_0(\Delta b)^2}\bigr)$
as in (\ref{eq:qhatfn}) then
gives (\ref{eq:LMWresum}).

% ---------------------------------------------------------------------------

\subsection{\boldmath$\Delta b$ dependence of logarithms at $O(\alphas)$}

\begin {figure}[t]
\begin {center}
  \includegraphics[scale=1.2]{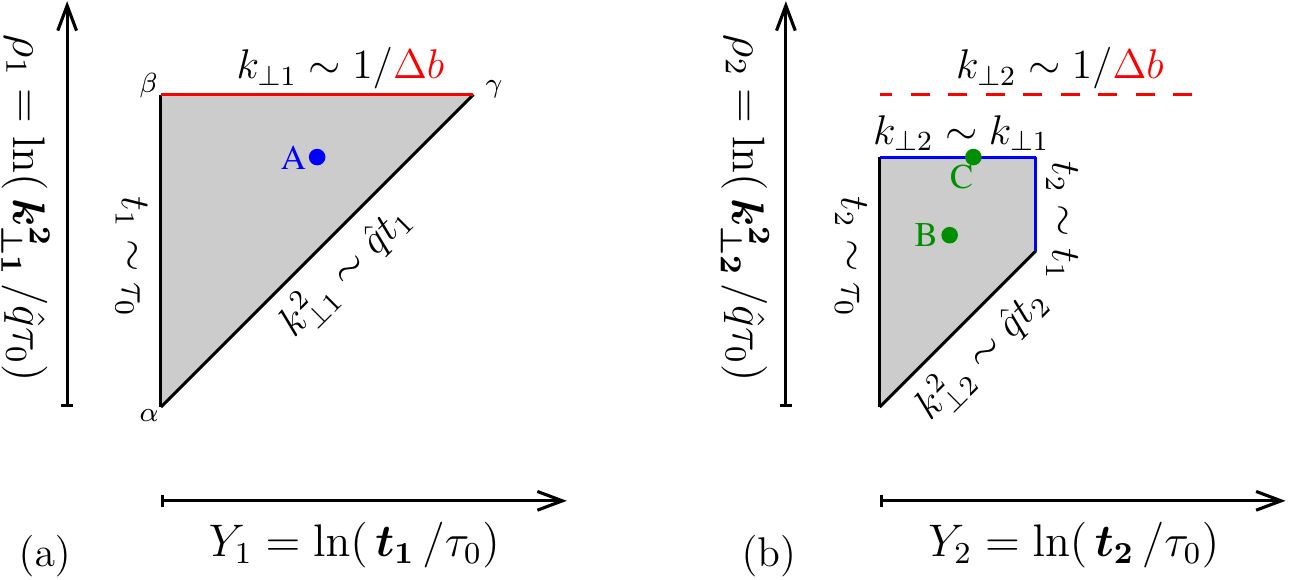}
  \caption{
     \label{fig:LMWregiontk}
     (a) The double-log region of fig.\ \ref{fig:LMWregion}
     in terms of the variables ($Y$,$\rho$) of (\ref{eq:Yrho}).
     (a+b) A depiction of the leading-log region at order $O(\alphas^2)$.
     In this figure, the extent of the $(Y_2,\rho_2)$ region is drawn
     for the case where $(Y_1,\rho_1)$ is at point ``A.''
  }
\end {center}
\end {figure}

It will be useful to also review some of the qualitative aspects of
double and single logs at $O(\alphas)$.
The double log approximation corresponds to
the $n{=}1$ case of (\ref{eq:resum1}):
\begin {equation}
  \delta   \hat q_\eff(\Delta b)
  \approx
  \baralphas \hat q_{(0)}
  \int_{\tau_0}^{1/\qhat(\Delta b)^2} \frac{dt_1}{t_1}
    \int_{\qhat t_1}^{1/(\Delta b)^2} \frac{dk_{\perp 1}^2}{k_{\perp 1}^2} ,
\label {eq:dbltk}
\end {equation}
where we've used (\ref{eq:t1limit}).
A picture of the integration region is shown in fig.\ \ref{fig:LMWregiontk}a,
which is equivalent to the integration region previously depicted in
fig.\ \ref{fig:LMWregion}.
LMW analyzed the sub-leading, single logarithms as well at this order.
What will be important for
our discussion are qualitative characterizations of
the following parametric regions.

(i) Double logarithms are generated by integrating over the
interior of the shaded region,
\begin {equation}
  \tau_0 \ll t_1 \ll \frac{1}{\qhat(\Delta b)^2} \,,
  \qquad
  \qhat t_1 \ll k_{\perp 1}^2 \ll \frac{1}{(\Delta b)^2} \,,
\end {equation}
such as the point labeled ``A'' in fig.\ \ref{fig:LMWregiontk}a.
The double log will be proportional to the area of the shaded region
in the log-log coordinates of the figure.

\begin {figure}[t]
\begin {center}
  \includegraphics[scale=1.2]{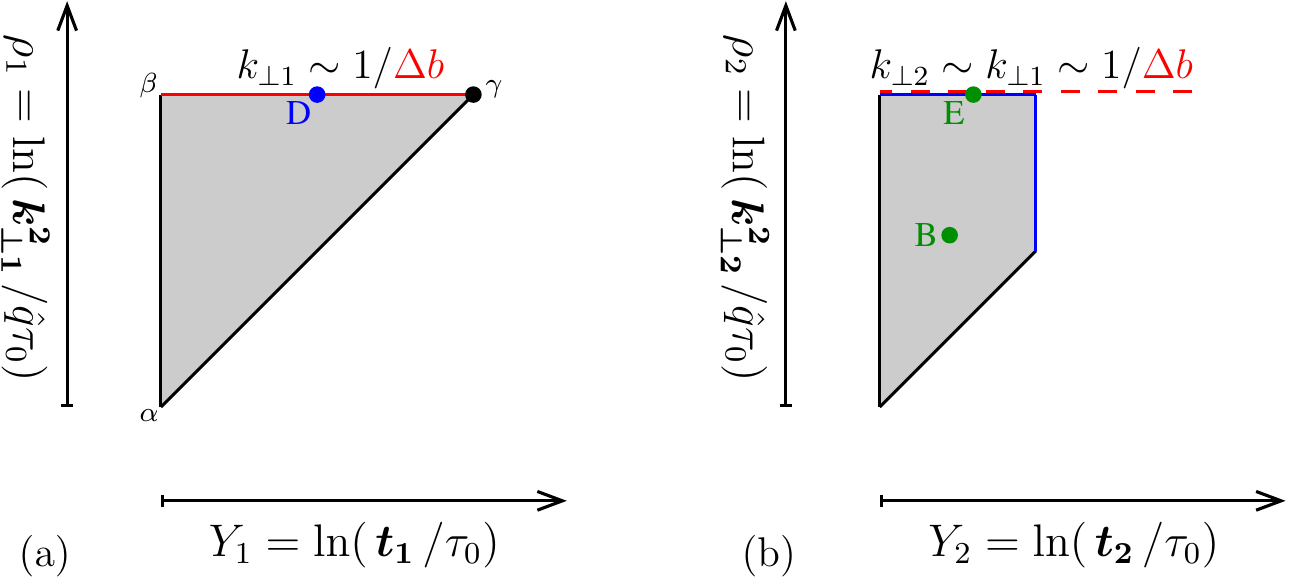}
  \caption{
     \label{fig:LMWregiontk2}
     Like fig.\ \ref{fig:LMWregiontk}, but here the extent of the
     $(Y_2,\rho_2)$ region is drawn
     for the case where $(Y_1,\rho_1)$ is at point ``D.''
  }
\end {center}
\end {figure}

(ii) Single logarithms arise from integrating along the edges, e.g.\ over
\begin {equation}
  \tau_0 \ll t_1 \ll \frac{1}{\qhat(\Delta b)^2} \,,
  \qquad
  k_{\perp 1}^2 \sim \frac{1}{(\Delta b)^2}
\label {eq:bedge}
\end {equation}
for the upper edge in fig.\ \ref{fig:LMWregiontk}a, which is
the edge most sensitive to the value of $\Delta b$.
Because $k_{\perp 1}^2 \sim 1/(\Delta b)^2$ in (\ref{eq:bedge}),
the red line representing this edge should be thought of as
having an $O(1)$ thickness in the log-log coordinates used
in the figure.
Similarly for the other edges.  In the limit of large logarithms,
the $O(1)$ thickness of the edges is parametrically small compared to
the size of the shaded, double-log region.
The point
labeled ``D'' in fig.\ \ref{fig:LMWregiontk2}a gives an example of how
we'll graphically indicate points contributing to the single log.

(iii) No logarithms are generated by the corners, such as
\begin {equation}
  t_1 \sim \frac{1}{\qhat(\Delta b)^2} \,,
  \qquad
  k_{\perp 1}^2 \sim \frac{1}{(\Delta b)^2} \,,
\end {equation}
which is labeled ``$\gamma$'' in the figure.

The single-log pieces can be thought of as
the dominant contribution to the difference of (a) the full integral
over all $(\omega_1,t_1)$ and (b) the double-log
approximation (\ref{eq:dbltk}).
It will be useful to give a name to the
integral that gives this difference.  We will call it
\begin {equation}
  \baralphas \hat q_{(0)}
  \int \frac{dt_1}{t_1}
    \int \frac{dk_{\perp 1}^2}{k_{\perp 1}^2}
    \, F_{\rm sl}(t_1,k_{\perp 1}^2) ,
\label{eq:Fsldef}
\end {equation}
where $F_{\rm sl}$ has support on the edges of the double log region and
falls rapidly towards zero as $(Y_1,\rho_1)$
moves away from those edges in fig.\ \ref{fig:LMWregiontk2}a.
The subscript ``sl'' stands for ``single log.''
Most details of $F_{\rm sl}$ will be unimportant.
The important property of $F_{\rm sl}$ is that it will be uniform along
each individual edge, by which we mean that integration of $F_{\rm sl}$ over the
direction perpendicular to an edge gives (to good approximation in
the large-log limit) the same result everywhere along that edge.
The uniformity of each edge
in this sense means that the contribution of each edge to
(\ref{eq:Fsldef}) will be proportional
to a single logarithm, with a coefficient depending on the
details of how $F_{\rm sl}$ behaves near that edge.

To give a concrete example of uniformity,
consider the edge (\ref{eq:bedge}) that is
sensitive to the physics of $k_{\perp} \sim 1/\Delta b$.
The (approximate) formula for $F_{\rm sl}$ along that edge may be extracted
from LMW \cite{LMW}
in terms of the variables $(t_1,\omega_1)$:%
\footnote{
  This comes from eq.\ (32) of ref.\ \cite{LMW}, where
  $S$ is $-\frac14 \qhat_\eff x_\perp^2 L$ and where there is an implicit
  $\Re\{\cdots\}$ on the right-hand side.
  Our $\delta\qhat_\eff$ then corresponds to integrating
  the right-hand side of their (32) with
  integral
  \[
      - \frac{4}{x_\perp^2 L} \int \frac{d\omega}{\omega} \,.
  \]
  Comparing to the $(\omega,t)$ version
  \[
    \baralphas \hat q_{(0)}
    \int \frac{dt_1}{t_1}
      \int \frac{d\omega_1}{\omega_1}
    \, F_{\rm sl}
  \]
  of our (\ref{eq:Fsldef}) then determines $F_{\rm sl}$, except that
  we must subtract away the double log piece already included in
  the $n{=}1$ version of (\ref{eq:resum2}), where the edge we are
  focused on is the upper limit $t_1/(\Delta b)^2$ of the $\omega_1$
  integration there.  That subtraction is implemented by the last
  term in our (\ref{eq:Fsl}).  We've written the argument of the
  $\theta$ function to match the $k_{\perp 1}^2 \le 1/(\Delta b)^2$
  condition in the (\ref{eq:resum1}) version of the leading-log resummation.
}
\begin {multline}
  F_{\rm sl} \simeq F_{\rm sl}^{\rm approx} =
  \Re\Biggl\{
  \frac13
   \left[
     \left( 1 + \frac{i\omega_1 (\Delta b)^2}{2t_1} \right)
         e^{i\omega_1 (\Delta b)^2/2t_1}
     + 2i \, \frac{(1 - e^{i\omega_1 (\Delta b)^2/2t_1})}{\omega_1 (\Delta b)^2/2t_1}
   \right]
\\
  - \theta\!\left( \frac{\omega_1(\Delta b)^2}{2t_1} < 1 \right)
  \Biggr\} .
\label {eq:Fsl}
\end {multline}
The detailed expression does not matter except to explicitly confirm
the important point that this edge's $F_{\rm sl}$
is a function of only $\omega_1(\Delta b)^2/2t_1$.
Since
$t_1 \approx 2\omega_1/k_{\perp 1}^2$ in this region of
vacuum-like emissions,
the $F_{\rm sl}(t_1,k_{\perp1}^2)$ of (\ref{eq:Fsldef}) is actually
a function of only $k_{\perp1}^2(\Delta b)^2$ near this (red) edge
of fig.\ \ref{fig:LMWregiontk2}a,
and $k_{\perp1}^2$ is the variable that parametrizes
the direction perpendicular to that edge.
This provides an example of how
$F_{\rm sl}$ is ``uniform'' along an edge, which in this case means
that
$F_{\rm sl}^{\rm approx}(t_1,k_{\perp1}^2)
 \simeq F_{\rm sl}^{\rm approx}\bigl((k_{\perp 1}\Delta b)^2\bigr)$
does not depend on $t_1$.

Because (\ref{eq:Fsl}) is localized near the edge, the limits of the
$dk_{\perp 1}^2$ integral in (\ref{eq:Fsldef}) that is perpendicular
to the edge (\ref{eq:bedge}) can be replaced (within the
large-log approximation) by 0 to $\infty$.
This gives
\begin {equation}
   \int_0^\infty \frac{dk_{\perp1}^2}{k_{\perp1}^2}
   F_{\rm sl}^{\rm approx}\bigl((k_{\perp 1}\Delta b)^2\bigr) =
   \int_0^\infty \frac{du}{u}
   F_{\rm sl}^{\rm approx}(u) =
   \mbox{an $O(1)$ constant independent of $\Delta b$}
\label {eq:intedge}
\end {equation}
for that edge.

Overall, the total result for double and single logs will have the form
\begin {equation}
  \qhat_\eff(\Delta b) = \qhat_{(0)} + \delta \qhat(\Delta b)
  \simeq \qhat_{(0)}
    \left\{
      1 +
      \frac{\baralphas}{2}
      \left[
         \ln^2\left( \frac{1}{\qhat \tau_0(\Delta b)^2} \right)
         +
         \kappa \ln\left( \frac{1}{\qhat \tau_0(\Delta b)^2} \right)
      \right]
    \right\} ,
\label {eq:qhatLMW2}
\end {equation}
where the single-log coefficient $\kappa$ is some constant%
\footnote{
   For details, see eq.\ (45) of LMW \cite{LMW}, where
   $\underline{x}$ and $l_0$ are our $\Delta b$ and $\tau_0$.
   Divide both sides of
   that equation by $L$ to get $\qhat_\eff$, and use the
   translation (\ref{eq:dbtranslate}) to replace the remaining
   occurrences of $L$ by $1/\qhat (\Delta b)^2$.
   Note that this replaces their $\ln(8m l_0/\underline{x}^2 \qhat L)$
   by a $\Delta b$-independent
   constant of $O(1)$.  The $m l_0$ and the integral in that formula
   arise from the boundary $t_1 \sim \tau_0$ in our
   fig.\ \ref{fig:LMWregiontk}a [what they call ``boundary (c)''].
   Since this boundary does not generate a logarithm with large dependence
   on the exact value of
   $\Delta b\sim{\cal B}_0$, we can ignore it in our analysis.
   We may
   also ignore the various complications in the analysis of this boundary,
   recently investigated by Ghiglieri and Weitz \cite{Jacopo}
   for the case of a quark-gluon plasma.
}
that is independent of $\Delta b$.
Eq.\ (\ref{eq:qhatLMW2}) refines (\ref{eq:qhatLMW}) to now include the single
log term.  This large single logarithm does not
generate any large $\Delta b$ dependence when included
in our earlier discussion of
section \ref{sec:dbDependence}.
That's because we were only interested in $\Delta b \sim {\cal B}_0$
as in (\ref{eq:Bzero}), and one may rewrite the single log
term in (\ref{eq:qhatLMW2}) as
\begin {equation}
  \kappa \ln\left( \frac{1}{\qhat \tau_0(\Delta b)^2} \right)
  =
  \kappa \ln\left( \frac{1}{\qhat \tau_0{\cal B}_0^2} \right)
  - \kappa \ln\left( \frac{(\Delta b)^2}{{\cal B}_0^2} \right) .
\end {equation}
On the right-hand side, the first term is a large logarithm but
does not depend on $\Delta b$, whereas
the second term depends on $\Delta b$ but is not a large logarithm
and so will not need to resummed.

% ---------------------------------------------------------------------------

\subsection{\boldmath$\Delta b$ dependence at NLLO and NNLLO}

Now move to the next order in $\alphas$ by considering the $n{=}2$
case of (\ref{eq:resum1}).  The corresponding leading-log region,
which generates an $O(\alphas^2 \log^4)$ contribution to $\qhat_\eff$,
corresponds to the combination of the shaded regions of
figs.\ \ref{fig:LMWregiontk}a and b.
The leading log is generated by points in the interior, such as the
combined pair AB in the figure.

The combination AC contributes at NLLO, which is $O(\alphas^2 \log^3)$
for $n=2$.
This combination corresponds to
\begin {equation}
   \frac{1}{\Delta b} \gg k_{\perp1} \sim k_{\perp 2} .
\end {equation}
If we continue on to yet higher orders in $\alphas$, the contributions
at NLLO order that involves a pair like AC will have
\begin {equation}
   \frac{1}{\Delta b} \gg k_{\perp1} \sim k_{\perp 2} \gg k_{\perp 3} \gg \cdots .
\end {equation}
None of the points will be sensitive to the exact value of $\Delta b$,
and so none of these contributions contribute to what we're interested
in, which is the $\Delta b$ dependence of resummed $\hat q_\eff$.

Now turn to the combination of figs.\ \ref{fig:LMWregiontk2}a and b,
with $(t_1,k_{\perp1}^2)$ along the edge $k_{\perp1} \sim 1/\Delta b$.
First, note that if $(t_1,k_{\perp1}^2)$ were at the vertex $\gamma$,
then we would lose both logs from the $(t_1,k_{\perp 1}^2)$ integration,
and so this would be a NNLLO contribution instead of an NLLO one.
So, at NLLO, we can replace the upper limit $1/\qhat(\Delta b)^2$
of (\ref{eq:t1limit}) on the $t_1$ integration by $1/\qhat{\cal B}_0^2$
--- a change which will only affect NNLLO.

So we should focus on combinations like DB, which correspond to
NLLO contributions with
\begin {equation}
   \frac{1}{\Delta b} \sim k_{\perp1} \gg k_{\perp 2} \gg k_{\perp 3} \gg \cdots .
\label {eq:scenarioDB}
\end {equation}
None of $(t_2,\omega_2)$, $(t_3,\omega_3)$, ... can be on an edge
because having placed $(t_1,\omega_1)$ on an edge (e.g.\ point D in the
figure) has already cost us a logarithm; having another point also on an
edge would move us from NLLO to NNLLO.
So we may use the leading-log approximation for all the
$(t_i,\omega_i)$ integrals except for $(t_1,\omega_1)$.
For the same reason,
the $k_{\perp 2}$ integration in (\ref{eq:resum1})
does not care about the exact value of
$k_{\perp 1}$ at this order, only its order of magnitude,
and so the upper limit $k_{\perp1}^2$
of integration can be replaced by $1/{\cal B}_0$ since
$k_{\perp1} \sim 1/{\Delta b} \sim 1/{\cal B}_0$ in
(\ref{eq:scenarioDB}).
Altogether, NLLO contributions of type (\ref{eq:scenarioDB})
then contribute
\begin {multline}
  \baralphas^{\,n} \hat q_{(0)}\!
  \int_{\tau_0}^{1/\qhat{\cal B}_0^2} \frac{dt_1}{t_1}
    \int_{k_{\perp 1}\sim 1/\Delta b} \frac{dk_{\perp 1}^2}{k_{\perp 1}^2} \>
    F_{\rm sl}(t_1,k_{\perp 1}^2)\!
  \int_{\tau_0}^{t_1} \frac{dt_2}{t_2}
    \int_{\qhat t_2}^{1/{\cal B}_0} \frac{dk_{\perp 2}^2}{k_{\perp 2}^2}
  \cdots\!\!
  \int_{\tau_0}^{t_{n-1}} \frac{dt_n}{t_n}
    \int_{\qhat t_n}^{k_{\perp,n-1}^2} \frac{dk_{\perp n}^2}{k_{\perp n}^2}
\\
  =
  \baralphas^{\,n} \hat q_{(0)}
  \int_{\tau_0}^{1/\qhat{\cal B}_0^2} \frac{dt_1}{t_1}
    \int_{k_{\perp 1}\sim 1/\Delta b} \frac{dk_{\perp 1}^2}{k_{\perp 1}^2} \>
    F_{\rm sl}(t_1,k_{\perp 1}^2)\,
    f_{n-1}\kern-1.5pt\Bigl(
      \ln\bigl(\tfrac{t_1}{\tau_0}\bigr),
      \ln\bigl(\tfrac{1}{\qhat\tau_0{\cal B}_0^2}\bigr)
    \Bigr)
\label {eq:DBscenario}
\end {multline}
to $\delta^n\qhat_\eff$ at NLLO.
$f_n$ is again defined by (\ref{eq:Yrho}) and (\ref{eq:resum3f}).
The $k_{\perp1}^2$ integral in (\ref{eq:DBscenario})
is the one presented in (\ref{eq:intedge})
and so is independent of $\Delta b$ (at this order in logs).
Since there is no other $\Delta b$ in (\ref{eq:DBscenario}), we see
that NLLO contributions from combinations like
DB are independent of $\Delta b$.

For a combination like DE in fig.\ \ref{fig:LMWregiontk2}, E would
be sensitive to $\Delta b$ since
$1/\Delta b \sim k_{\perp 1} \sim k_{\perp 2}$.  But this is an NNLLO
contribution since both points are on edges.

We've now addressed the interesting cases.  We conclude that
NLLO does not generate any $\Delta b$ dependence not already included
in the LLO result (\ref{eq:resum1}), which sums to the
formula (\ref{eq:LMWresum}) used in the main text.
Our analysis above suggests
that additional $\Delta b$ dependence will appear at NNLLO,
but that is beyond the scope of what is needed for this paper.

% --------------------------------------------------------------------------

\subsection{A loose end: the prefactor of eq.\ (\ref{eq:qeff1})}
\label{app:prefactor}

In the main text, we ignored a prefactor when discussing the
$\Delta b$ dependence of the leading-log resummation.
The leading term in the large-argument expansion of $I_1$
in (\ref{eq:LMWresum}) actually gives
\begin {equation}
  \qhat_\eff(\Delta b)
  \approx
  \qhat_{(0)}
  \left( \frac{1}{\qhat \tau_0(\Delta b)^2} \right)^{\!2\sqrt{\baralphas}}
  \times
  \frac{1}{\sqrt{4\pi}}
  \left[ \sqrt{\baralphas}
      \ln\bigl( \frac{1}{\qhat \tau_0(\Delta b)^2} \bigr) \right]^{-3/2}
\label {eq:qeff1b}
\end {equation}
instead of (\ref{eq:qeff1}).  Including the full prefactor then changes
(\ref{eq:qeff2}) and (\ref{eq:sqrtExpansion}) to
\begin {multline}
  \qhat_\eff(\Delta b)
  \approx \qhat_{(0)}
  \left( \frac{1}{\qhat \tau_0\Bzero^2} \right)^{\!2\sqrt{\alphas}}
  \left[
     1
     - 2\sqrt{\baralphas}
        \ln\left( \frac{(\Delta b)^2}{\Bzero^2} \right)
  \right]
\\
  \times
  \frac{1}{\sqrt{4\pi}}
  \left[ \sqrt{\bar\alphas}
      \ln\Bigl( \frac{1}{\qhat \tau_0\Bzero^2} \Bigr) \right]^{-3/2}
  \left[
     1
     - \frac{ 3 \ln\bigl( (\Delta b)^2/\Bzero^2 \bigr) }
            { 2 \ln( 1/\qhat \tau_0\Bzero^2) }
  \right]
\label {eq:qeff2b}
\end {multline}
and
\begin {equation}
  \qhat_\eff(\Delta b)
  =
  \qhat_\eff(\Bzero) \,
  \left\{
    1 + O(\sqrt{\alphas}\,)
      + O\Bigl(
           \frac{1}{ \ln(1/\qhat \tau_0\Bzero^2) }
        \Bigr)
  \right\} .
\label {eq:sqrtExpansionb}
\end {equation}
Now remember that, when making the large-argument expansion of $I_1$ in
(\ref{eq:qeff1}), we were taking the large-logarithm limit where
$\alphas \ln^2\bigl( 1/\qhat \tau_0(\Delta b)^2 \bigr) \sim
 \alphas \ln^2( 1/\qhat \tau_0\Bzero^2 )$
is $\gg 1$.
So the $O(1/\log)$ term in (\ref{eq:sqrtExpansionb}) can be ignored
compared to the $O(\sqrt{\alphas}\,)$ term, leaving
us with (\ref{eq:sqrtExpansion}).

% ===========================================================================

\section{Asymptotic behavior of \boldmath$\hat\eps_\LO(\hat z)$}
\label {app:asymp}

In this appendix, we will derive the asymptotic fall-off of the
energy stopping distribution $\eps_\LO(z)$
for large $z$.  We follow a procedure similar to that used
in ref.\ \cite{qedNfstop} for the fall-off of the leading-order
charge distribution $\rho_\LO(z)$ at large $z$.%
\footnote{
  Specifically, see appendix B of ref.\ \cite{qedNfstop}.
}
In that case, the conclusion was that
\begin {equation}
   \rho_\LO(z) \sim e^{-\Gamma_\LO(E_0) \, z}
\label {eq:rhoLO}
\end {equation}
for large $z$, where $\Gamma_\LO$ is the total leading-order rate for
the relevant splitting process $e\to e\gamma$.
In our case, however, the total rate for $g{\to}gg$ in $\qhat$ approximation
is infinite because of the $x^{-3/2}$ [or symmetrically $(1{-}x)^{-3/2}$]
IR divergence of eq.\ (\ref{eq:LOrate0})
for $[d\Gamma/dx]^\LO$, and so (\ref{eq:rhoLO}) suggests that the
fall-off of our $\eps_\LO(z)$ must be faster than simple exponential decay.
We'll find that our large-$z$ tail is approximately Gaussian.

Start from the leading-order energy deposition equation (\ref{eq:epseqLO}):
\begin {equation}
  \frac{\partial\hat\eps_\LO(\hat z)}{\partial\hat z}
  =
  \int_0^1 dx \> x \biggl[\frac{d\hat\Gamma}{dx}\biggr]^\LO
  \bigl\{ x^{-1/2}\,\hat\eps_\LO(x^{-1/2}\hat z)
          - \hat\eps_\LO(\hat z) \bigr\} .
\label {eq:epseqLO9}
\end {equation}
Note that the $x{\to}0$ contribution to the integration converges
because (i) $x [d\hat\Gamma/dx]^\LO \sim x^{-1/2}$ and
(ii) $\eps_\LO(z')$ should fall to zero faster than, for example,
$(z')^{-1/2}$ as $z'\to\infty$.
The $x{\to}1$ contribution to the integration converges because
(i) $x [d\hat\Gamma/dx]^\LO \sim (1{-}x)^{-3/2}$ and
(ii) there is a cancellation between the two terms inside the braces:
\begin {equation}
  \bigl\{ x^{-1/2}\,\hat\eps_\LO(x^{-1/2}\hat z)
          - \hat\eps_\LO(\hat z) \bigr\}
  \sim 1{-}x ~\mbox{as $x \to 1$.}
\label {eq:x1cancel}
\end {equation}

Now rewrite $\hat\eps_\LO(\hat z)$ in the WKB-inspired form
\begin {equation}
   \hat\eps_\LO(\hat z) \equiv e^{-\calW(\hat z)} ,
\label {eq:WKB}
\end {equation}
where, asymptotically, $\calW(\hat z)$ should be an increasing function
of $\hat z$ so that $\eps_\LO(z) \to 0$ as $z \to \infty$.
Plugging (\ref{eq:WKB}) into the leading-order energy deposition equation
(\ref{eq:epseqLO9}) gives
\begin {equation}
  \calW^{\,\prime}(\hat z) =
  \int_0^1 dx \> x \biggl[\frac{d\hat\Gamma}{dx} \biggr]^\LO \,
  \bigl\{ 1 - x^{-1/2} e^{\calW(\hat z)-\calW(x^{-1/2}\hat z)} \bigr\} .
\label {eq:Weq1}
\end {equation}
Let's more carefully examine the cancellation (\ref{eq:x1cancel}) as
$x {\to} 1$, now in the language of (\ref{eq:Weq1}).
For this limit, we define $\delta \equiv 1{-}x \ll 1$, which gives
\begin {equation}
  \calW(\hat z) - \calW(x^{-1/2}\hat z) \simeq
  - \tfrac12 \hat z \, \calW^{\,\prime}(\hat z) \, \delta
\end {equation}
and so
\begin {equation}
  \bigl\{ 1 - x^{-1/2} e^{\calW(\hat z)-\calW(x^{-1/2}\hat z)} \bigr\} 
  \simeq 1 - (1{-}\delta)^{-1/2} e^{- \frac12 \hat z \, \calW^{\,\prime}(\hat z) \, \delta}
  .
\label {eq:Wstep0}
\end {equation}
$\hat z\calW^{\,\prime}(\hat z)$ will be large for large $\hat z$.
There are then two regions of small $\delta$ to consider.
For $x$ extremely close to 1, such that
\begin {equation}
  \delta \ll \frac{1}{\hat z\,\calW^{\,\prime}(\hat z)} \, \ll 1 ,
\label {eq:Wregion0}
\end {equation}
(\ref{eq:Wstep0}) gives
\begin {equation}
  \bigl\{ 1 - x^{-1/2} e^{\calW(\hat z)-\calW(x^{-1/2}\hat z)} \bigr\} 
  \simeq \tfrac12 \bigl[ \hat z \calW^{\,\prime}(\hat z) - 1 \bigr] \delta ,
\label {eq:Wstep1}
\end {equation}
which vanishes linearly as $\delta \to 0$ and describes the
cancellation (\ref{eq:x1cancel}).
In contrast, in the other small-$\delta$ region
\begin {equation}
  \frac{1}{\hat z\,\calW^{\,\prime}(\hat z)} \ll \delta \ll 1 ,
\label {eq:Wregion}
\end {equation}
where $x$ is close but not arbitrarily close to 1,
the exponential term in (\ref{eq:Wstep0}) will be suppressed, so that
\begin {equation}
  \bigl\{ 1 - x^{-1/2} e^{\calW(\hat z)-\calW(x^{-1/2}\hat z)} \bigr\} 
  \simeq 1 .
\end {equation}
That means that the $\delta^{-3/2}$ divergence of
$x [d\hat\Gamma/dx]^\LO$ will not be moderated
in the integration region (\ref{eq:Wregion}), and so
(when $\hat z$ is large) the
integral in (\ref{eq:Weq1}) is dominated
\begin {equation}
  \delta \sim \frac{1}{ \hat z \, \calW^{\,\prime}(\hat z) } \ll 1 ,
\end {equation}
which is the transition between the
lower end of region (\ref{eq:Wregion}) and region (\ref{eq:Wregion0}).
We may therefore
approximate the {\it full}\/ integral (\ref{eq:Weq1})
by approximating $\delta \ll 1$ in the integrand, which corresponds to
the approximation (\ref{eq:Wstep0}).
It's convenient to use that $\delta \ll 1$ approximation to
also rewrite
\begin {equation}
  (1-\delta)^{-1/2} \simeq e^{\delta/2},
  \qquad
  x \biggl[\frac{d\hat\Gamma}{dx} \biggr]^\LO \simeq
    \frac{1}{\pi \delta^{3/2}} ,
\end {equation}
and so (\ref{eq:Weq1}) becomes
\begin {equation}
  \calW^{\,\prime}(\hat z) \simeq
  \int_0^\infty \frac{d\delta}{\pi\delta^{3/2}} \,
  \bigl\{ 1 - e^{-\tfrac12 [ \hat z \, \calW^{\,\prime}(\hat z) - 1 ] \delta} \bigr\} .
\label {eq:Weq2}
\end {equation}
Note that we've replaced the upper limit of integration by $\infty$,
which introduces negligible relative error in the large-$\hat z$
limit for the same reason that $\delta \ll 1$ dominated over $\delta \sim 1$.
The integral gives
\begin {equation}
  \calW^{\,\prime}(\hat z) \simeq
  \sqrt{
     \tfrac{2}{\pi}
     \bigl[ \hat z \, \calW^{\,\prime}(\hat z) - 1 \bigr]
  } .
\label {eq:Weq3}
\end {equation}

Before solving (\ref{eq:Weq3}), we can simplify a bit by
again remembering our expectation that
$\hat z\,\calW'(\hat z) \gg 1$ in the large $\hat z$
limit, so that (\ref{eq:Weq3}) becomes
\begin {equation}
  \calW^{\,\prime}(\hat z) \simeq
  \sqrt{
     \tfrac{2}{\pi} \,
     \hat z \, \calW^{\,\prime}(\hat z)
  } .
\label {eq:Weq4}
\end {equation}
Solving for $\cal W$ gives
\begin {equation}
  \calW(\hat z) \simeq \frac{\hat z^2}{\pi}
\end {equation}
at large $\hat z$, which is equivalent to the asymptotic behavior
quoted in (\ref{eq:zasymp}):
\begin {equation}
   \hat\eps_\LO(\hat z) \sim e^{-\hat z^2/\pi} .
\label {eq:zasymp2}
\end {equation}

There is a short-cut that we might have taken to determine (\ref{eq:zasymp2}).
Once we had completed enough of the argument to realize that the calculation
of ${\cal W}(\hat z)$ would be dominated by $\delta \ll 1$, we could
have replaced $[d\Gamma/dx]^\LO$ by the BIM \cite{BIM1} model rate
(\ref{eq:BIMrate}), which agrees
with $[d\Gamma/dx]^\LO$ in the limits $x\to0$ and $x\to1$.
Then we could have extracted (\ref{eq:zasymp2}) from the energy
deposition distribution (\ref{eq:BIMeps}) of the BIM model.

With some work, one could refine our leading large-$\hat z$ approximation
to $\calW$ to compute $O(\hat z)$ corrections to the exponent in
(\ref{eq:zasymp2}) and even further to find
power-law prefactors to the exponential.%
\footnote{
  We do not expect these corrections to be the same as the BIM model
  result (\ref{eq:BIMeps}).
}
However, we find in practice that (\ref{eq:zasymp2}) by itself is
adequate to get good numerical convergence of our results in the
large-$\hat z_\max$ limit.

% ===========================================================================

\section{\boldmath$\eps_\LO(\hat z)$ in the BIM model}
\label {app:BIM}

Using the formula $P_{g{\to}gg}(x) = 2 \CA (1 - x + x^2)^2/x(1-x)$
for the DGLAP splitting function, the LO splitting rate
(\ref{eq:LOrate0}) can be rewritten as
\begin {equation}
  \left[ \frac{d\Gamma}{dx} \right]^\LO
  =
    \frac{\CA\alphas(1{-}x{+}x^2)^{5/2}}{\pi[x(1{-}x)]^{3/2}}
    \sqrt{\frac{\qhatA}{E}} \,.
\label {eq:LOrate9}
\end {equation}
Blaizot, Iancu, and Mehtar-Tani (BIM) \cite{BIM1} realized that if
one replaces the leading-order splitting rate (\ref{eq:LOrate9}) by the
simpler function
\begin {equation}
  \left[ \frac{d\Gamma}{dx} \right]_\BIM
  =
    \frac{\CA\alphas}{\pi[x(1{-}x)]^{3/2}}
    \sqrt{\frac{\qhatA}{E}} \,,
\label {eq:BIMrate}
\end {equation}
then it is possible to solve leading-order shower development
analytically.  We will refer to this as the BIM model of shower
development.  The BIM rate (\ref{eq:BIMrate}) is equal to the
actual LO rate in the limit that
one of the two daughters is soft, i.e.\ $x(1{-}x)\ll 1$.
But for perfectly democratic splitting $x=0.5$, the BIM rate overestimates
the LO BDMPS-Z rate by a factor of $(4/3)^{5/2} \simeq 2$.
In our notation, their analytic solution for the time development of
the gluon density in $x$ is
\begin {equation}
  \hat n_\BIM(x,\hat t\,) =
  \frac{ \hat t \, e^{-\hat t^{\kern1pt 2}/\pi(1{-}x)} }{ \pi[x(1{-}x)]^{3/2} }
  \quad{\rm for}~ x>0 ,
\label {eq:NBIM}
\end {equation}
with $\hat t \equiv t/\ell_0$,
and $\ell_0$ defined by (\ref{eq:ell0}).

In general, the energy which is still moving ($x > 0$) at time $t$ is
\begin {equation}
  E_{\rm moving}(t) = \int_{0^+}^1 dx \> x E_0 \, n(x,E_0,t) .
\end {equation}
The moving energy decreases at the rate that energy is deposited into
the medium, and so
\begin {equation}
  \eps(z) = - \frac{dE_{\rm moving}}{dt} \biggl|_{t=z}
  = - \left[ \frac{d}{dt} \int_{0^+}^1 dx \> x E_0 \, n(x,E_0,t) \right]_{t=z} .
\end {equation}
Switching to dimensionless variables (\ref{eq:hatvars}) and plugging
in the BIM solution (\ref{eq:NBIM}) yields%
\footnote{
  One way to do the $x$ integral is to switch integration variable
  to $u \equiv \sqrt{x/(1{-}x)}$, which leads to a simple
  Gaussian integral in $u$.
}
\begin {equation}
  \hat\eps(\hat z) = -\frac{d}{d\hat z} \, e^{-\hat z^2/\pi}
  = \frac{2\hat z}{\pi} \, e^{-\hat z^2/\pi} .
\label {eq:BIMeps}
\end {equation}
The corresponding stopping distance is
\begin {equation}
  \hat\ell_\stop^{\,\BIM} = \langle\hat z\rangle_\BIM = \frac{\pi}{2} ,
\end {equation}
and the shape function (\ref{eq:shape}) is then
\begin {equation}
  S_\BIM(Z) =
  \frac{\pi Z}{2} \, e^{-\pi Z^2/4} .
\label {eq:SBIM}
\end {equation}

The BIM stopping distance $\langle \hat z \rangle_\BIM \simeq 1.571$ is
shorter than the LO stopping distance
$\langle \hat z \rangle_\LO \simeq 2.1143$ of table \ref{tab:moments}
because the BIM rate (\ref{eq:BIMrate}) overestimates the splitting
rate for democratic splittings.
Other moments of the BIM energy stopping distribution are
\begin {equation}
  \langle \hat z^n \rangle_\BIM =
  \pi^{n/2} \, \Gamma\bigl(1 + \tfrac{n}{2}\bigr) .
\end {equation}

% ===========================================================================

\section {Energy conservation for eq.\ (\ref{eq:Devolve})}
\label{app:energy}

To see that the evolution equation (\ref{eq:Devolve}) for
$D(\zeta,E_0,t)$ conserves energy, integrate both sides of the
equation over $\zeta$ and then switch the order of integration
on the right-hand side to get
\begin {multline}
  \frac{dE_{\rm total}}{dt}
  =
  \int_0^1 dx
  \int_0^1 d\zeta \>
  \biggl\{
    \theta(x>\zeta)
    \left[
       \frac{d\Gamma}{dx} \bigl(\tfrac{\zeta E_0}{x},x\bigr)
     \right]_{\rm net}
     D\bigl( \tfrac{\zeta}{x}, E_0, t \bigr)
\\
    -
    x \left[ \frac{d\Gamma}{dx} (\zeta E_0,x) \right]_{\rm net}
    D(\zeta,E_0,t)
  \biggr\} .
\label {eq:Econserve}
\end {multline}
The $\zeta$ integral of the first term can be rewritten as
\begin {equation}
   \int_0^x d\zeta 
    \left[
       \frac{d\Gamma}{dx} \bigl(\tfrac{\zeta E_0}{x},x\bigr)
     \right]_{\rm net}
     D\bigl( \tfrac{\zeta}{x}, E_0, t \bigr)
  =
   \int_0^1 d\zeta'
    x
    \left[
       \frac{d\Gamma}{dx} (\zeta' E_0,x)
     \right]_{\rm net}
     D( \zeta', E_0, t) ,
\end {equation}
where $\zeta' \equiv \zeta/x$.
The first term of (\ref{eq:Econserve}) then cancels the second term, giving
$dE_{\rm total}/dt = 0$.

%%%%%%%%%%%%%%%%%%%%%%%%%%%%%%%%%%%%%%%%%%%%%%%%%%%%%%%%%%%%%%%%%%%%%%%%%%%%%%%

%%%%%%%%%%%%%%%%%%%%%%%%%%%%%%%%%%%%%%%%%%%%%%%%%%%%%%%%%%%%%%%%%%%%%%%%%%%%%%%

\begin{thebibliography}{}

\bibitem{LP1}
  L.~D.~Landau and I.~Pomeranchuk,
  ``Limits of applicability of the theory of bremsstrahlung electrons and
  pair production at high-energies,''
  Dokl.\ Akad.\ Nauk Ser.\ Fiz.\  {\bf 92} (1953) 535.

\bibitem{LP2}
  L.~D.~Landau and I.~Pomeranchuk,
  ``Electron cascade process at very high energies,''
  Dokl.\ Akad.\ Nauk Ser.\ Fiz.\  {\bf 92} (1953) 735.

\bibitem{Migdal}
  A.~B.~Migdal,
  ``Bremsstrahlung and pair production in condensed media at high-energies,''
   Phys.\ Rev.\  {\bf 103}, 1811 (1956);

\bibitem{LPenglish}
  L. Landau,
  {\sl The Collected Papers of L.D. Landau}\/
  (Pergamon Press, New York, 1965).

\bibitem{Blaizot}
  J.~P.~Blaizot and Y.~Mehtar-Tani,
  ``Renormalization of the jet-quenching parameter,''
  Nucl.\ Phys.\ A {\bf 929}, 202 (2014)
  [arXiv:1403.2323 [hep-ph]].
  %%CITATION = ARXIV:1403.2323;%%

\bibitem{Iancu}
  E.~Iancu,
  ``The non-linear evolution of jet quenching,''
  JHEP \textbf{10}, 95 (2014)
  [arXiv:1403.1996 [hep-ph]].
  %%CITATION = ARXIV:1403.1996;%%

\bibitem{Wu}
  B.~Wu,
  ``Radiative energy loss and radiative $p_{\bot}$-broadening of
    high-energy partons in QCD matter,''
  JHEP \textbf{12}, 081 (2014)
  [arXiv:1408.5459 [hep-ph]].
  %%CITATION = ARXIV:1408.5459;%%

\bibitem{finale}
  P.~Arnold, O.~Elgedawy and S.~Iqbal,
  ``Are gluon showers inside a quark-gluon plasma strongly coupled?
    a theorist's test,''
  [arXiv:2212.08086 [hep-ph]].

\bibitem{2brem}
  P.~Arnold and S.~Iqbal,
  ``The LPM effect in sequential bremsstrahlung,''
  JHEP \textbf{04}, 070 (2015)
  [{\it erratum} JHEP \textbf{09}, 072 (2016)]
  %doi:10.1007/JHEP09(2016)072, 10.1007/JHEP04(2015)070
  [arXiv:1501.04964 [hep-ph]].
  %%CITATION = doi:10.1007/JHEP09(2016)072, 10.1007/JHEP04(2015)070;%%
  %9 citations counted in INSPIRE as of 20 Sep 2016

\bibitem{seq}
  P.~Arnold, H.~C.~Chang and S.~Iqbal,
  ``The LPM effect in sequential bremsstrahlung 2: factorization,''
  JHEP \textbf{09}, 078 (2016)
  [arXiv:1605.07624 [hep-ph]].
  %%CITATION = ARXIV:1605.07624;%%

\bibitem{dimreg}
  P.~Arnold, H.~C.~Chang and S.~Iqbal,
  ``The LPM effect in sequential bremsstrahlung: dimensional regularization,''
  JHEP \textbf{10}, 100 (2016)
  %doi:10.1007/JHEP10(2016)100
  [arXiv:1606.08853 [hep-ph]].
  %%CITATION = ARXIV:1606.08853;%%
  %10 citations counted in INSPIRE as of 25 May 2020

\bibitem{4point}
  P.~Arnold, H.~C.~Chang and S.~Iqbal,
  ``The LPM effect in sequential bremsstrahlung: 4-gluon vertices,''
  JHEP \textbf{10}, 124 (2016)
  %doi:10.1007/JHEP10(2016)124
  [arXiv:1608.05718 [hep-ph]].
  %%CITATION = doi:10.1007/JHEP10(2016)124;%%
  %4 citations counted in INSPIRE as of 10 Jan 2018

\bibitem{QEDnf}
  P.~Arnold and S.~Iqbal,
  ``In-medium loop corrections and longitudinally polarized gauge bosons
    in high-energy showers,''
  JHEP \textbf{12}, 120 (2018)
  %doi:10.1007/JHEP12(2018)120
  [arXiv:1806.08796 [hep-ph]].
  %%CITATION = doi:10.1007/JHEP12(2018)120;%%

\bibitem{qcd}
  P.~Arnold, T.~Gorda and S.~Iqbal,
  ``The LPM effect in sequential bremsstrahlung:
    nearly complete results for QCD,''
  JHEP \textbf{11}, 053 (2020)
  [{\it erratum} JHEP \textbf{05}, 114 (2022)]
  %doi:10.1007/JHEP11(2020)053, 10.1007/JHEP05(2022)114
  [arXiv:2007.15018 [hep-ph]].

\bibitem{qcdI}
  P.~Arnold, T.~Gorda and S.~Iqbal,
  ``The LPM effect in sequential bremsstrahlung:
    incorporation of ''instantaneous'' interactions for QCD,''
  [arXiv:2209.03971 [hep-ph]].

\bibitem{DeepLPM}
  P.~B.~Arnold and C.~Dogan,
  ``QCD Splitting/Joining Functions at Finite Temperature in
    the Deep LPM Regime,''
  Phys. Rev. D \textbf{78}, 065008 (2008)
  %doi:10.1103/PhysRevD.78.065008
  [arXiv:0804.3359 [hep-ph]].

\bibitem{BDMPS3}
  R.~Baier, Y.~L.~Dokshitzer, A.~H.~Mueller, S.~Peigne and D.~Schiff,
  ``Radiative energy loss and $p_\perp$-broadening of high energy partons in
    nuclei,''
  {\it ibid.}\ {\bf 484} (1997)
  [arXiv:hep-ph/9608322].
  %%CITATION = NUPHA,B484,265;%%

\bibitem{Peshier}
  A.~Peshier,
  ``QCD running coupling and collisional jet quenching,''
  J. Phys. G \textbf{35}, 044028 (2008).
  %doi:10.1088/0954-3899/35/4/044028

\bibitem{BIM1}
  J.~P.~Blaizot, E.~Iancu and Y.~Mehtar-Tani,
  ``Medium-induced QCD cascade: democratic branching and wave turbulence,''
  Phys.\ Rev.\ Lett.\  {\bf 111}, 052001 (2013)
  %doi:10.1103/PhysRevLett.111.052001
  [arXiv:1301.6102 [hep-ph]].
  %%CITATION = doi:10.1103/PhysRevLett.111.052001;%%
  %75 citations counted in INSPIRE as of 13 Sep 2018

\bibitem{BIM2}
  J.~P.~Blaizot and Y.~Mehtar-Tani,
  ``Energy flow along the medium-induced parton cascade,''
  Annals Phys.\  {\bf 368}, 148 (2016)
  %doi:10.1016/j.aop.2016.01.002
  [arXiv:1501.03443 [hep-ph]].
  %%CITATION = doi:10.1016/j.aop.2016.01.002;%%
  %15 citations counted in INSPIRE as of 13 Sep 2018

\bibitem{BDMPS1}
  R.~Baier, Y.~L.~Dokshitzer, A.~H.~Mueller, S.~Peigne and D.~Schiff,
  ``The Landau-Pomeranchuk-Migdal effect in QED,''
  Nucl.\ Phys.\  B {\bf 478}, 577 (1996)
  [arXiv:hep-ph/9604327];

\bibitem{qedNfstop}
  P.~Arnold, S.~Iqbal and T.~Rase,
  ``Strong- vs. weak-coupling pictures of jet quenching: a dry run using QED,''
  JHEP \textbf{05}, 004 (2019)
  %doi:10.1007/JHEP05(2019)004
  [arXiv:1810.06578 [hep-ph]].

\bibitem{BDMPS2}
  R.~Baier, Y.~L.~Dokshitzer, A.~H.~Mueller, S.~Peigne and D.~Schiff,
  ``Radiative energy loss of high-energy quarks and gluons in a
    finite volume quark - gluon plasma,''
  Nucl.\ Phys.\  B {\bf 483}, 291 (1997) [arXiv:hep-ph/9607355].
  %%CITATION = NUPHA,B483,291;%%

\bibitem{Zakharov1}
 B.~G.~Zakharov,
 ``Fully quantum treatment of the Landau-Pomeranchuk-Migdal effect in
   QED and QCD,''
 JETP Lett.\  {\bf 63}, 952 (1996)
 [arXiv:hep-ph/9607440].

\bibitem{Zakharov2}
 B.~G.~Zakharov,
 ``Radiative energy loss of high-energy quarks in finite size nuclear matter
   and quark-gluon plasma,''
 JETP Lett.\  {\bf 65}, 615 (1997)
 [Pisma Zh.\ Eksp.\ Teor.\ Fiz.\  {\bf 63}, 952 (1996)]
 [arXiv:hep-ph/9607440].
 %%CITATION = JTPLA,63,952.%%

\bibitem{BDMS}
  R.~Baier, Y.~L.~Dokshitzer, A.~H.~Mueller and D.~Schiff,
  ``Medium induced radiative energy loss:
    Equivalence between the BDMPS and Zakharov formalisms,''
  Nucl. Phys. B \textbf{531}, 403-425 (1998)
  %doi:10.1016/S0550-3213(98)00546-X
  [arXiv:hep-ph/9804212 [hep-ph]].

\bibitem{Zakharov3}
  B.~G.~Zakharov,
  ``Light cone path integral approach to the Landau-Pomeranchuk-Migdal effect,''
  Phys. Atom. Nucl. \textbf{61}, 838-854 (1998)
  [arXiv:hep-ph/9807540 [hep-ph]].

\bibitem{simple}
  P.~B.~Arnold,
  ``Simple Formula for High-Energy Gluon Bremsstrahlung in a Finite,
    Expanding Medium,''
  Phys. Rev. D \textbf{79}, 065025 (2009)
  %doi:10.1103/PhysRevD.79.065025
  [arXiv:0808.2767 [hep-ph]].

\bibitem{logs2}
  P.~Arnold,
  ``Universality (beyond leading log) of soft radiative corrections
    to $ \hat{q} $ in p$_\perp$ broadening and energy loss,''
  JHEP \textbf{03}, 134 (2022)
  %doi:10.1007/JHEP03(2022)134
  [arXiv:2111.05348 [hep-ph]].

\bibitem{logs}
  P.~Arnold, T.~Gorda and S.~Iqbal,
  ``The LPM effect in sequential bremsstrahlung:
    analytic results for sub-leading (single) logarithms,''
  JHEP \textbf{04}, 085 (2022)
  %doi:10.1007/JHEP04(2022)085
  [arXiv:2112.05161 [hep-ph]].

\bibitem{LMW}
  T.~Liou, A.~H.~Mueller and B.~Wu,
  ``Radiative $p_\bot$-broadening of high-energy quarks and gluons in
    QCD matter,''
  Nucl.\ Phys.\ A {\bf 916}, 102 (2013)
  [arXiv:1304.7677 [hep-ph]].
  %%CITATION = ARXIV:1304.7677;%%

\bibitem{run1}
  E.~Iancu and D.~N.~Triantafyllopoulos,
  ``Running coupling effects in the evolution of jet quenching,''
  Phys. Rev. D \textbf{90}, no.7, 074002 (2014)
  %doi:10.1103/PhysRevD.90.074002
  [arXiv:1405.3525 [hep-ph]].

\bibitem{run2a}
  P.~Caucal and Y.~Mehtar-Tani,
  ``Anomalous diffusion in QCD matter,''
  Phys. Rev. D \textbf{106}, no.5, L051501 (2022)
  %doi:10.1103/PhysRevD.106.L051501
  [arXiv:2109.12041 [hep-ph]].

\bibitem{run2b}
  P.~Caucal and Y.~Mehtar-Tani,
  ``Universality aspects of quantum corrections to transverse momentum
    broadening in QCD media,''
  JHEP \textbf{09}, 023 (2022)
  %doi:10.1007/JHEP09(2022)023
  [arXiv:2203.09407 [hep-ph]].

\bibitem{stop}
  P.~B.~Arnold, S.~Cantrell and W.~Xiao,
  ``Stopping distance for high energy jets in weakly-coupled
    quark-gluon plasmas,''
  Phys.\ Rev.\ D {\bf 81}, 045017 (2010)
  %doi:10.1103/PhysRevD.81.045017
  [arXiv:0912.3862 [hep-ph]].

\bibitem{Mathematica}
  Wolfram Research, Inc., Mathematica (various versions),
  Champaign, IL (2018--2021)

\bibitem{Jacopo}
  J.~Ghiglieri and E.~Weitz,
  ``Classical vs quantum corrections to jet broadening in a
    weakly-coupled Quark-Gluon Plasma,''
  JHEP \textbf{11}, 068 (2022)
  %doi:10.1007/JHEP11(2022)068
  [arXiv:2207.08842 [hep-ph]].

\end{thebibliography}
\end {document}